\documentclass{aa}  

\usepackage{natbib}
\bibpunct{(}{)}{;}{a}{}{,} 
\usepackage{graphicx}
\usepackage{multirow}
\usepackage{amsmath}
\usepackage{newtxtext,newtxmath}
\usepackage{lscape}
\usepackage{subcaption}
\usepackage{gensymb}
\usepackage{placeins}
\usepackage{tabularx}

\usepackage{float}

\usepackage[separate-uncertainty=true,multi-part-units=single]{siunitx}
\DeclareSIUnit[]{\angstrom}{\textup{\AA}}
\DeclareSIUnit[]{\Rsun}{\text{\ensuremath{R_{\sun}}}}
\DeclareSIUnit[]{\Msun}{\text{\ensuremath{M_{\sun}}}}
\DeclareSIUnit[]{\parsec}{pc}

\usepackage{pgfplots}
\usepackage{color}
\usepackage[colorinlistoftodos,prependcaption,textsize=tiny,color=red]{todonotes}

\usepackage[]{hyperref}
\hypersetup{
  colorlinks   = true, 
  urlcolor     = blue, 
  linkcolor    = blue, 
  citecolor   = blue 
}

\begin{document}

   \title{Now you see me, now you don't\dots}
   \subtitle{Deep, long-lived obscuration events of white dwarfs\\ by dust and debris from disintegrating planetesimals}

   \author{J.~van~Roestel \inst{\ref{ista},\ref{uva}}\thanks{email: jcjvanroestel.astro@gmail.com} 
        \and S.~Bhattacharjee \inst{\ref{caltech}}
        \and J.A.~Guidry \inst{\ref{BU}}
        \and I.~Caiazzo \inst{\ref{ista}}
        \and T.~Cunningham \inst{\ref{cfa}}
        \and J.J.~Hermes \inst{\ref{BU}}
        \and S.~Torres \inst{\ref{tenerife}}
        \and Z.~Vanderbosch \inst{\ref{caltech}}
        \and A.J.~Drake \inst{\ref{caltech}}
        \and M.~Hollands \inst{\ref{warwick}}
        \and B.T.~Gaensicke \inst{\ref{warwick}}
        \and K.~De \inst{\ref{columbia}}
        \and T.~Prince \inst{\ref{caltech}}
        \and A.C.~Rodriguez \inst{\ref{caltech}}
        \and L.B.~Ould~Rouis \inst{\ref{BU}}
        \and \\ A.A.~Mahabal \inst{\ref{caltech},\ref{CD3}}
        \and E.C.~Bellm \inst{\ref{DIRAC}} 
        \and M.W.~Coughlin \inst{\ref{UM}} 
        \and M.J.~Graham \inst{\ref{caltech}} 
        \and D.~Hale \inst{\ref{coo}} 
        \and G.~Helou \inst{\ref{ipac}} 
        \and J.C.~Jaimes \inst{\ref{caltech}} %
        \and M.~Kasliwal \inst{\ref{caltech}} 
        \and R.R.~Laher \inst{\ref{ipac}} 
        \and R.L.~Reed \inst{\ref{coo}} 
                  }
   \institute{
   Institute of Science and Technology Austria, Am Campus 1, 3400 Klosterneuburg, Austria\label{ista} 
   \and Anton Pannekoek Institute for Astronomy, University of Amsterdam, 1090 GE Amsterdam, The Netherlands\label{uva}
   \and Cahill Center for Astrophysics, California Institute of Technology, Pasadena, CA 91125, USA\label{caltech}
   \and Department of Astronomy \& Institute for Astrophysical Research, Boston University, Boston, MA 02215, USA\label{BU}
   \and Center for Astrophysics, Harvard \& Smithsonian, 60 Garden St., Cambridge, MA 02138, USA\label{cfa}
   \and Instituto de Astrofisica de Canarias, E-38205 La Laguna, Tenerife, Spain\label{tenerife}
   \and Department of Physics, University of Warwick, Coventry, CV4 7AL, United Kingdom\label{warwick}
   \and Columbia University, 538 West 120th Street 704, MC 5255, New York, NY 10027, USA\label{columbia}
   \and Center for Data Driven Discovery, California Institute of Technology, Pasadena, CA 91125, USA \label{CD3}
   \and DIRAC Institute, Department of Astronomy, University of Washington, 3910 15th Avenue NE, Seattle, WA 98195, USA\label{DIRAC}
   \and School of Physics and Astronomy, University of Minnesota, Minneapolis, MN 55455, USA\label{UM}
   \and IPAC, California Institute of Technology, 1200 E. California Blvd, Pasadena, CA 91125, USA\label{ipac}
   \and Caltech Optical Observatories, California Institute of Technology, Pasadena, CA 91125, USA\label{coo}
   }
   \date{Received September 8, 2026; accepted XX XX, XXXX}

\titlerunning{White dwarfs with disintegrating planetesimals}

  \abstract
  {
  Transiting dust and debris around white dwarfs provide a direct view of planetary disruption. We report six previously non-variable white dwarfs that underwent single, deep obscuration events, including ZTF~J2327+0019, which fully disappeared for more than a year. These discoveries raise the number of confirmed debris-transit white dwarfs to 20. In addition, the systematic search of negative alerts from the Zwicky Transient Facility (ZTF), followed by photometric and spectroscopic observations, also identified three hot ($T_\mathrm{eff}\approx55\,000$--$92\,000$ K) irregular variables as candidate debris hosts.

  The six confirmed events last weeks to years, reach transit depths of 25--100\%, and show no recurrence over the $\approx8$-yr ZTF baseline. They are therefore much deeper and longer lived than the recurring, minute-scale transits typical of previously known systems. The observed reddening is consistent with dust extinction, and high-speed photometry reveals minute-scale structure in several systems, indicating rapidly moving material close to the white dwarf. ZTF~J2327+0019 also shows a simultaneous \textit{WISE} infrared brightening, the first observed coincidence of infrared brightening and optical obscuration in a white dwarf.

  Two of the six hosts, one DAZ and one DBZ, are strongly metal enriched, with accretion rates among the highest known for their spectral types and minimum accreted masses of $3.5\times10^{19}$ and $2\times10^{24}$ g. We interpret the one-off transit events as collisions between planetesimal remnants and pre-existing circumstellar debris that trigger collisional cascades: short-timescale variability traces dust-producing fragments crossing the star, while slower changes trace the build-up and clearing of an optically thick dusty structure. In contrast, ZTF~J0436+5728 shows a coherent 5.910-h period, consistent with a rocky body shedding material near the tidal-disruption radius.
  }

   \maketitle

%


\section{Introduction}\label{sec:intro}
The majority of stars host planetary systems, and most of those stars end their stellar evolution as a white dwarf. These planetary systems are altered by the transformation of a main-sequence star into a white dwarf, but are expected to survive it \citep{sackmann1993, schroder2008, adams2013, martin2020, maldonado2021, chamandy2024}.
Several independent lines of evidence confirm that planetary material persists around white dwarfs: photospheric enrichment (also called metal pollution) from the accretion of planetary debris \citep{zuckerman2003,koester2014,ouldrouis2024,williams2024,lebourdais2025,farihi2026}; infrared excess and metal emission lines from dusty and gaseous debris discs \citep{zuckerman1987,gansicke2006,bonsor2017,wilson2019,xu2020}; and transits of both giant planets and planetesimal debris \citep{vanderburg2015,vanderburg2020}. Further evidence comes from directly imaged giant planets \citep{mullally2024}, infrared excess from giant planets \citep{limbach2024,limbach2025}, planetesimals detected via their effect on gas discs \citep{manser2019}, and evaporating close-in planets \citep{gansicke2019,schreiber2019}. White dwarfs that accrete material from disrupted rocky bodies (planetesimals, asteroids, comets, moons, or planets) must therefore be common \citep[][]{debes2002, jura2008}, although many aspects of this picture remain uncertain; we summarise it briefly below and refer the reader to the overviews by \citet{veras2021} and \citet{brouwers2022}.

The general picture is that rocky planetesimals are gravitationally scattered onto high-eccentricity orbits ($e\gtrsim0.9$) that take them inside the tidal disruption radius of the white dwarf. For young, hot white dwarfs ($\lesssim 100$ Myr), this is driven by dynamical instabilities triggered by the late evolution of the progenitor, typically within the first tens of Myr of cooling \citep[e.g.][]{mustill2012, veras2016, bonsor2012, frewen2014, smallwood2018, xing2025}. For older white dwarfs ($\gtrsim 300$ Myr), debris is instead supplied by long-term secular interactions in multi-planet systems \citep[e.g.][]{veras2015,mustill2018} and by perturbations from wide companions, Galactic tides, or stellar encounters acting on distant reservoirs such as exo--Oort clouds \citep[][]{payne2016,caiazzo2017, antoniadou2019,oconnor2023,pham2024,veras2024a,torres2025,veras2025}.
Whatever their origin, the bodies are tidally disrupted into fragments on eccentric, crossing orbits, which collide and grind down into a compact circumstellar dust and debris disc \citep{debes2012, veras2014, veras2015, brown2017, malamud2020a,brouwers2022} that evolves further \citep{metzger2012, kenyon2017a, kenyon2017, miranda2018}. Observational estimates of disc lifetimes are $\sim10^4-10^6$\,yr \citep{girven2012, cunningham2021}, but these discs are dynamically active rather than static \citep{guidry2024}, as shown by time-variable emission lines from the gaseous component \citep{wilson2014, manser2016, dennihy2018, gentilefusillo2021,rogers2025} and variable infrared emission from the warm dust \citep{xu2014, swan2019, wang2019, swan2020, debes2025, noor2025}. The dust is eventually ground down and sublimates into a gas disc that reaches the stellar surface \citep{kenyon2017}, where it is accreted and enriches the white dwarf atmosphere \citep[and is possibly detectable in X-rays,][]{cunningham2022}.

Dust and debris orbiting close to the white dwarf have recently been detected through transits. The first such system, \object{WD~1145+017}, was found using K2 data and shows many shallow recurring transits with an orbital period of $\approx$\qty{4.5}{hr}, close to the tidal-disruption (Roche) radius \citep{vanderburg2015}. It also shows strong photospheric metal contamination, infrared excess, and circumstellar gas absorption \citep{xu2016, fortin-archambault2020}, and intense monitoring has revealed debris on a range of orbital periods whose depth varies dramatically over months to years \citep{gansicke2016, gary2017, rappaport2018,aungwerojwit2024}.
At least 14 white dwarfs are now known to show debris transits, and they vary greatly in amplitude, duration, recurrence timescale, and orbital period. The transits in ZTF~J013906.17+524536.89 (hereafter ZTF\,J0139+5245) occur on much longer timescales and possibly recur every $\simeq$\qty{107}{d} \citep{vanderbosch2020}, although they have become shallower in recent years. \citet{guidry2021} used the Zwicky Transient Facility (ZTF) and \textit{Gaia} data to identify five further candidates, one of which, ZTF~J032833.52\textminus121945.27 (hereafter ZTF~J0328$-$1219), shows a \qty{9.94}{h} period that is likely orbital \citep{vanderbosch2021}. High-speed follow-up of metal-enriched stars yielded WD~1054\textminus226, with a \qty{25}{h} period \citep{farihi2022}, and most recently \citet{bhattacharjee2025} used ZTF light curve statistics to identify six new systems, among which WD~J1944+4557 hosts highly active debris on a \qty{4.9704}{hr} orbit \citep{guidry2025}. \textit{TESS} photometry of SBSS~1232+563 further revealed a coherent \qty{14.842}{hr} signal, tentatively identified as the dominant orbital period of its debris \citep{hermes2025}.

Collectively, these systems show variability on all timescales and with a wide range of light curve morphologies. Individual transits last minutes, the time for debris to cross the white dwarf, and recur on the orbital timescale of the debris, typically hours to days, with multiple chunks around a single star spanning a range of periods \citep{vanderburg2015,vanderbosch2021,farihi2022,guidry2025}. Recurring transits can change shape, drift in phase, and appear or disappear on timescales of days to years \citep{gansicke2016,aungwerojwit2024,hermes2025,guidry2025}. Finally, there are dimming events lasting days to months, such as ZTF~J0347\textminus1802 \citep{guidry2021}, SBSS~1232+563 \citep{hermes2025}, WD~J1237+5937, and WD~J1013\textminus0427 \citep{bhattacharjee2025}, which could result either from debris on a long orbit or from dust that temporarily engulfs the white dwarf.

In this paper, we present six white dwarfs newly discovered to show transiting debris and dust obscuration events, including one that completely disappears for a year, as well as three white dwarfs showing persistent, irregular variability that may also be caused by orbiting debris. To identify these events, we systematically searched the ZTF alerts associated with white dwarfs and visually inspected the full light curves (Sect.~\ref{sec:targetselection}). Sections~\ref{sec:data} and \ref{sec:methods} describe the archival and follow-up data and our analysis methods, and Sect.~\ref{sec:analysis} presents the analysis of each object. The results are presented in Sect.~\ref{sec:results} and in Sect.~\ref{sec:discussion} we discuss our interpretation of the events, compare the new objects with previously known systems, and discuss occurrence rates and the potential for future discoveries. We summarise the paper in Sect.~\ref{sec:conclusion}.


\section{Target selection}
\label{sec:targetselection}

To find new white dwarfs that show dimming events, we searched data from the Zwicky Transient Facility \citep{graham2019,bellm2019,masci2019,dekany2020}. ZTF uses the Palomar 48-inch (P48) telescope to monitor the entire visible sky with a cadence of 2--3 days, mostly in the $g$ and $r$ bands, reaching a median limiting magnitude of \mbox{$\approx20.5$} AB-mag in \qty{30}{s}. Images are processed automatically and PSF photometry is measured for any source detected in both the reference and science images \citep{masci2019}. In addition, automated image differencing reports any change larger than five standard deviations as an `alert' \citep{patterson2019}. Although alerts are primarily intended for the rapid identification of extra-Galactic transients, they are generated for both positive and negative differences and can therefore be used to study any kind of photometric variability, including variable stars. Of the properties reported for each alert\footnote{See \url{https://zwickytransientfacility.github.io/ztf-avro-alert/schema.html}.}, the most important in this work are \textsc{magpsf}, the magnitude of the alert, and \textsc{magnr}, the magnitude of the source in the reference image used to construct the difference image.

In this work, we used the ZTF alerts to identify white dwarfs that have decreased in brightness. We queried\footnote{\url{https://github.com/dmitryduev/penquins}} the Kowalski alert database\footnote{\url{https://github.com/skyportal/Kowalski}} \citep{vanderwalt2019,coughlin2023} and collected all alerts within 3\arcsec\ of objects in the white dwarf candidate catalogue by \citet{gentilefusillo2021a} (based on \textit{Gaia} eDR3 data). We did not apply additional filtering and included all negative alerts, even those with a low real-bogus score \citep{mahabal2019,duev2019}; cuts on the real-bogus value and their effect on completeness are discussed in Sect.~\ref{sec:discussion}.
The query, last executed on 2025-09-12, returned \num{602926} alerts associated with \num{26466} of the $\approx$\num{273000} white dwarf candidates detected in the ZTF reference images. Of these, \num{13391} have at least one negative alert and \num{7753} show only negative alerts (Fig.~\ref{fig:negalerts}). For each white dwarf candidate with at least one negative alert, we constructed a combined light curve from the alerts (based on difference images) and the PSF photometry (based on the science images). We then visually inspected the light curves, together with other archival data, of the \num{4027} white dwarfs with three or more negative alerts, excluding known periodic variables, to identify astrophysically variable white dwarfs of interest. Most of these candidates are cataclysmic variables, image artefacts, or high proper motion stars; the vetting is described in Appendix~\ref{sec:app_vetting}.

\begin{figure}
    \centering
    \includegraphics{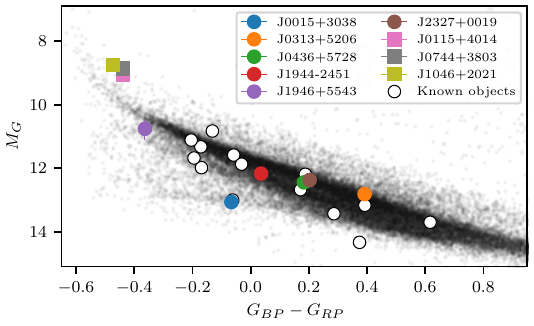}
    \caption{The colour-magnitude diagram based on the \textit{Gaia} DR3 data and distance estimates \citep{bailer-jones2021} for white dwarfs with transiting debris (dots), including the three hot, irregularly variable white dwarfs from this paper that are candidates. The black background shows stars within \qty{200}{pc}. The coloured markers show the nine objects presented in this paper. The white dots show 14 other white dwarfs (of which we recovered eight) with transiting debris.}
    \label{fig:HR}
\end{figure}

A total of 19 white dwarfs of interest remained after visual inspection. Of these, eight are previously known transiting debris white dwarfs \citep{vanderburg2015,guidry2021,vanderbosch2021,bhattacharjee2025,hermes2025}, listed in Table~\ref{tab:knownobject_overview}. Of the remainder, one was already known as a variable white dwarf and is the subject of ongoing work (ZTF~J0850+1956; Vanderbosch et al., in prep.), and another is of an entirely different nature and will be presented separately (ZTF~J2022+4637; Van Roestel et al., in prep.). The nine remaining new white dwarfs show long-duration transits or irregular long-timescale variability\footnote{For real-time light curves, see \url{https://janvanroestel.github.io/WD-transitingdebris-alerts/}}; they are shown in the \textit{Gaia} colour-magnitude diagram in Fig.~\ref{fig:HR} and listed in Table~\ref{tab:overview}.

\begin{table*}[]
    \small 
    \centering
    \caption{New variable white dwarfs discovered with ZTF alerts that potentially show transiting debris. The top part of the table lists white dwarfs that are strong candidates of systems showing transiting debris and/or fading due to the disintegration of a planetesimal. The bottom part lists white dwarfs with persistent irregular variability whose nature is uncertain, but could be due to transiting debris.}
    \renewcommand{\arraystretch}{1.2}
    \begin{tabular}{lllllllll}
    Name & ZTF alert ID & RA & Dec  & G-mag & $T_\mathrm{eff}$ & $\log g$ & Distance & Sp-type  \\
         &              & hh:mm:ss & dd:mm:ss  & Vega-mag & \unit{K} & \unit{cm.s^{-2}} & \unit{\parsec} &  \\
    \hline
    \hline
    \rule{0pt}{1ex} \\
    \multicolumn{9}{c}{White dwarfs that show single dimming events} \\
    \hline
    \object{ZTF~J0015+3038} & ZTF22acawcsg & 00:15:39.19 & +30:38:14.76 & 19.15 & $14280\pm230$ & $8.82\pm0.03$ & $164^{+10}_{-9}$ & DA\\
    \object{ZTF~J0313+5206} & ZTF20aaaejmy & 03:13:00.54 & +52:06:46.31 & 18.93 & $\phantom{0}9720\pm80$ & $8.27\pm0.06$ & $167^{+8}_{-8}$ & DA\\
    \object{ZTF~J0436+5728} & ZTF20abykyba & 04:36:35.46 & +57:28:40.30 & 19.38 & $12580\pm110$ & $7.95\pm0.03$ & $244^{+25}_{-16}$ & DA\\
    \object{ZTF~J1944\textminus2451} & ZTF24aapzpen & 19:44:15.16 & \textminus24:51:31.12 & 19.32 & $\phantom{0}9570 \pm 70$ & $7.84\pm0.06$ & $269^{+30}_{-20}$ & DBZ\tablefootmark{a} \\
    \object{ZTF~J1946+5543} & ZTF25aanmfif & 19:46:28.67 & +55:43:04.48 & 19.63 & $21250\pm120$ & $8.34\pm0.02$ & $515^{+58}_{-14}$ & DA \\
    \object{ZTF~J2327+0019} & ZTF22aanygxe & 23:27:24.86 & +00:19:42.18 & 19.07 & $10170\pm30$ & $8.30\pm0.03$& $219^{+14}_{-12}$ & DAZ\tablefootmark{b}\\
    \rule{0pt}{1ex} \\
    \multicolumn{9}{c}{Hot, persistent and irregularly variable white dwarfs} \\
    \hline
    \object{ZTF~J0115+4014} & ZTF19abatwrl & 01:15:44.83 & +40:14:28.73 & 17.18 & \qty{67000\pm2000}{} & \qty{8.1\pm0.2}{} & $423^{+16}_{-16}$ & DA \\ 
    \object{ZTF~J0744+3803} & ZTF18aaegptk & 07:44:58.75 & +38:03:01.70 & 17.77 & \qty{55000\pm1000}{} & \qty{7.93\pm0.08}{} & $604^{+41}_{-33}$ & DA \\ 
    \object{ZTF~J1046+2021} & ZTF20aalyafj & 10:46:16.01 & +20:21:27.99 & 16.22 & \qty{92000\pm2000}{} & \qty{7.66\pm0.06}{} & $312^{+7}_{-5}$ & DA \\
    \end{tabular}
    \tablefoot{
    \tablefoottext{a}{ZTF~J1944\textminus2451 shows strong Ca-\textsc{ii} lines and also Mg-\textsc{i}, Na-\textsc{i}, and Fe-\textsc{i} lines.}
    \tablefoottext{b}{ZTF~J2327+0019 shows only Ca-\textsc{ii} absorption lines, in addition to the Balmer lines.}
    }
    \label{tab:overview}
\end{table*}

\section{Archival and follow-up data}\label{sec:data}

\subsection{Archival time series photometry}
To better characterise the variability of the objects, we collected additional time series photometry from other surveys. This not only increases the time baseline further into the past, but is also helpful in filling in any gaps in the ZTF light curve. We obtained data from LINEAR \citep[][]{stokes2000}, CRTS \citep{drake2009}, PTF/iPTF \citep{law2009,rau2009}, and ATLAS \citep{tonry2018,shingles2021}. Finally, we also obtained \textit{WISE/neoWISE} \citep{wright2010,mainzer2011} time-resolved data, but only ZTF~J2327+0019 shows any significant change over time. For further details, see Appendix~\ref{sec:app_lcs}.

\subsection{Spectral energy distribution}\label{sec:archivaldata}
For each object in the sample, we collect multi-wavelength data obtained from multiple surveys: UV data from \textit{GALEX} \citep{bianchi2017}; optical data from \textit{Gaia} eDR3 \citep{gaiacollaboration2021}, Pan-STARRS \citep{chambers2016}, and SDSS \citep{gunn2006}, near-infrared data from the 2MASS survey \citep{cutri2003}, and mid-infrared data from \textit{WISE} \citep{eisenhardt2020,marocco2021}. We searched for the nearest object in each catalogue, with a maximum distance of 5\arcsec. For each object, we checked if there was no nearby object that could affect the photometry. We used \textit{Vizier} and \textit{astroquery} to cross-match and collect the data. The SEDs are shown in Fig.~\ref{fig:SEDs}.

\subsection{High-speed photometry}
We obtained high-speed photometry of five targets: with CHIMERA \citep[][]{harding2016} for ZTF~J0313+5206, ZTF~J0436+5728, ZTF~J1944\textminus2451, and ZTF~J2327+0019; with ULTRACAM \citep{dhillon2007} for ZTF~J1944\textminus2451 and ZTF~J2327+0019; and with PRISM for ZTF~J1946+5543. The instruments, observations, and data reduction are described in Appendix~\ref{sec:app_highspeed} and shown in Figs.~\ref{fig:chimera_J0313}--\ref{fig:hs_J2327}.

\subsection{Identification spectra}\label{sec:followup}
\begin{figure*}[]
    \centering
    \includegraphics[width=\textwidth,height=0.9\textheight]{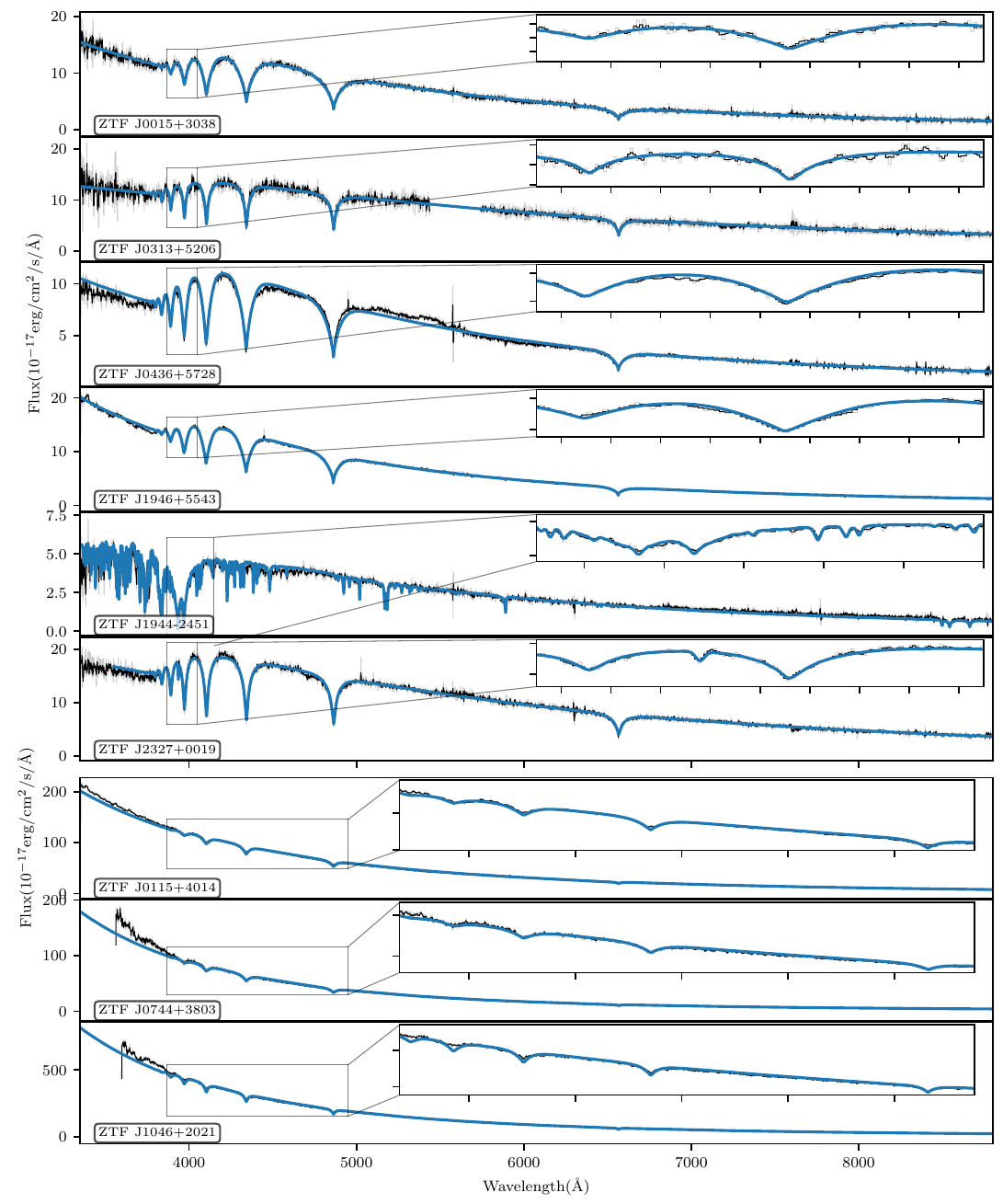}
    \caption{Identification spectra of all nine of the white dwarfs. The top six show the white dwarfs with single dimming events, the bottom three show the three white dwarfs with persistent variability. The data are shown in black and the best fitting model is overplotted in blue. The inset panel shows the blue part of the spectrum that contains the Ca H\&K doublet, typically the strongest metal lines in white dwarf spectra.}
    \label{fig:spectra_all}
\end{figure*}

We obtained identification spectra of all nine objects, either new spectra or spectra from archives. New spectra were obtained with LRIS at Keck \citep{oke1995,mccarthy1998}, DBSP at the Hale telescope \citep{oke1982}, and DeVeny at the Lowell Discovery Telescope. We obtained archival spectra from the SDSS \citep{york2000}, SDSS-V~DR19 \citep{sdsscollaboration2025}, DESI~DR1 \citep{desicollaboration2025}, and LAMOST archives \citep{cui2012}. The best (highest signal-to-noise) spectra are shown in Fig.~\ref{fig:spectra_all}, with the six white dwarfs that show single events at the top, and spectra for the three persistently variable white dwarfs at the bottom. The spectra are listed in Table~\ref{tab:spec_overview}. For more details, see Appendix~\ref{sec:app_spectra}.

\section{Methods}\label{sec:methods}

\subsection{Light curve analysis}
For analysis of the light curves, we normalised each optical light curve by dividing by the median brightness. Where possible, we fit dimming events (using \textit{emcee}; \citealt{foreman-mackey2013}) with an asymmetric hyperbolic secant function \citep[AHS, see][]{rappaport2016,xu2019,bhattacharjee2025}. The goal is to get a numerical estimate of the duration and depth of each event and the best-fit ingress and egress timescales are listed in Table~\ref{tab:fit_pars}. We also searched for periodic signals in the light curve using both the Lomb-Scargle periodogram \citep{lomb1976,scargle1982} and the Box Least Squares method \citep{kovacs2002}\footnote{As implemented by \url{https://github.com/johnh2o2/cuvarbase}.}. We searched the entire light curve, but we also searched only for the time frame in which the white dwarf showed activity. The light curve and analysis for each individual object is discussed in Sect.~\ref{sec:analysis}.

We determined if the dimming events show a colour dependence by using a similar method as in \citet{hillenbrand2022}, see also \citealt{farihi2022,hermes2025}. First, we matched every $g$-band point in the ZTF light curve to an $r$-band measurement if it was closer than 4 hours (the typical time difference is 2 hours). We removed any outliers where the $g$ or $r$ measurement is significantly brighter than the average brightness. 
We then fit the ZTF $g$ versus $r$ magnitudes (with the median subtracted) with a simple linear trend ($r=C\times g + D$, where $C$ and $D$ are free parameters), taking into account both the uncertainties on the $g$ and $r$ colours. In addition, the points were weighted inversely proportional to the time between the $g$ and $r$ measurements. We minimise the log-likelihood and include an additional fit parameter to account for any underestimate of the uncertainties \citep[e.g.][]{hogg2010}. To estimate the uncertainty in the parameters, we use \textit{emcee}. In addition, we use 20 bootstraps of the data (randomly using a subset of the observations), which makes the trend estimate more robust to outliers. The final result is shown in Fig.~\ref{fig:colourtrends}. In the case of grey extinction, a 1:1 correlation ($C=1$) is expected, if the extinction also causes reddening, a correlation of $0<C<1$ is expected. Finally, we also apply this method to the CHIMERA data of ZTF~J1944-2451, which has simultaneous $g$ and $r$ data, and also shows large amplitude variability.

In the context of interstellar dust \citep[e.g.][]{green2019a}, the relation between extinction and reddening is typically expressed as $A_g = R_g\ E(g-r)$, where $A_g$ is the extinction in the $g$ band (in magnitudes), $E(g-r)$ the amount of the reddening (in magnitude difference), and $R_g$ is the linear coefficient between the two. The value of $R_g$ is determined by the dust particle size distribution and for interstellar dust in our Galaxy, $R_g$ is estimated to be $\approx$3.5, but shows significant variance \citep{green2019a}. The $R_g$ and the correlation between $r$ and $g$ are directly related as $R_g = (1-C)^{-1}$.

\subsection{Fit to the spectra}
We fitted (again using \textit{emcee}) the optical spectra of each object with a DA white dwarf model \citep{koester2009} to measure the temperature ($T_\mathrm{eff}$) and surface gravity ($\log g$). We multiplied the spectral model with a third-order polynomial to make the model more flexible and to account for trends due to slit losses and systematics in the calibration. Once the temperature and surface gravity are determined, we used models that included calcium to measure the Ca/H ratio. For most objects, no calcium is detected and we report an upper limit (99 percentile limit).

ZTF~J1944\textminus2451 is a DBZ that shows many metal lines and to model this spectrum we use a DBZ atmosphere model which is described in more detail in Sect.~\ref{sec:ZTFJ1944}. The results are shown in Figs.~\ref{fig:spectra_all} 
and listed in Tables~\ref{tab:fit_pars} and \ref{tab:J1944_parameters}.

\subsection{The metal accretion rate and minimum accreted mass}
\label{sec:method_metals}
A secular metal accretion rate can be inferred from a spectroscopically measured atmospheric abundance of a given metal, and a model-dependent diffusion timescale which governs the gravitational sedimentation of said metal \citep[e.g.][]{koester2009,williams2024,bhattacharjee2025,coutu2019,hollands2018,koester2020,cunningham2025,dupuis1992}.
To do so, we assume a steady state of the metal abundance where the accretion rate ($\dot{M}^{\rm acc}$) is equal to the convective zone (cvz) diffusion rate ($\dot{M}^{\rm cvz}_Z$). We then estimate the diffusion rate as:
\begin{equation}
{\dot{M}_Z^{\rm cvz} = M_{\rm WD} \cdot 10^{q_{\rm cvz}} \cdot 10^{[Z/{\rm H}]} \cdot A_Z/A_{\rm H}\ / \tau_Z}
\end{equation}
where $M_{\rm WD}$ is the mass of the white dwarf, $q_{\rm cvz}$ is the mass fraction of the convective zone, [Z/H] the metal abundance ratio relative to hydrogen, and $A$ is the atomic mass.

In addition, the metal enrichment can also be used to calculate the amount of a given metal currently present in the atmosphere of the white dwarf, which represents the minimum amount of mass that must have been accreted:
\begin{equation}
    M_\mathrm{min} = \sum_Z \dot{M}_Z \cdot \tau_Z
\end{equation}
where, again, $\dot{M}_Z$ and $\tau_Z$ are the metal accretion rate and the diffusion timescale for metal $Z$ \citep[see][]{koester2009,izquierdo2021,williams2024}.

We used diffusion rates ($\tau_Z$) and convective zone mass fractions ($q_{\rm cvz}$) based on models from \citet{koester2020}. The diffusion rates depend on which element is considered and are a strong function of temperature and type of atmosphere (DA, DB, or DZ). To illustrate; DB white dwarf ZTF~J1944\textminus2451 has an estimated diffusion timescale of $\tau_\mathrm{Ca}\approx$\qty{5E6}{yr}, for DA white dwarf ZTF~J0313+5206 $\tau_\mathrm{Ca}\approx$\qty{1000}{yr}; and for the hot white dwarfs in our sample ($T_\mathrm{eff}\approx$\qtyrange{55000}{92000}{K}), the atmosphere is fully radiative, so gravitational settling is not governed by convective-zone mixing and diffusion timescales are extremely short, of order days or less. Finally, to estimate the total amount of accreted material, we assume an Earth-like Ca-fraction of 1.6\% \citep[1/62.5, see][]{allegre1995}.  Diffusion rates also depend on the inclusion of enhanced mixing processes. For hydrogen atmosphere white dwarfs with convective atmospheres (i.e. those with effective temperatures below 18\,000\,K), inclusion of convective overshoot may increase mixed masses and diffusion timescales by two orders of magnitude \citep{cunningham2019}. For hydrogen-atmosphere white dwarfs warmer than $\approx$\qty{10000}{K}, it has been suggested that thermohaline mixing may drive an even larger increase in the mixed mass \citep{bauer2019,buchan2025}.

The assumption of a steady state accretion rate is in general reasonably well-justified in the case of hydrogen-atmosphere white dwarfs due to their short diffusion timescales (thus allowing the system to reach a steady state within a short period of time). The assumption is in general less valid for helium-atmosphere white dwarfs, in which the diffusion timescales can reach Myr \citep{brouwers2022,buchan2025,koester2009}. In the case that the system is not in a steady state (i.e. the accretion recently began or recently ceased), the steady state assumption will underestimate the instantaneous accretion rate. Thus, our assumption of steady state accretion to derive a \textit{minimum} accreted mass remains valid, regardless of the actual accretion state of the system.

\subsection{SED fit and infrared excess}
The spectral energy distribution of each white dwarf was fitted with the same DA models as the spectra. To calculate the flux per passband, we convolved the spectrum with the filter response curves. The main goal is to determine if there is any infrared excess due to a dust disc around the white dwarf. We fit the available archival photometry and the \textit{Gaia} parallax, except the \textit{WISE} infrared data, using the \textit{emcee} package. For the white dwarf, we assume masses of \qtyrange{0.5}{0.8}{\Msun}, corresponding to radii of \qty{0.014}{\Rsun} to \qty{0.010}{\Rsun}. The SEDs are shown in Fig.~\ref{fig:SEDs}. 

\subsection{Analysis of high-speed light curves}
We analysed the high-speed photometry using Gaussian process regression with the goal of determining a typical variability timescale. To do so, we used the \textit{celerite} python package\footnote{\url{https://celerite.readthedocs.io/en/stable/}} \citep{foreman-mackey2017}. We used a Mat\'ern 3/2 kernel and a white kernel to model each colour band individually. We optimised the kernel parameters using the L-BFGS-B method as implemented in \textsc{scipy.optimize} \citep{virtanen2020}. This model allows us to determine if there is any variability, and at what typical timescale. Figs.~\ref{fig:chimera_J0313}--\ref{fig:hs_J2327} show the light curves and the Gaussian process models.

\section{Analysis}\label{sec:analysis}

\subsection{Objects with single obscuration events}
We first analyse the long term and high cadence light curves, spectra, and SED of the six white dwarfs that show a single dimming event in the ZTF light curve. We briefly summarise the observational properties of the six objects in general, before discussing each one in greater detail in the rest of the Section. 

The long term light curve characteristics differ between objects, but all white dwarfs have in common that they are initially non-variable, and that at some point in the ZTF light curve, they show a sudden dimming event (Table~\ref{tab:fit_pars} presents the approximate depth and duration of the events). A common feature for all white dwarfs is that they show a $r$ and $g$ correlation between 0--1: the white dwarfs become redder as they become fainter (Fig.~\ref{fig:colourtrends}). 
High-speed photometry shows that some of the white dwarfs show short timescale variability (see Appendix~\ref{sec:appendix_B} for the light curves). For most objects, the (detectable) presence of short time variability seems to be anti-correlated with the brightness of the white dwarfs; as the white dwarfs fade, the amplitude of short timescale variability increases.  
Only two objects show detected metal lines in their spectra, the DBZ white dwarf ZTF~J1944\textminus2451 and the DAZ white dwarf ZTF~J2327+0019 (Fig.~\ref{fig:spectra_all}). The analysis of the data shows that the metal accretion rate must be high for metals to be detected for these white dwarfs.
Finally, ZTF~J2327+0019 is the only object with detected infrared excess, which is also time-variable (Fig.~\ref{fig:SEDs}). For ZTF~J0313+5206, the SED also suggests some IR excess, but this is most likely contamination from a star a few arcsec away.

\subsubsection{ZTF~J0015+3038}
\begin{figure}
    \centering
    \includegraphics{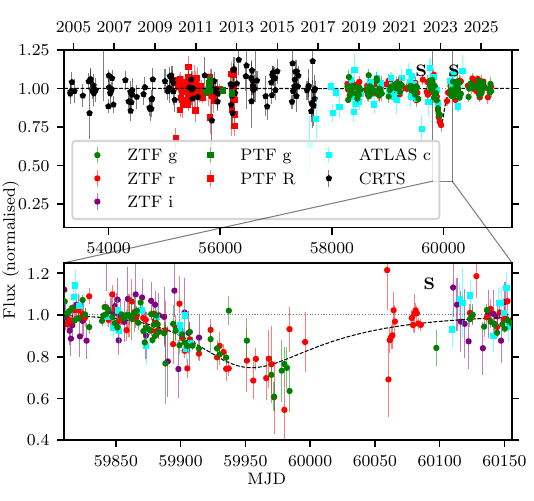}
    \caption{The light curve of ZTF~J0015+3038. The light curve shows one dimming event which lasts about 200 days and is about 25\% deep. The dashed line shows a simple model we used to approximate the event. The measurements show no other (irregular) variability. The \textbf{S} shows when spectra were taken.}
    \label{fig:ZTFJ0015}
\end{figure}

The light curve of ZTF~J0015+3038 (Fig.~\ref{fig:ZTFJ0015}) shows a dimming event with a depth of $\approx$25\% and the duration of approximately 200--250 days. It is only partially observed; there is a 50-day gap in the data during the second half of the event. The available data show a slow and gradual dimming and re-brightening and no significant irregular variability as seen in some other events. No other significant dimming events are seen in the rest of the available data stretching back 20 years. However, the signal-to-noise ratio of the archival data is lower and the sampling sparser, and could hide events with an amplitude similar to the detected transit event. Finally, there is a clear correlation between ($g$-band) brightness and $g-r$ colour (Fig.~\ref{fig:colourtrends}).

The LRIS spectrum shows a typical DA-type spectrum with a best DA-model fit with parameters of $T_\mathrm{eff} = $ \qty{14280 \pm 230}{K} and $\log g= $ \qty{8.82 \pm 0.03},  suggesting a high mass of \qty[]{1.11\pm0.02}{\Msun} (calculated using the mass-radius relation by \citealt{verbunt1988}).
This is consistent with the photometric mass estimate of \citet{gentilefusillo2019}; \qty[]{1.15\pm0.07}{\Msun} based on the \textit{Gaia} data. In the spectrum, no Ca or other metal lines are detected, and the limit to the abundance of Ca is $\log_{10} [\mathrm{Ca/H}] < -5.5$. 
The diffusion timescale of Ca is expected to be on the order of days \citep{koester2020}, or as much as half a year with the inclusion of convective overshoot models \citep{cunningham2019}.
We obtained the LRIS spectrum at the tail-end of the obscuration event, \qty{150}{d} after the deepest point in the event. Therefore, if there was a large amount of (near) instantaneous enrichment of the white dwarf atmosphere, we likely would have detected it before it disappeared. The white dwarf was not detected in the infrared, with only upper limits for the \textit{WISE} W1 and W2 bands. 

\subsubsection{ZTF~J0313+5206}
\begin{figure}
    \centering
    \includegraphics{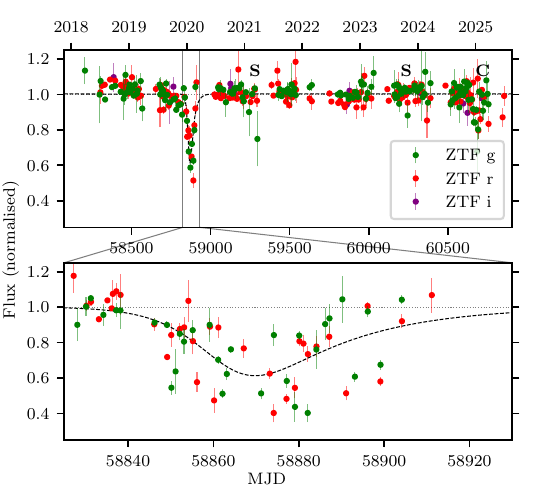}
    \caption{The ZTF light curve of ZTF~J0313+5206, with a dashed line as an approximation of the event that lasts about 70 days. The bottom panels show that there is a lot of variability during the event. However, the data are too sparse to resolve the shorter timescale variability or determine a period. The \textbf{S} indicates when a spectrum was taken, and \textbf{C} indicates when CHIMERA data were obtained.}
    \label{fig:ZTFJ0313}
\end{figure}

For ZTF~J0313+5206 (Fig.~\ref{fig:ZTFJ0313}), we observe a dimming event in the ZTF light curve that lasts about 70 days. The white dwarf fades to on average 60\% of its nominal brightness, but the event is not smooth and shows a large amount of irregular variability, rapidly alternating between 40\% and 90\% of the nominal brightness. We were unable to determine any periodicity in the variability during the dimming event, possibly because of the sparse sampling or simply because the signal is not periodic. 
The light curve indicates that the time scale of variability is short; the transit depth changes from 60\% to 20\% in one hour. Although the model suggests that the ingress time is a few weeks, the sampling is poor, and it is possible that the onset of the event was instantaneous (as seen in other events). We also note that the white dwarf is slightly fainter after the event compared to before the event, by about 2--9\%. There is also a clear correlation between brightness and $g-r$ colour.
Finally, we obtained a high-cadence light curve a few years after the main event, but it shows no variability (see Fig.~\ref{fig:chimera_J0313}).

A DBSP spectrum was obtained almost 4 years after the dimming event. It shows that ZTF~J0313+5206 is a mostly unremarkable DA white dwarf with a typical effective temperature but with a somewhat high surface gravity (\qty{9720\pm80}{K}, \qty{8.27\pm0.06}). The spectrum shows no signs of metal lines, with an upper limit to the Ca abundance of $\log_{10} [\mathrm{Ca/H}] < -7.0$. The diffusion timescale for metals is typically \qty{1000}{yr} \citep{koester2020}, suggesting that there has not been a lot of accretion in the recent past.

There seems to be a hint of infrared excess for this white dwarf in the \textit{WISE} W1 and W2 bands, but there is a faint ($i=20.8$), red point-source nearby (2.6\arcsec). Because the \textit{WISE} PSF is $\approx$8\arcsec, the other source is blended together with ZTF~J0313+5206, and therefore we do not consider this to be a detection of infrared excess of the white dwarf.

\subsubsection{ZTF~J0436+5728}
\begin{figure*}
    \centering
    \includegraphics{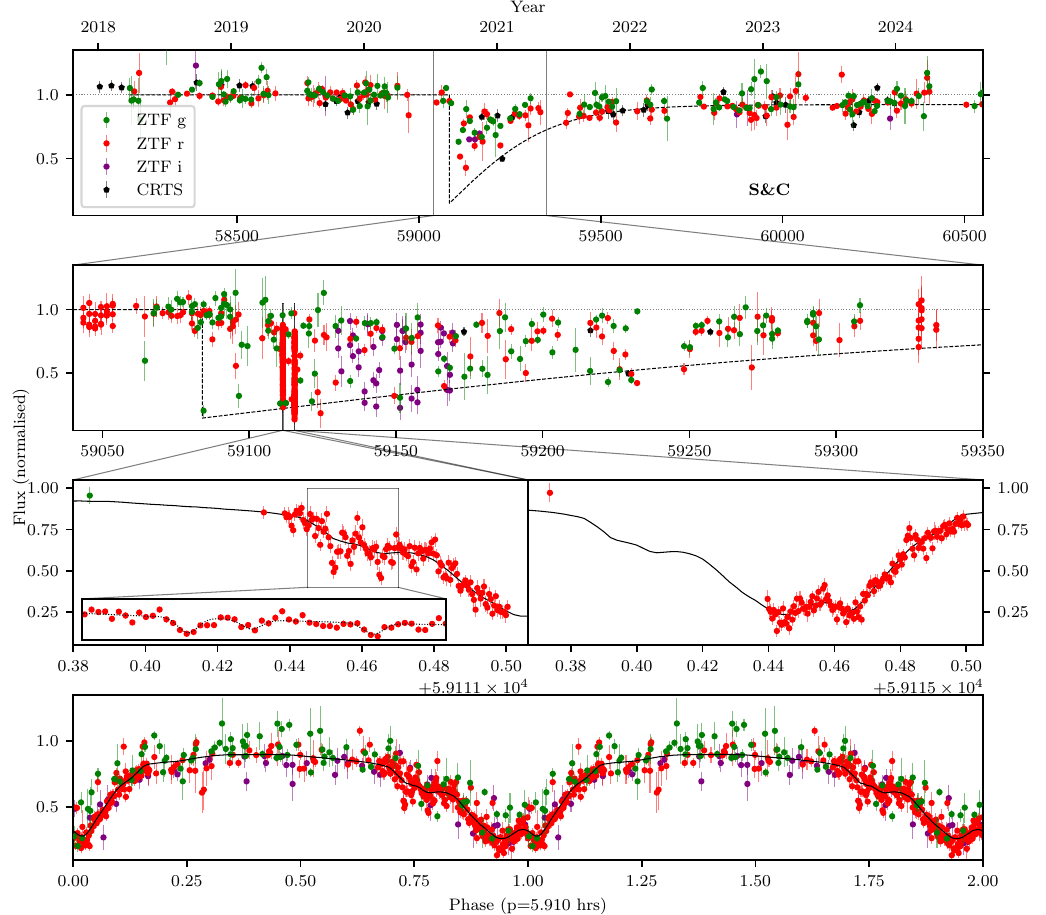}
    \caption{The ZTF light curve of ZTF~J0436+5728. It shows a single event with a lot of shorter timescale variability. A BLS period search of the data shows a clear periodic signal of 5.910 hours. The bottom panel shows the light curve folded to this period. About 20 days after the first ZTF alert, ZTF obtained continuous observations that are shown in the third row. The solid line shows the periodic SuperSmoother model, and the dotted line shows a local linear fit with 3 Gaussians with a FWHM of \qty{170}{s}. The dashed line in the top two panels shows the amplitude of the periodic SuperSmoother model (assuming an exponential decay of the periodic signal). The \textbf{S} indicates when a spectrum was taken, and \textbf{C} indicates when CHIMERA data were obtained.
}
    \label{fig:ZTFJ0436}
\end{figure*}

For ZTF~J0436+5728 (Fig.~\ref{fig:ZTFJ0436}) we observe an event that lasts approximately \qty{320}{days}. It is different from other white dwarfs with transiting debris as it starts abruptly, with the white dwarf dropping in brightness by 80\% within a day or even less (on MJD=59084.43). This star happened to be observed in continuous observing mode by ZTF for two consecutive nights \citep{kupfer2021} which shows that the brightness decreases and increases by tens of percent in the span of an hour. The curve is not smooth, and shows some discontinuities and variability on a few minute timescales. The ZTF light curve is highly variable during the entire dimming event.

We determined that the variability is periodic with a period of 5.910 hours. The folded light curve shows a mostly smooth signal that is coherent in phase over $\approx$\qty{120}{d}, corresponding to 500 orbits (bottom panel of Fig.~\ref{fig:ZTFJ0436}). The depth of the transits seems to decrease with time. To model this effect, we (1) capture the shape of the periodic signal with a periodic SuperSmoother model \citep{friedman1984}.
In the next step, (2) we take this template and fit the overall light curve with this periodic model with an exponentially decreasing transit depth over time. As can be seen in Fig.~\ref{fig:ZTFJ0436}, the amplitude decrease is reasonably approximated by our exponential model.

Finally, we note that the long term brightness has decreased $7.0\pm0.9$\% in the $g$ and $r$ filters. We confirm this by comparing the flux level from before and after the event (mid 2020 and 2022). The yearly average flux shows that the $r$-band flux is still increasing by about 1.5 percentage points per year after 2022. In addition, there is likely a correlation between brightness and $g-r$ colour, but the measurement might be somewhat affected by rapid variability.

We obtained a CHIMERA high-speed light curve two years after the start of the activity. However, the data do not show any variability (see Fig.~\ref{fig:chimera_J0436}). An LRIS spectrum, also obtained about 2 years after the event, shows that ZTF~J0436+5728 is unremarkable and looks like a common DA type white dwarf without any metal lines. The temperature and surface gravity (\qty{12580\pm110}{K}, \qty{7.95\pm0.03}{}) of the white dwarf are also typical of the sample of \textit{Gaia} white dwarfs. We determined a limit on the Ca abundance $\log_{10} [\mathrm{Ca/H}] < -5.5$. The diffusion timescale for metals is typically 1--200 days, which means that we can expect the absorption lines to vary on human timescales.

\subsubsection{ZTF~J1944\textminus2451}\label{sec:ZTFJ1944}
\begin{figure}
    \centering
    \includegraphics{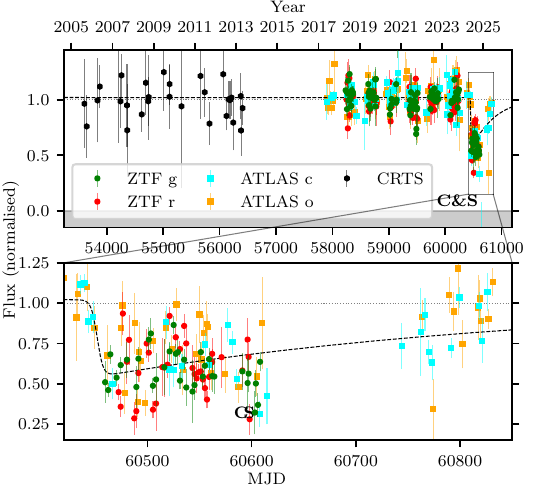}
    \caption{The ZTF light curve of ZTF~J1944\textminus2451. The CRTS and ZTF data show that the star has a steady brightness for years, except for the last two years when the brightness decreased to as low as 25\% of the pre-transit brightness. \textbf{S} indicates when a spectrum was taken, \textbf{C} and \textbf{U} indicate when CHIMERA and ULTRACAM high-speed photometry were obtained.}
    \label{fig:ZTFJ1944}
\end{figure}

The ZTF, ATLAS, and CRTS light curve of ZTF~J1944\textminus2451 (Fig.~\ref{fig:ZTFJ1944}) shows that it is non-variable for 5 years, and possibly 20 years. However, in 2024 the ZTF data show a sudden onset of activity with a dimming of about 45\% on average. The AHS model confirms this, and shows that the star is slowly returning to its nominal brightness. A visual inspection of the light curve shows that the brightness does fluctuate from 20\% to 100\% of its nominal level, but it seems that the timescale and amplitude are not as short and high compared to e.g. ZTF~J0313+5206. However, sampling by ZTF and ATLAS of this event is sparse, and a quantitative measurement is challenging. In addition, we are unable to detect any period in the variability. Only a small amount of nearly simultaneous $g$ and $r$ data are available, which means that no reliable colour measurement is possible.

The CHIMERA high-speed light curve (Fig.~\ref{fig:chimera_J1944}), obtained during the event, shows a large amount of variability ($\approx 70\%$) on timescales of hours to as short as a few minutes. However, the baseline of \qty[]{1}{hr} is too short to determine any periodicity. ULTRACAM observations a year later show that the white dwarf is again non-variable.

Spectroscopically, ZTF~J1944\textminus2451 is by far the most feature-rich: it is a DBZ white dwarf that does not show any hydrogen lines but shows He-I lines and many metal lines (Fig.~\ref{fig:spectra_all}). 
The strongest metal absorption lines are the Ca-\textsc{ii} (H\&K) lines that are almost 90\% deep compared to the continuum. The second deepest lines are the Mg-\textsc{i} b lines at \qty{5175}{\angstrom} that are 50\% deep. Other visible lines are the Ca-\textsc{ii} triplet at \qtyrange{8500}{8700}{\angstrom}, an Mg-\textsc{ii} line, and the Na-\textsc{i} lines at \qty{5800}{\angstrom}, and finally a lot of Fe-\textsc{i} and a few Fe-\textsc{ii} lines. All of these metal lines are often seen in DBZ white dwarfs \citep[e.g.][]{hollands2018}. We modelled the LRIS spectrum with DBZ spectral models to determine the relative atmospheric abundances as in \citet{hollands2017}. Based on the measured metallicities, the estimated accretion rate is $\log{\dot{M}} = 9.7 - 10.1$ [\unit{g.s^{-1}}] \citep[there is a small difference depending on the method, see ][]{bhattacharjee2025}. The typical diffusion timescale for metals in a helium-dominated atmosphere at the measured temperature is \qty{5E6}{yr}. Based on the Ca accretion rate, we estimate a minimum accreted mass of \qty{2E24}{g}, about twice the mass of Ceres. 
The parameters of the spectrum fit are reported in Table~\ref{tab:J1944_parameters}. 

\subsubsection{ZTF~J1946+5543}
\begin{figure}
    \centering
    \includegraphics[]{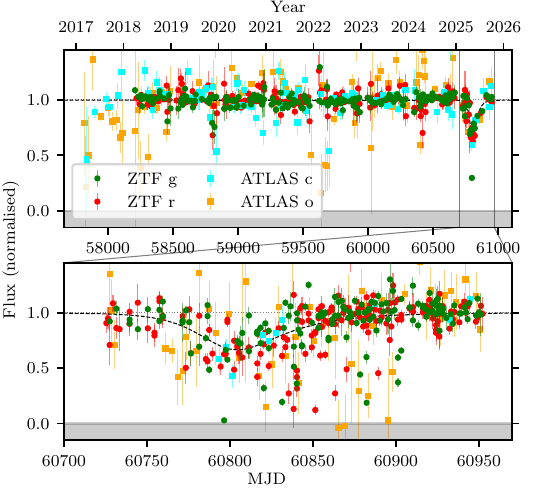}
    \caption{The ZTF light curve of ZTF~J1946+5543. The ZTF and ATLAS data show that the star has a steady brightness for years, except for the event in 2025. The dimming event has a slow overall evolution, but is characterised by rapid brightness changes of close to 100\% in just an hour. }
    \label{fig:ZTFJ1946}
\end{figure}

The behaviour of the light curve (Fig.~\ref{fig:ZTFJ1946}) is similar to ZTF~J0313+5206; the white dwarf does not show any variability in the past and only a single obscuration event shows high-amplitude and irregular variability. The model fit that tracks the average behaviour shows a dimming of 30\% over a couple of weeks. However, a close inspection of the light curve shows that, within 1 hour, the brightness of the white dwarf changes from almost fully visible to completely obscured. As was the case for ZTF~J0313+5206, the sampling of ZTF is not good enough to determine the exact timescale of the variability and if it is periodic. Although there are a few outliers, likely due to the rapid variability, the overall trend suggests again there is a correlation between colour and brightness. 

No archival spectra are available for this white dwarf. We obtained follow-up spectra with LRIS and DeVeny, which confirm that ZTF~J1946+5543 is a DA white dwarf, showing only Balmer lines and no metal absorption lines, consistent with the expectation from the \textit{Gaia} colours, distance, and SED that this white dwarf is relatively hot. A fit to the Balmer lines gives $T_\mathrm{eff}=21250\pm120$~K, $\log g = 8.34\pm0.02$, and a spectroscopic distance of $534\pm34$~pc, in good agreement with the \textit{Gaia} parallax distance. Given the absence of any detected metals, we place an upper limit on the calcium abundance of $\log_{10} [\mathrm{Ca/H}] \lesssim -5.5$, and we can only place an upper limit on the accretion rate; the estimated sedimentation timescale is $\approx$\qty{7.5}{d} \citep{koester2009}, which means that the metals rapidly diffuse out of the outer envelope of the white dwarf. This corresponds to an accretion rate limit of $\log \dot{M} \lesssim 9.34$ [\unit{g.s^{-1}}], which is not constraining.

\subsubsection{ZTF~J2327+0019}
\begin{figure*}
    \centering
    \includegraphics{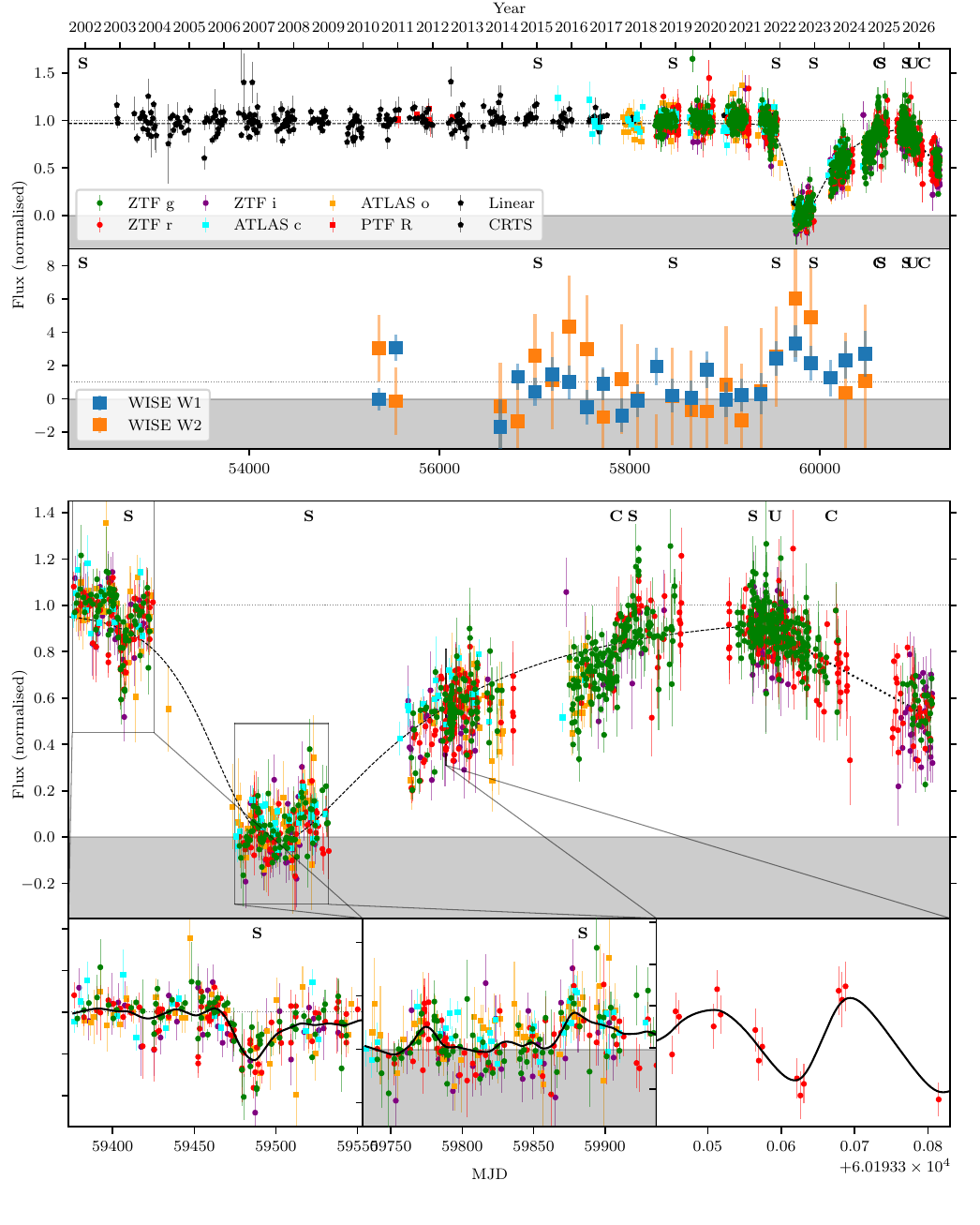}
    \caption{The light curve of ZTF~J2327+0019. The optical ZTF and ATLAS light curve shows one deep transit that reduces the flux to zero and slowly increases again to the pre-transit level over the course of three years. The dashed line shows a simple approximation with an AHS model. The dotted line shows a linear fit to the data since the renewed fading in 2025 (MJD~60875), showing the white dwarf getting fainter by $\approx$3\% per month. The second panel shows the \textit{WISE} W1 and W2 data, which shows a subtle but significant flare at the same time as the optical flux dropped. The bottom panels show three zoomed in views, showing variability at various timescales. The black line shows a Gaussian Process model using a Mat\'ern 3/2 kernel. The \textbf{S} indicates when a spectrum was taken, \textbf{C} and \textbf{U} indicate when CHIMERA and ULTRACAM high-speed photometry were obtained.}
    \label{fig:ZTFJ2327}
\end{figure*}

ZTF~J2327+0019 shows the most extreme obscuration event, both in depth and duration (Fig.~\ref{fig:ZTFJ2327}). The CRTS and ZTF light curves show that this object has not shown significant variability for 19 years. However, at the start of the 2022 observing season, the white dwarf decreased in brightness to 0\%, disappearing (almost) completely throughout the observing season. A year later, the object was back at 50\% of normal brightness and was slowly brightening further. Another year later, the brightness was initially at 60\% and slowly returned to 90\% nominal brightness, $\approx$3 years after the start of the event. To guide the eye, we fitted the AHS model. In 2025, the system started to fade again, and is currently still getting fainter. We fitted the trend here, showing the white dwarf is getting fainter by $\approx$3\% per month.

The \textit{NEOWISE} infrared light curve is noisy, but does show a clear increase in brightness at the same time as the optical flux decreases. The infrared brightness after the start of the optical event is $2.3\pm0.45$ and $2.9\pm1.4$ times brighter than before, corresponding to an increase by \qty{12\pm4}{\mu Jy} and \qty{12\pm3}{\mu Jy} in the $W1$ and $W2$ bands respectively. For both filters, although the brightening is significant, the S/N is not enough to establish if the luminosity has reached the same values as before the event.

We fit the average increase in infrared brightness with a simple model (see Fig.~\ref{fig:SEDs}). For a blackbody, the temperature and the equivalent radius of the emitting body are strongly anti-correlated and cannot be constrained independently: reproducing the observed \qty{12}{\mu Jy} excess requires an equivalent radius of $\approx$\qty{0.1}{\Rsun} at \qty{1200}{K}, but $\approx$\qty{0.9}{\Rsun} at \qty{550}{K}. The comparable excess in the two bands ($W2/W1 = 1.0\pm0.4$) breaks this degeneracy, because a blackbody with equal $W1$ and $W2$ flux densities has $T\approx$\qty{1200}{K}: the coolest solutions are excluded, as they would produce a $W2$ excess several times larger than observed (\qty{36}{\mu Jy} at \qty{550}{K}). The emission is therefore consistent with a blackbody of $T\gtrsim$\qty{900}{K} and a corresponding equivalent radius of \qtyrange{0.05}{0.2}{\Rsun}.
We also tested a slightly more complex disc model \citep{jura2003}, (see also \citealt{xu2018} for examples of IR fits to IR variable white dwarfs). This model suggests that the inner edge of the disc is close to the white dwarf, at a distance of \qtyrange{0.1}{0.3}{\Rsun} and with temperatures of \qtyrange[]{600}{900}{K} \citep{chiang1997}. Technically, with this simple model fully edge-on solutions ($i\gtrsim$ \qty{85}{\degree}) are excluded, but this model is overly simplistic and the best-fit values should be interpreted with caution; see e.g. \citet{bhattacharjee2025a}.

In a closer inspection of the optical light curve, variability can be seen on the time scales of weeks, days, and hours. We note that the year before the dramatic brightness drop, there is a 20\% deep transit-like event that lasts about 30 days (the bottom panels in Fig.~\ref{fig:ZTFJ2327}). There are also minor fluctuations visible during the deepest phase of the dimming event, where the white dwarf is visible for a few days at a $\approx$10\% level. Finally, we observe short timescale (minute) variability of tens of percent when the white dwarf was rising in brightness again (bottom right panel). This is also seen in the CHIMERA and ULTRACAM data that were obtained later when the white dwarf was almost back to its nominal brightness (Fig.~\ref{fig:hs_J2327}). The CHIMERA data clearly show irregular brightness drops of 10\% on timescales as short as \qty[]{2}{min}. The ULTRACAM data, taken a year later when the white dwarf was back to 100\% brightness, still show some minor fluctuations. A few months later the white dwarf is a bit fainter again, and the CHIMERA data show 20\% deep, short-duration dips, demonstrating an increase in activity (bottom two panels).

Archival spectra from various survey telescopes are available before and during the obscuration event, but all have low SNR. All spectra show broad Balmer lines, typical for DA white dwarf stars. The spectrum that was taken just after the minimum of the light curve by SDSS-V is no different from the pre-event spectra. Three and four years after the event, we obtained two higher signal-to-noise spectra with LRIS. The LRIS spectra show, in addition to the Balmer lines, a shallow Ca-\textsc{ii} H\&K and triplet absorption line. We do not observe any wavelength offset between the Ca and Balmer lines, which suggests that both lines share the same gravitational redshift and therefore originate from the white dwarf surface and are not interstellar. We also do not observe any significant change in the depth of the Ca lines between the two LRIS spectra. Based on the depth of the H\&K lines, we estimate that the atmospheric Ca abundance is $\log_{10} [\mathrm{Ca/H}] = -6.34 \pm 0.15$. If we assume an Earth-like abundance, we estimate that the total required accretion rate is $\log \dot{M} = 10.05 \pm 0.15$ [\unit{g.s^{-1}}], with a diffusion timescale of $\approx$\qty{100}{yr} \citep{koester2020}.
The estimated minimum amount of mass accreted is \qty{3.5E19}{g}.

\subsection{Hot, irregularly variable white dwarfs}\label{sec:analysis_irregularvariables}
In our search of the white dwarfs that trigger alerts, we also found three white dwarfs that show persistent irregular variability. All three white dwarfs are hot and are almost in the same location in the \textit{Gaia} colour-magnitude diagram (Fig.~\ref{fig:HR}). The spectra (Fig.~\ref{fig:spectra_all}) and SED (Fig.~\ref{fig:SEDs}) do not show the presence of a binary companion (they are not Roche-lobe accreting). The irregular variability could be due to transiting debris \citep[similar to WD~J0923+7326 and WD~J1302+1650, see][]{bhattacharjee2025}. However, the signature is less clear, and since there is no evidence of infrared excess or metals, we consider these white dwarfs candidate debris white dwarfs.

We fit the spectra of the three white dwarfs using the same methods as for the other white dwarfs. All three initially returned \qty{60000}{K}, the upper limit of the temperature range of our default model grid, so we repeated the fits using a grid extending to higher temperatures; the effective temperatures and surface gravities reported in Table~\ref{tab:overview} are those of the extended fits. We also observe that ZTF~J0115+4014 has filled-in Balmer line cores, suggesting a weak, narrow Balmer emission component may be present.

\begin{figure}
    \centering
    \includegraphics{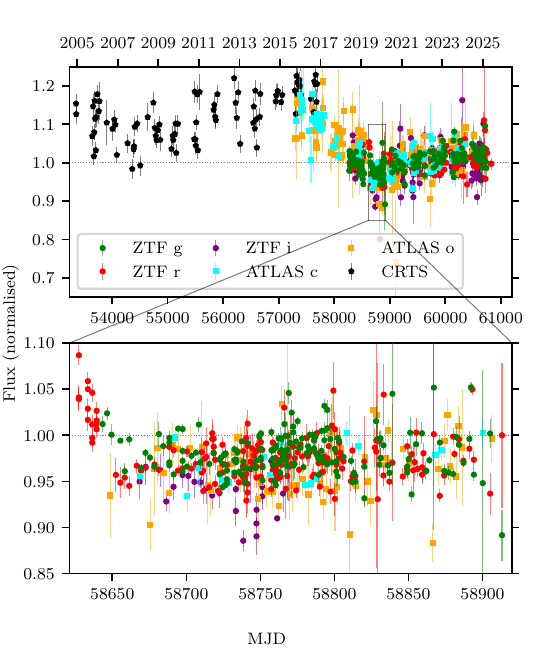}
    \caption{The light curve of ZTF~J0115+4014. It shows long timescale decreases and increases in brightness. The bottom panel shows a 10\% drop in brightness over four weeks time that triggered the first ZTF alerts, followed by a gradual re-brightening over a year. The \textbf{S} indicates when a spectrum was obtained.
    }
    \label{fig:ZTFJ0115}
\end{figure}

\begin{figure}
    \centering
    \includegraphics{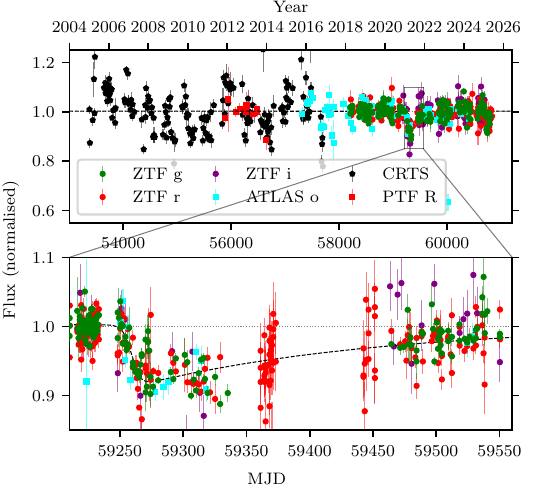}
    \caption{The light curve of ZTF~J0744+3803. The archival data suggest some long term variability. The ZTF data show a `dipping' event that lasts about 300 days and no obvious short timescale variability. The \textbf{S} indicates when a spectrum was obtained.}
    \label{fig:ZTFJ0744}
\end{figure}

\begin{figure}
    \centering
    \includegraphics{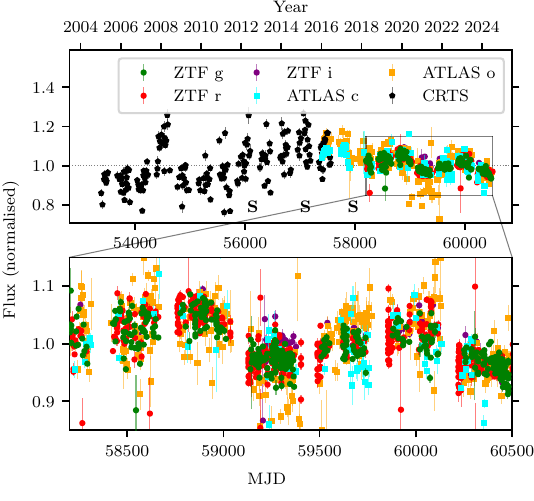}
    \caption{The light curve of ZTF~J1046+2021 shows persistent variability with an amplitude of $\approx$10\% in the ZTF light curve. The \textbf{S} indicates when a spectrum was obtained. }
    \label{fig:ZTFJ1046}
\end{figure}

\subsubsection{ZTF~J0115+4014}
ZTF~J0115+4014 (Fig.~\ref{fig:ZTFJ0115}) shows long time-scale variability, alternating between brightening and dimming over several years. The dominant variability is slow (years), but it does also show variability on a shorter timescale: the bottom panel in Fig.~\ref{fig:ZTFJ0115} shows a drop in brightness by 10\% on the order of 2 weeks. An analysis of the colour and brightness shows a significant correlation, although the correlation is not as strong as for most other objects (Fig.~\ref{fig:colourtrends}).

\subsubsection{ZTF~J0744+3803}
For ZTF~J0744+3803 (Fig.~\ref{fig:ZTFJ0744}) we observe a transit event in the ZTF light curve that lasts about 220 days. The ZTF light curve fades by about 10\% and does not show much scatter before or after the dimming event. The CRTS data preceding the ZTF data show more scatter than expected from the uncertainties. This star does not have any nearby neighbours that could affect the photometry quality. This object also shows a strong correlation between brightness and $g-r$ colour, suggesting that the obscuring material consists of small particles (dust).

\subsubsection{ZTF~J1046+2021}
ZTF~J1046+2021 shows (Fig.~\ref{fig:ZTFJ1046}) a small amount of variability of $\approx$ 5\% that appears to be correlated on a timescale of years. During the years that both ZTF and ATLAS observed this star, they show the same general trends; however, the ATLAS data are noisier compared to ZTF. The CRTS data also show that this white dwarf slowly increases and decreases in brightness. This target does not have any nearby neighbour that could have affected the photometry, and we conclude that the variability is most likely real. A colour-brightness analysis of the light curve does show a significant correlation, but the correlation is the smallest of all objects analysed in this paper.


\section{Results}\label{sec:results}
We searched the ZTF alerts for white dwarfs that show obscuration events by dust and/or debris (Sect.~\ref{sec:targetselection}). We identified 19 white dwarfs of interest: six new white dwarfs that show single obscuration events; three hot white dwarfs that show persistent and irregular variability (see Table~\ref{tab:overview}) and finally we recovered eight previously known objects (Table~\ref{tab:knownobject_overview}). The two other white dwarfs we identified are of a different nature and will be published in other papers.

We characterised the nine new white dwarfs that potentially show transiting dust and debris by analysing archival light curves, spectra, and the spectral energy distribution. We measured dimming event properties such as depth, duration, and other characteristics (see Sect.~\ref{sec:analysis} and Tables~\ref{tab:overview} and ~\ref{tab:fit_pars}). The six white dwarfs that show single dimming events show various behaviours: one event is smooth, while others show irregular variability, some show a sudden onset, while others are more gradual, and one system shows persistent periodic variability during the dimming event. The three hot white dwarfs ($T_\mathrm{eff}\approx$\qtyrange{55000}{92000}{K}) show persistent long-term irregular variability that could be the result of persistent transiting dust or debris, but there could be other possibilities, and we consider these candidates only. Based on statistical analyses, all variability is likely chromatic and slightly deeper in $g$ than in $r$; with a typical correlation between the $g-r$ colour and brightness in $g$ of \numrange{0.06}{0.78}. 

From the spectra, we determined the white dwarf spectral type for all nine white dwarfs. Most objects are DA white dwarfs, but there are two exceptions; one object is a DAZ and another is a DBZ white dwarf (Figs.~\ref{fig:spectra_all} and Table~\ref{tab:fit_pars}). We used the metal lines to measure the metal fraction and used this to infer the secular metal accretion rates. For the other white dwarfs, we estimated an upper limit to the amount of metals (calcium) in the white dwarf atmosphere.

Analysis of the spectral energy distribution shows that there is no infrared excess in any of the white dwarfs except ZTF~J2327+0019 (Fig.~\ref{fig:SEDs}). ZTF~J2327+0019 initially did not show an infrared excess, but the \textit{WISE} IR light curve features an IR-flare at the same time as the optical light drops to zero.

\section{Discussion}\label{sec:discussion}

\subsection{Engulfed in dust and debris or a cloud passing by?}
\begin{figure}
    \centering
    \includegraphics{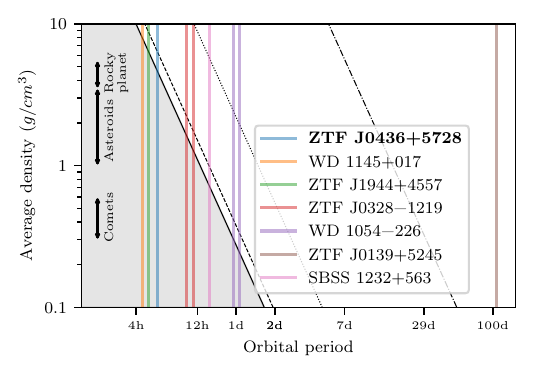}
    \caption{Orbital period versus the average density of object orbiting white dwarfs, adapted from \citet{vanderbosch2021}. The solid line shows the minimum orbital period for a given average density (and no tensile strength). The dashed, dotted, and dashed-dotted line show the limits for eccentricities of 0.1, 0.5, and 0.9. Vertical lines show the orbital periods for different white dwarfs with transiting debris.}
    \label{fig:period_density}
\end{figure}

The light curves of the six white dwarfs presented here suggest material (dust and/or larger debris) passing between the observer and the star, resulting in significant obscuration. To explain these events, there are two distinct physical geometries: (1) the white dwarf being engulfed by a global, geometrically thick ring or shell of dust and debris, or (2) a distant, extended cloud of dust and debris on a long, eccentric orbit passing through our line of sight only once. For ZTF~J0436+5728, a \qty{5.910}{h} periodic signal in the light curve is present that we interpreted as the orbital period (see also Fig.~\ref{fig:period_density}), placing the obscuring material close to the white dwarf ($\approx$\qty{1}{\Rsun}). However, as can be seen in the Figure, ZTF~J0139+5245 seems to show a much longer recurrence time of events, and if it is the orbital period, it would put the dust cloud much further away. In the rest of this subsection, we consider which of the two scenarios is the correct interpretation for the five white dwarfs presented in this paper that show only a single detected event.

To get some more constraints, we can inspect the minimal timescale of variability, which is likely set by the crossing time of the material; how long it takes for the obscuring material to cross in front of the white dwarf. This does assume that the dust and debris do not change significantly on these same timescales.
The crossing timescale is:
\begin{equation}\label{eq:crossingtime}
    \tau_\mathrm{cross}\approx\dfrac{2R_\mathrm{WD}}{v}
\end{equation} 
with $v$ the relative velocity of the transiting material. If we assume a Keplerian velocity and a circular orbit, the timescale is \qty{62}{s} for orbital periods of $\approx$6 hours, and \qty{20}{minutes} for objects on a 5-year orbit ($\tau_\mathrm{cross}\propto P^{1/3}$). In the high-speed light curves, we detected typical correlation timescales of \qtyrange{100}{200}{s} (see Figs.~\ref{fig:chimera_J0313}--\ref{fig:hs_J2327}), which suggests that the material is moving fast and is therefore close to the white dwarf. However, light curves are too chaotic to detect transits of individual fragments of material, as is seen in other white dwarfs with transiting debris \citep[see e.g.][]{gansicke2016}. We also note that for long period but very eccentric orbits, the material also has high velocities at the pericenter. So based on the variability timescale, we can conclude that the material must be moving quite fast, is close to the white dwarf, either in a short period circular orbit, or an eccentric longer period orbit.

For ZTF~J2327+0019, we can also consider the infrared brightening coincident with the optical dimming. The fact that they are coincident in time would support that the white dwarf is engulfed by dust due to a disruption event (possibly tidal disruption event) close to the white dwarf. For the alternative scenario, where a dust and debris cloud appeared further away from the white dwarf, the infrared brightening would have most likely appeared before the optical dimming. An example of this is ASASSN-21qj \citep{kenworthy2023}, where a collision occurred \qtyrange{2}{16}{au} away from a main-sequence star that generated a cloud of material that started emitting in the IR, and later, as the cloud expanded or moved in its orbit, obscured the star, resulting in a dip in the optical light curve.

To summarise: we cannot fully exclude the scenario where a distant debris cloud passes in front of the white dwarf once (2), but given indirect evidence, the scenario where the white dwarf is engulfed by dust and debris (1), seems more likely. We will assume that this interpretation is correct for the rest of the discussion section.

\subsection{Understanding the light curve properties and underlying physical process}

\subsubsection{ZTF~J0436+5728: a rock shedding its outer layers}
First, we consider ZTF~J0436+5728, whose clear periodic signal we assume to be an orbital period. The dimming event starts suddenly and the signal stays coherent in phase for $\approx$500 cycles, although its amplitude decreases over time, and the phase-folded light curve shows short-duration dips that suggest multiple small chunks of material generating dust and obscuring the white dwarf. The coherence is the key constraint: a fully disrupted body would spread into a ring within $\approx$20 orbits \citep{veras2017}, and any periodic signal would quickly disappear. The parent body must therefore have remained largely intact while actively shedding its outer layers \citep[similar to active asteroids in the Solar System, see][]{jewitt2012}. The `V'-shape of the periodic signal further suggests that dust and debris both lead and trail the parent body, with mass lost through the first and second Lagrange points \citep[e.g. see Fig. 6 in][]{veras2017}. Some material also lingers, since years after the event the white dwarf is still a few percent fainter than before, as supported by the subtle but significant colour change as a function of brightness.

These observations are all consistent with the model proposed by \citet{veras2017} for WD~1145+017, and with the other white dwarfs with transiting debris that show periodic transits: a differentiated rocky body in a near-circular orbit close to the tidal radius, intermittently shedding its lower-density outer layers and increasing its average density as it slowly spirals closer to the white dwarf. Assuming the body has no tensile strength, we can infer its average density from
\begin{equation}\label{eq:perioddensity}
P_\mathrm{min} \simeq 12.4 [\mathrm{h}]
\left( \dfrac{\bar{\rho}}{[\mathrm{g\ cm^{-3}}]} \right) ^{-1/2}
\end{equation}
where $\bar{\rho}$ is the average density and $P_\mathrm{min}$ the shortest period \citep[see][]{rappaport2013,rappaport2021,vanderbosch2021}; for a more detailed definition, see Eq. 9 in \citet{veras2016}. Because Eq.~\ref{eq:perioddensity} gives the shortest period at which a strengthless body of a given density can survive, an observed period constrains the density from below only: a denser body could equally well orbit outside its Roche limit. For a circular orbit we therefore obtain $\bar{\rho}\gtrsim$\qty{4.4}{g.cm^{-3}} (\qty{4.6}{g.cm^{-3}} if the classical fluid Roche coefficient is used instead). This is above the \qtyrange{3}{4}{g.cm^{-3}} of most stony meteorites, a range exceeded only by the stony-irons \citep{britt2003}, and no silicate rubble pile can reach it because macroporosity only lowers the bulk density further. The progenitor of the current remnant was therefore iron-rich rather than an undifferentiated asteroid or a comet, consistent with the exposed interior of a differentiated body. We note that this limit assumes no tensile strength; a body held together by internal strength could survive at this period with a lower density \citep[as inferred for the planetesimal orbiting SDSS~J1228+1040 by][]{manser2019}.

\subsubsection{One-off dimming events: collisional cascades}
The ZTF light curves of the other five white dwarfs show two types of variability: high-amplitude variability on short timescales ($\lesssim$ hours), and slower, smoother variability lasting weeks to months. We interpret the short-timescale variability as transits of a large number of active, dust-generating chunks of debris (see also Eq.~\ref{eq:crossingtime}), varying in activity level over hours to days and occupying a range of orbits. The slower component is then the obscuration of the white dwarf by the circumstellar dust cloud or disc that this debris generates, supported by the reddening observed as the white dwarfs fade. Because the chunks generate dust continuously, the total amount of dust builds up over time into a geometrically thick ring, increasing the opacity. Viewed at a high inclination, such an event shows both components, whereas at an intermediate inclination only the slower one is visible; ZTF~J0313+5206 and ZTF~J0015+3038 are examples of the two cases.

The ingress duration is set by how long it takes to grind the parent body down into enough dust to obscure the white dwarf, and some of the light curves suggest that this happens almost instantaneously, pointing to a catastrophic disruption rather than gradual erosion. This fits into the scenario described by \citet{kenyon2017a} and \citet{brouwers2022}: a large rocky body was tidally disrupted in the past, leaving remnants on eccentric orbits; collisions between these remnants built up a dust and debris disc close to the white dwarf, consistent with the metal enrichment detected in two of our objects and plausibly present in all of them; and the event we observe is a later collision between a remnant still on an eccentric orbit and a chunk of debris already in the disc, which triggered a collisional cascade that produced a large amount of new dust and debris. The shallow, short-duration precursor event in the light curve of ZTF~J2327+0019 may mark that initial collision, with the full obscuration following a few months later.

Once the dust and debris has engulfed the white dwarf, the cascade is predicted to flatten the debris into a thin disc and the dust to begin sublimating, while active debris can continue to add new dust; the combination of these processes sets how long the dimming event lasts. Comparing with the numerical predictions of \citet{kenyon2017a}, the time for the vertical scale height to change is of order days, too short to explain the timescales we observe, whereas the time for the disc mass density to decrease by an order of magnitude is days to years and matches them better, although an ongoing collisional cascade could impede the gravitational settling. For example, ZTF~J2327+0019 takes about three years to return to its nominal brightness, which according to \citet{kenyon2017a} corresponds to an object of $\approx$\qtyrange[]{3}{10}{km}, broadly consistent with the lower end of our mass estimates (Sect.~\ref{sec:mass}).

\subsection{Reddening}
\begin{figure}
    \centering
    \includegraphics[]{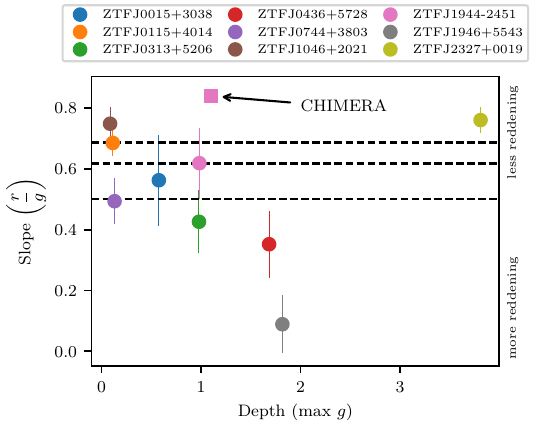}
    \caption{The correlation between the $r$ and $g$ magnitudes as function of transit depth. A slope of 1 corresponds to no reddening, and a slope of 0 corresponds to infinite reddening. When fitting the slope, the model assumes that the ZTF $g$ and $r$ were taken simultaneously, which is not correct. We bootstrapped the data to try and account for this, but the result should still be interpreted with caution. The CHIMERA measurement of ZTF~J1944-2451 is the exception and does have simultaneous $g$ and $r$ data (square). The horizontal dashed lines indicate the typical amount of reddening for ISM-like material.}
    \label{fig:reddening}
\end{figure}

In order to show that the white dwarfs are obscured by dust and to infer the typical dust particle size, we attempted to measure the amount of reddening (again, see Fig.~\ref{fig:colourtrends}). The main difficulty here is that the ZTF $g$ and $r$ measurements are not obtained simultaneously, but typically \qtyrange{1}{2}{h} apart. As we can see from the ZTF and especially the high-cadence light curves, some white dwarfs can change their brightness on much shorter timescales. For example, ZTF~J1946+5543 shows many short duration dips (Fig.~\ref{fig:ZTFJ1946}), which results in a large amount of scatter in the $g$ versus $r$ colour plot, which makes the colour measurement unreliable. On the other hand, the three hot white dwarfs tend to be dominated by longer term variability, and the colour trend measurement seems reasonably robust. 

Caveats aside, in Fig.~\ref{fig:colourtrends} we observe that all white dwarfs redden as they fade, which is consistent with dust extinction. The amount of reddening is also approximately similar to the amount of reddening observed for the ISM, indicated with the dashed lines (reddening of $R_V$=2.1, 3.1, and 4.1). One object stands out with substantially stronger reddening, ZTF~J1946+5543 ($\dfrac{r}{g}=0.09\pm0.09$), with ZTF~J0436+5728 and ZTF~J0313+5206 also falling below the $R_V$=2.1 line. However, as discussed above, for objects whose variability is large and on short timescales these measurements might not be reliable. For ZTF~J1944\textminus2451 the simultaneous CHIMERA measurement (Fig.~\ref{fig:reddening}) lies much closer to 1 than the non-simultaneous ZTF value, illustrating this effect. However, it is possible that the dust properties (size distribution) and therefore amount of reddening is not the same for dust that is responsible for the short and long timescale obscuration.

\subsection{Metal enrichment and secular accretion rate}
\label{sec:accretionrate}
\begin{figure}
    \centering
    \includegraphics{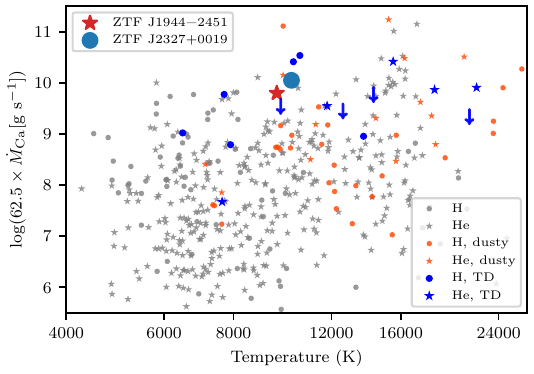}
    \caption{The inferred accretion rate based on the metal absorption line strength versus the white dwarf temperature.  The factor of 62.5 is chosen to account for the typical calcium abundance. The objects from this paper with measured abundances are indicated with a blue dot and red star, while the four objects with upper limits are shown with blue arrows. Other markers show metal-enriched white dwarfs from MWDD \citep{dufour2017}. `TD' stands for other white dwarfs with transiting debris, and `dusty' stands for objects with detected infrared excess. This figure is adapted from \citet{bhattacharjee2025}.}
    \label{fig:accretion}
\end{figure}

Metal enrichment of the white dwarf atmosphere is commonly observed for white dwarfs with orbiting debris. Of the 14 known systems, only three show no evidence of metal enrichment (and those three also happen to be of type DA). Among the remaining 11 white dwarfs showing metal enrichment, 6 have hydrogen atmospheres (type DAZ), and 5 have helium atmospheres (types DBZ or DZ);
see also Table~\ref{tab:knownobject_overview}. 
For objects discovered in this paper, the ratio is much lower: out of the six new white dwarfs with clear planetesimal disruption events and spectra, only two show metal enrichment. For comparison, 11--15\% of all white dwarfs show metal lines (Ca) in the optical \citep{obrien2024,lopez-sanjuan2024} and 40\% in UV spectra, mostly detected via Si absorption lines \citep{koester2014,ouldrouis2024}. 
Although it is possible that there has not been enough time for the white dwarfs' atmospheres to be enriched, this is most likely a data quality limitation. A calculation of the Ca accretion rate shows that with the obtained spectra, we can only detect the most metal-enriched white dwarfs $\log\dot{M} \gtrsim 9.3$, Fig.~\ref{fig:accretion} \citep[see also Fig. 8 of][]{koester2014}. 

If we compare the accretion rate of the two new white dwarfs to all other metal-enriched white dwarfs, they have some of the highest accretion rates. With a $\mathrm{\log{Ca/He}}\approx -7$, ZTF~J1944\textminus2451 is one of the most metal-enriched white dwarfs known, especially considering its temperature \citep[see Fig. 11 in ][]{hollands2017}.
However, compared to white dwarfs with transiting debris, the accretion rate is typical. The fact that these white dwarfs are strongly enriched suggests that plenty of planetesimal material has already made it into the white dwarf atmosphere in the recent past, and the disruption events we have detected are not the first for these white dwarfs. Given the relatively high upper limits of the four other white dwarfs and the fact that most white dwarfs with transiting debris are metal-enriched; we expect metals to be present in the atmospheres of most of our systems, but higher S/N spectra are needed to reveal the metal enrichment.

\subsection{Lack of metal emission lines}\label{sec:noemission}
In some white dwarfs with orbiting debris, the material can sublimate and form a gas disc and can be detected as metal emission lines, most commonly in Ca~\textsc{ii} \citep{gansicke2006,manser2020,bhattacharjee2025,xu2024}. In addition, \citet{rogers2025} recently showed that emission lines are not always persistent and can appear and disappear on short timescales. None of the white dwarfs presented in this paper show metal emission lines. This suggests that only a small amount of gas is present or generated as a result of the disruption events. In addition, we also do not detect any transient emission lines for objects where we have multiple spectra, even for ZTF~J2327+0019 where we have a spectrum just after the most extreme point of the obscuration event. This suggests that either not enough dust sublimates into gas, or if it does, it quickly disappears. We note that the signal-to-noise ratio of many of the available spectra (especially the ones not obtained with Keck or Palomar) is low, and only strong emission lines would be detectable.

\subsection{Infrared excess}

\begin{figure}
    \centering
    \includegraphics[]{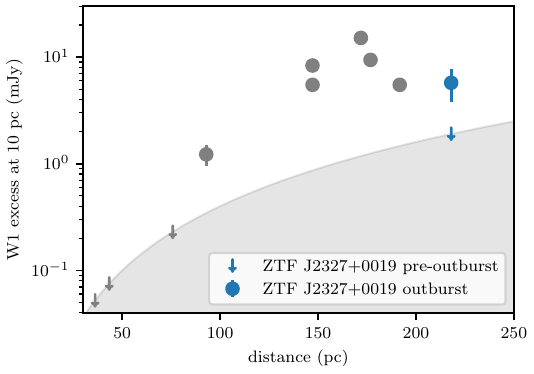}
    \caption{The intrinsic infrared excess (scaled to \qty{10}{pc}) in the W1 band for white dwarfs with transiting debris as a function of distance. The grey arrows show the upper limit for IR-excess, and the grey region shows the approximate region for which the S/N is too low to detect IR-excess. The blue arrow shows the limit on IR excess pre-event, and the blue dot shows the peak IR-excess during the event of ZTF~J2327+0019.}
    \label{fig:IRexcess}
\end{figure}

As mentioned in the introduction, some white dwarfs show an infrared excess from circumstellar dust discs, including some of the white dwarfs with transiting debris. To better understand the IR brightening of ZTF~J2327+0019 and the lack of any IR detection of the other systems, we compare the objects from this paper to other white dwarfs. To do so, we applied the same SED fitting method to all white dwarfs with known transiting debris. After fitting the SED, we found that six show infrared excess in the W1 band (excluding any white dwarfs that have nearby, bright IR sources). For two, the IR-excess was already known: WD~1145+017 \citep{xu2018a}, SBSS~1232+563 \citep{hermes2025}. For the others the infrared excess was not identified earlier: SDSS~J0107+2107, ZTF~J0139+5245, ZTF~J0923+4236, and WD~J1237+5937. Fig.~\ref{fig:IRexcess} shows an overview of white dwarfs with transiting debris and their excess infrared in the W1 band.

The figure shows that ZTF~J2327+0019 is the most distant (\qty{220}{pc}) white dwarf with transiting debris that has an excess in the W1 IR-band. The intrinsic amount of IR excess is \qty{5}{mJy} when scaled to \qty{10}{pc}; similar to the amount of IR excess detected for some other white dwarfs with transiting debris. However, white dwarfs with dusty discs that do not show transiting debris, usually present an IR excess of the order of \qty{10}{mJy} \citep[e.g.][]{xu2020}, about a factor of two more than ZTF~J2327+0019, and Fig.~\ref{fig:IRexcess} shows that the two populations overlap. 

Although the amount of infrared excess of ZTF~J2327+0019 is not unusual compared to other white dwarfs with infrared excess (with or without transiting debris), the fact that it varies over time is novel for white dwarfs with transiting debris. However, infrared variability of white dwarfs in general is certainly not new \citep[e.g.][]{xu2014,xu2018,swan2019,swan2020,guidry2024,noor2025}. We can compare the outburst of ZTF~J2327+0019 to the outburst of WD~0145+234, a white dwarf with a small amount of infrared excess and Ca-\textsc{ii} emission lines that showed an outburst in \textit{WISE} \citep{wang2019}. It brightened by $\approx$\qty{1}{mag} in the \textit{WISE} bands within half a year and was still rising thereafter; relative to its quiescent \textit{AllWISE} flux ($W1=13.82$\,mag at \qty{29.4}{pc}, from the Gaia DR3 parallax) that is an increase of $\approx$\qty{1.4}{mJy} in $W1$. Scaled to a common distance of \qty{10}{pc}, the two flares are of comparable size: $\approx$\qty{12}{mJy} for WD~0145+234 against \qty{5.8\pm1.9}{mJy} for ZTF~J2327+0019. The two events differ in an important respect, however. \citet{wang2019} find that WD~0145+234 already hosted a dust disc in its quiescent state, and interpret the outburst as the replenishment or redistribution of dust that was already there. For ZTF~J2327+0019 the infrared flux before the event is consistent with the white dwarf photosphere alone, with no detectable excess, so the excess indicates the production of new dust, as the optical light curve also suggests. \citet{wang2019} also report no significant optical variability during that outburst; ZTF~J2327+0019 is the first white dwarf in which an infrared brightening and an optical obscuration event are seen together.

To summarise, the total amount of IR excess is not unusually large compared to other white dwarfs, but the increase in IR excess is comparable to one of the largest IR-flares detected around white dwarfs. We therefore suggest that the total amount of dust around ZTF~J2327+0019 is correspondingly modest compared to some of the white dwarfs with large dust discs, but the large IR flare does suggest that a significant amount of warm dust was produced. This dust could slowly settle in an optically thick disc, decreasing the amount of IR excess (although \textit{WISE} stopped observing to confirm this). We also note that the inclination of this disc is likely high, which could explain why the amount of IR light is relatively small compared to some of the other white dwarfs. 

\subsection{How massive are the disrupted rocky bodies?}\label{sec:mass}

\begin{table}[]
    \centering
    \footnotesize
    \caption{Mass estimates for ZTF~J1944\textminus2451 and ZTF~J2327+0019 from the three methods described in the text. Note that the estimate based on the Ca abundance is the minimum mass accreted over the Ca-diffusion timescale, \qty{5E6}{yr} and \qty{100}{yr} for ZTF~J1944\textminus2451 and ZTF~J2327+0019, respectively.}
    \label{tab:mass_methods}
    \begin{tabular}{>{\raggedright\arraybackslash}p{3.3cm}>{\raggedright\arraybackslash}p{1.9cm}>{\raggedright\arraybackslash}p{1.9cm}}
        Method (assumptions) & M (J1944\textminus2451) & M (J2327+0019) \\
        \hline
        Extinction, cylinder & \qty{5e17}{g} & \qty{3e18}{g} \\
        Extinction, full sphere & \qty{5e19}{g} & \qty{3e20}{g} \\
        Extinction, replenished & \qty{3e20}{g} & \qty{4e21}{g} \\
        Ca-metallicity (minimum) & \qty{2e24}{g} & \qty{3.5e19}{g} \\
        IR flare (SED) & -- & \qtyrange{5e16}{1e18}{g} \\
    \end{tabular}
\end{table}

To estimate the masses of the parent bodies and the amount of material accreted by the white dwarfs, we use three methods: (1) extinction to infer the amount of dust and debris around the white dwarfs, (2) white dwarf atmospheric metallicity to infer the accreted mass, and (3) infrared emission to infer the amount of warm circumstellar dust. We limit this discussion to only ZTF~J1944\textminus2451 and ZTF~J2327+0019 which show the most extreme events, are the only two with detected metal enrichment of the atmospheres, and of the two, ZTF~J2327+0019 shows an infrared flare (ZTF~J1944\textminus2451 does not). Table~\ref{tab:mass_methods} summarises the resulting mass estimates, which span nearly seven orders of magnitude depending on the method and assumptions.

\begin{figure}
    \centering
    \includegraphics[]{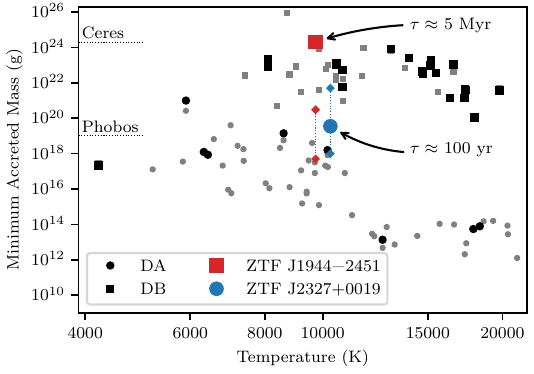}
    \caption{The minimum accreted mass as function of white dwarf temperature for ZTF~J1944\textminus2451 and ZTF~J2327+0019 based on the Ca metallicity. For comparison, we show other DA and DB white dwarfs (dots and squares) from the PEWDD catalogue \citep{williams2024}. Black points are estimates using the Ca accretion rate, and grey dots are mass estimates based on any and all detected metal lines.
    The diamonds connected by dotted lines show the range of dust mass-estimates for the observed events based on the amount of extinction.}
    \label{fig:accreted_mass}
\end{figure}

To estimate the mass in the transiting dust and debris we use the same method as in \citet{gansicke2016} \citep[but see also ][]{chen2001,rappaport2012,vanderburg2015,vanderbosch2021}:
\begin{equation}
M_\mathrm{dust} = A \tau \kappa^{-1}
\end{equation}
where $A$ is the area of the dust and debris obscuring the white dwarf,  $\kappa$=\qty{1000}{cm^2.g^{-1}} is the extinction coefficient \citep{ossenkopf1992}, and $\tau$ is the optical depth. For ZTF~J1944\textminus2451, we assume $\tau=0.5$ (39\% absorption), and for ZTF~J2327+0019 $\tau=3$ (95\% absorption). Using conservative assumptions, a cylindrical dust distribution with a radius of \qty{1.25}{\Rsun} (corresponding to an orbital period of \qty{5}{hr}) and height of $2R_{WD}$, we take $A$ as the lateral surface area of the cylinder\footnote{For comparison, bright Sun-grazing (Kreutz-group) comets such as C/1882 R1 have estimated masses of up to $\approx$\qty{e19}{g} \citep{knight2010}.}. If instead the dust and debris fully engulf the white dwarf (a spherical shell of the same radius), we take $A$ as the surface area of the sphere, which raises the mass estimate by roughly two orders of magnitude. Finally, the mass estimate increases by a further order of magnitude if we assume that the dust lifetime is $\approx$ a month \citep[e.g.][]{vanderbosch2021} and is constantly replenished for 0.5/1 years (for ZTF~J1944\textminus2451/ZTF~J2327+0019); under these assumptions, and assuming a bulk density of \qty{3}{g.cm^{-3}}, the dust masses are equivalent to asteroids with radii of \qty{29}{km} and \qty{68}{km}.

As already shown in Sect.~\ref{sec:method_metals}, we can also estimate the minimum amount of mass that was accreted onto the white dwarf based on the amount of metal enrichment. As calculated in Sect.~\ref{sec:analysis}, the minimum accreted mass for ZTF~J1944\textminus2451 is up to about twice the mass of Ceres, and for ZTF~J2327+0019 a few times the mass of Phobos. The large difference between the estimates is due to the fact that the diffusion timescale of the convective zone for ZTF~J1944\textminus2451 is about $\approx$\qty{5E6}{yr}, while for ZTF~J2327+0019 it is only $\approx$\qty{100}{yr}. As can be seen in Fig.~\ref{fig:accreted_mass}, these are some of the highest estimates compared to other DAZ and DBZ white dwarfs \citep{williams2024}.

If we assume that the mass accretion rate and conversion rate of planetesimals to dust is the same, we can calculate how often the detected events should occur. We use the cylindrical and spherical extinction estimates here, and exclude the replenished estimate, which already folds in an assumed event duration. For ZTF~J1944\textminus2451, mass-estimates of \qtyrange{5E17}{5E19}{g} combined with an accretion rate of $\log\dot{M}=9.7$--$10.1$ [\unit{g.s^{-1}}] suggest a recurrence time of \qtyrange{1.3}{320}{years}. For ZTF~J2327+0019, mass-estimates of \qtyrange{3e18}{3e20}{g} combined with an accretion rate of $\log\dot{M}=10.05\pm0.15$ [\unit{g.s^{-1}}] suggest a recurrence time of \qtyrange{6}{1200}{yr}. Based on the light curves, we can exclude the shortest recurrence time estimates, suggesting that the lowest mass estimates are too low. We stress that the width of these ranges is set almost entirely by the assumed dust geometry: the cylindrical and spherical cases differ by a factor of $\approx$100, whereas the measured spread in the accretion rate contributes only a factor of $\approx$2.

For ZTF~J2327+0019 we detected an IR flare and we can use that to estimate the amount of warm dust around the white dwarf. However, the signal-to-noise ratio is too low for any SED modelling and only a basic analysis is feasible. We assume that the infrared outburst is similar to the outburst of WD~0145+234 and follow the same procedure as in \citet{wang2019} to estimate the dust mass. In short, the observed infrared flux depends on the total mass of the dust by:
\begin{equation}
    f_\nu = \dfrac{3M_\mathrm{dust}}{4\rho d_L^2} \left< \dfrac{Q_\nu}{a} \right > B_\nu (T)
\end{equation}
where $f_\nu$ is the observed flux, $M_\mathrm{dust}$ the total dust mass, $\rho$ the internal density of the grains, $d_L$ the distance, $Q_\nu$ the absorption efficiency which is $\approx 1$ \citep{laor1993}, and $a$ is the particle size. We evaluate this expression for the observed \qty{12}{\mu Jy} excess at \qty{219}{\parsec}, adopting the dust temperature of \qty{1200}{K} implied by the $W2/W1$ colour (Sect.~\ref{sec:analysis}). The assumption $Q_\nu\approx1$ is only valid for grains larger than $\approx\lambda/2\pi\approx$\qty{0.5}{\micro\metre}, so we adopt \qtyrange{1}{10}{\micro\metre} for the particle size and \qtyrange{2}{4}{g.cm^{-3}} for the grain density, which gives a dust mass of \qtyrange{5e16}{1e18}{g}. As a consistency check, at fixed $T$, $a$, and $\rho$ the dust mass scales as $M_\mathrm{dust}\propto f_\nu d_L^2$, so this flare implies $\approx$0.5 times the dust mass of the outburst of WD~0145+234 (\qty{1.4}{mJy} at \qty{29.4}{\parsec}): the greater distance almost exactly compensates the fainter flare. 

The resulting dust mass estimate (Table~\ref{tab:mass_methods}) is one to four orders of magnitude lower than the metallicity estimate and the less conservative extinction estimates; only the most conservative (cylindrical) extinction estimate is comparable. We speculate that the mass-estimate from the IR is underestimated because the dust forms an optically thick disc, instead of an optically thin shell (which is what is assumed in the calculation). In addition, the disc is likely viewed from a high inclination angle. The combination of these effects can significantly decrease the amount of detected IR emission (see, for example, \citealt{bhattacharjee2025a}). We conclude that this makes infrared emission and the assumptions of an optically thin shell not suitable to estimate the mass. More complex SED modelling is needed to get a more accurate mass estimate \citep[e.g.][]{ballering2022}.

\subsection{Completeness, event rate and recurrence rate}
Because our search relies on alerts, a transit must make the white dwarf fade by at least five standard deviations on at least three occasions, so we are sensitive only to deep events; the cadence, seasonal gaps, and signal-to-noise limitations (Fig.~\ref{fig:negalerts}) make short-duration or shallow transits unlikely to be detected. This is borne out by the systems of \citet{bhattacharjee2025}, five of the six of which we did not recover even though some triggered alerts (Table~\ref{tab:knownobject_overview}); WD~J0923+7326 triggered almost 100 alerts, but its irregular variability was not initially recognised as due to debris transits \citep{guidry2021}. Of the white dwarfs showing deep, long-duration debris transits, however, we recovered all of them.

We detect only single events for the six white dwarfs, whereas other white dwarfs with transiting debris show recurring activity over years. A repeat of a short event such as that of ZTF~J0015+3038 or ZTF~J0313+5206 could in principle have fallen in a seasonal gap, but this is unlikely over the \qty{7.6}{yr} ZTF baseline, so we conclude that these events have recurrence times of $\gtrsim$\qty{7.6}{yr}.

\begin{figure}
    \centering
    \includegraphics{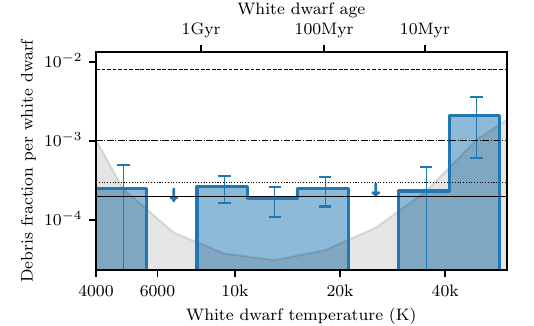}
    \caption{Transiting debris occurrence rate around white dwarfs as a function of temperature. Error bars show the Poisson uncertainty only, with the grey region showing the approximate detection limit, and down-pointing arrows show the 99\% confidence interval upper limit. The two rightmost, hashed bins rest entirely on the three hot, irregularly variable white dwarfs presented in this work. These are debris \textit{candidates} only: none of them shows photospheric metal lines or an infrared excess (Sect.~\ref{sec:analysis_irregularvariables}). If they do not host debris, these two bins are upper limits and the apparent excess at the highest temperatures disappears.
    Horizontal lines show the occurrence fraction of debris: the solid line shows the estimate from this work, the dotted line shows the estimate from \citet{bhattacharjee2025} based on ZTF data but using a different selection method, the dashed-dotted line shows the rate based on the closest white dwarf with detected debris transits \citep[see][]{aungwerojwit2024}, and the dashed line shows the rate of metal-enriched white dwarfs with detectable debris transits \citep{robert2024}. }
    \label{fig:agehist}
\end{figure}

ZTF has observed \num{70907} white dwarfs ($G<20$, $P_\mathrm{WD}>0.75$ from \citealt{gentilefusillo2021a}{}, and $>$300 epochs), of which the 14 with debris transits found in our search (six new and eight previously known) give a ratio of $\approx 2 \times 10^{-4}$. This is slightly below the $\approx 3 \times 10^{-4}$ found by \citet{bhattacharjee2025}, who also used ZTF but with a different selection method, consistent with our lower sensitivity to low-amplitude transits. Higher values follow from the distance to the nearest known system (WD~1054\textminus226 at \qty{36}{pc}, Table~\ref{tab:knownobject_overview}; $\approx 1 \times 10^{-3}$, \citealt{aungwerojwit2024}) and from the more precise TESS data ($\approx 8 \times 10^{-3}$, \citealt{robert2024}). Both are a factor of a few to $\approx$40 larger than our value, because debris is not always active and because ZTF is less precise than Kepler and TESS. ZTF therefore detects only the 2.5--20\% of white dwarfs with the most significant debris transits, the tip of the iceberg.

Fig.~\ref{fig:agehist} shows the occurrence rate of white dwarfs with transiting debris as a function of temperature. The rate is approximately flat, suggesting that remnant planetary systems remain dynamically active as the white dwarf cools and that debris survives long enough to be detected. There is a possible excess at the highest temperatures ($\sim$\qtyrange{55000}{92000}{K}), but only if the three irregularly variable objects identified in this work do host debris, in which case it may reflect short-lived dust production or strong evaporation near young, UV-bright white dwarfs. The lower but non-zero rates at cooler temperatures are consistent with the longer-term delivery mechanisms discussed in Sect.~\ref{sec:intro}.

Finally, we estimate an event rate. Assuming events are distributed randomly among white dwarfs, and that each is observed on average 50\% of the time because of seasonal visibility gaps, the effective ZTF exposure is $0.5 \times \qty{7.6}{yr} \times \num{70907} \approx \num{2.7E5}$ white-dwarf years. For $G<20$, events deeper than 50\% can be detected (7 systems; Fig.~\ref{fig:negalerts}), giving $\mathcal{R}_\mathrm{G<20,depth>50\%}=$\qty{2.6e-5}{yr^{-1}} per white dwarf. For $G<19.5$ the effective exposure is $\approx$\num{1.6E5} white-dwarf years (using the ratio of white dwarfs with $G<19.5$ versus $G<20$ in the \citealt{gentilefusillo2021a} catalogue) and events deeper than 25\% are detectable (13 systems), giving $\mathcal{R}_\mathrm{G<19.5,depth>25\%}=$\qty{8e-5}{yr^{-1}} per white dwarf. These rates are averages: event probabilities are likely higher in already active and/or metal-enriched systems.

\subsection{Alerts as a discovery tool and prospects for future surveys}
The image differencing method and alerts are designed primarily to find astrophysical transients \citep[e.g. supernovae, see][]{goldstein2015,zackay2016,bellm2019,duev2019,duev2021,carrasco-davis2021}. In this paper, we systematically analyse negative ZTF alerts to identify fading or disappearing objects for the first time \citep[for related work, see][]{gallay2025}, a capability that is of interest in a range of astrophysical contexts \citep[e.g.][]{kochanek2008,kenworthy2015,villarroel2020,solano2022,kenworthy2023,tzanidakis2025}.

Because alerts arrive within minutes of an image being taken, the same search can be run in real time: about 10 white dwarfs per month reach three negative alerts with a real/bogus score \citep{duev2019} of $\mathrm{RB}>0.9$, a cut that would have recovered all of our targets. The method is not perfect, however, and its main limitation is purity: most surviving candidates are high proper motion stars or subtraction artefacts (Appendix~\ref{sec:app_vetting}). The simplest improvement is to extend the real/bogus training set to include fading stars and high proper motion stars. Ideally, surveys would run real-time forced photometry on all stars, with a proper motion correction, but this is computationally expensive; a hybrid approach in which forced photometry \citep[e.g.][]{masci2023} is performed only at alert positions, as the Vera C. Rubin Observatory \citep{ivezic2019} does, is almost as effective. While preparing this manuscript, ZTF~J2327+0019 triggered Rubin alerts, showing that such events are recoverable from the Rubin alert stream.

The prospects for finding more such systems are good. Given the estimated event rate, ZTF should detect $\approx1$ event per year as it continues to operate, and other surveys of similar depth and cadence, such as ATLAS \citep{tonry2018}, GOTO \citep{steeghs2022}, BlackGEM \citep{groot2024}, LS4 \citep{miller2025}, and the Argus Array \citep{law2022}, are poised to find more. Rubin will reach a much deeper limiting magnitude ($\approx$24) and observe an order of magnitude more white dwarfs \citep{fantin2020}; although its cadence is slower (one observation every 3 days), the events presented here last weeks to years and should be readily recognisable, so we expect Rubin to detect a few per year.

Identifying these events while they are ongoing is what makes them valuable. High-cadence photometry can measure orbital periods or characteristic timescales, constraining the orbital separation and the mean density of the disrupted body; multi-colour optical and infrared photometry can constrain grain sizes, dust temperatures, and the dust mass and geometry; and time-resolved spectroscopy can test for gas production and, for systems with diffusion timescales of months to years, for variability in the atmospheric metal lines.

\subsection{The nature of the hot, persistently and irregularly variable white dwarfs}\label{sec:hotWDnature}
Finally, we also reported the discovery of three hot white dwarfs that show persistent, irregular variability (Sect.~\ref{sec:analysis_irregularvariables}). Although irregular variability can be a sign of Roche lobe overflow accretion, both the spectra and SED do not show any signs of a (close) stellar companion or accretion disc. There is also no known physical mechanism that could cause the white dwarf to vary intrinsically as observed. Since the light curve of these white dwarfs is somewhat similar to that of some white dwarfs with transiting debris (ZTF~J0328\textminus1219, WD~1145+017, WD~J1013\textminus0427), we suggest that this variability is the result of debris orbiting the white dwarf and the longer-term variability is due to changes in debris activity \citep{aungwerojwit2024}. 

It is interesting that these three white dwarfs are all hot. It is possible that this is a selection effect; only the brightest white dwarfs trigger alerts for the detected level of variability. However, we speculate that the strong UV radiation could result in sublimation of the debris orbiting close to the white dwarf. As shown by \citet{veras2022} (their Fig. 2), the sublimation radius of a $\approx$\qty{40000}{K} white dwarf already exceeds the tidal disruption radius. Writing the sublimation radius for blackbody grains as $r_\mathrm{sub}=(R_\mathrm{WD}/2)(T_\mathrm{eff}/T_\mathrm{sub})^2$, a \qty{40000}{K} white dwarf with $R_\mathrm{WD}=$\qty{0.012}{\Rsun} gives $r_\mathrm{sub}=$\qty{2.4}{\Rsun} for a grain sublimation temperature $T_\mathrm{sub}=$\qty{2000}{K}, and \qty{4.3}{\Rsun} for \qty{1500}{K}. This is a factor of \numrange{1.5}{2.7} beyond the tidal radius of $\approx$\qty{1.6}{\Rsun} for a body with $\bar{\rho}=$\qty{3}{g.cm^{-3}}. Since the sublimation radius scales as $T_\mathrm{eff}^2$ (their Eq.~10), this effect is even stronger for our three hot white dwarfs, which are considerably hotter still ($T_\mathrm{eff}\approx$\qtyrange{55000}{92000}{K}). It could be possible that a large planetesimal got tidally disrupted close to the white dwarf, and the remnants, now in a close orbit around the white dwarf, are being evaporated by the strong UV radiation, resulting in near-constant debris activity (see also Fig. 5 and Eq. 10 in \citealt{brouwers2022}). We note that we have not detected any metal emission lines (Sect.~\ref{sec:noemission}), which would be expected if dust was continuously sublimated into gas. This scenario is only speculation, and more observations are needed to show that there is indeed debris present close to the white dwarf. Because these white dwarfs are too hot to detect metal lines in optical spectra, UV spectra are needed to detect (ongoing) metal enrichment of the white dwarf \citep{bruhweiler1981,barstow2003,preval2019}. We caution, however, that at these temperatures radiative levitation can support metals in the atmosphere without any ongoing accretion \citep[e.g.][]{barstow2003,koester2014}, so a UV metal detection alone would not establish that debris is currently being accreted; abundances in excess of the levitation predictions would be required.

\section{Summary and conclusion}\label{sec:conclusion}

We summarise our paper as follows:
 \begin{enumerate}
 \setlength{\itemsep}{3pt}

\item \textit{New white dwarfs with transiting debris:} Using ZTF alert data, we identified six previously non-variable white dwarfs that suddenly dimmed for weeks to years, with depths from 25\% to 100\%. In the most extreme case, ZTF~J2327+0019, the white dwarf disappeared entirely for a year and took three years to return to its nominal brightness (Fig.~\ref{fig:ZTFJ2327}). This brings the total number of confirmed white dwarfs with debris transits to 20, with three further candidates, and shows that obscuration by dust and debris can be sudden, intermittent, and far more extreme than previously observed.

\item \textit{Optical transit characteristics:} The events are varied: some are smooth, others show strong short-timescale variability, and one is strictly periodic (Figs.~\ref{fig:ZTFJ0015}--\ref{fig:ZTFJ2327}). High-speed photometry (Figs.~\ref{fig:chimera_J0313}--\ref{fig:hs_J2327}) reveals minute-scale variability in some objects, implying high transverse velocities and orbital periods of hours. All white dwarfs redden as they fade (Fig.~\ref{fig:colourtrends}), by an amount comparable to ISM-like dust ($R_V$ of \numrange{2}{4}), with ZTF~J1946+5543 reddening substantially more strongly; the non-simultaneous ZTF $g$ and $r$ sampling makes these values uncertain, but the simultaneous CHIMERA photometry of ZTF~J1944\textminus2451 confirms mild but significant reddening. Some objects remain slightly fainter long after the event, suggesting that dust lingers.

\item \textit{Metal enrichment of the white dwarf:} Of the six white dwarfs with single dimming events, two are metal enriched (one DAZ, one DBZ), with accretion rates among the highest known for metal-enriched white dwarfs, although typical for white dwarfs with transiting debris (Fig.~\ref{fig:accretion}); the limits on the Ca abundance of the four DA white dwarfs are not constraining. No metal emission lines were detected, indicating little or no gas production, or gas that dissipated quickly.

\item \textit{Infrared excess and variability:} None of the objects initially showed an infrared excess, most likely because they are too distant for all but the largest excesses to be detectable. ZTF~J2327+0019 is the exception: its \textit{WISE} W1 and W2 light curve brightens by a factor of a few coincident with the optical dimming, the first white dwarf in which an infrared brightening and an optical obscuration event are seen together (Figs.~\ref{fig:ZTFJ2327} and \ref{fig:IRexcess}).

\item \textit{Disruption of rocky bodies and collisional cascades:} We interpret these events as the sudden appearance of dust and debris rather than the passage of a pre-existing cloud: a rocky body was tidally disrupted close to the white dwarf in the past, and the event we observe is a later collision between a remnant on an eccentric orbit and debris already present in the resulting disc, triggering a collisional cascade. The shallow precursor dip a year before the main event of ZTF~J2327+0019 may mark that initial collision. The short-timescale variability is then the transit of dust-producing chunks of debris, and the longer-timescale variability the build-up and subsequent clearing of a dust disc or shell, whose duration matches the predicted disc-clearing timescale, although not the much shorter vertical settling timescale. The masses involved span a wide range (Table~\ref{tab:mass_methods}): \qtyrange{3e18}{4e21}{g} of dust and debris for ZTF~J2327+0019, and minimum accreted masses of \qty{3.5e19}{g} (a few times Phobos) and \qty{2e24}{g} (up to twice the mass of Ceres, accumulated over $\approx$5\,Myr) for the two metal-enriched white dwarfs. Combined with the measured accretion rates, these masses imply that such events recur every \qtyrange{1.3}{320}{yr} for ZTF~J1944\textminus2451 and \qtyrange{6}{1200}{yr} for ZTF~J2327+0019 (ranges dominated by the unknown dust geometry rather than by measurement uncertainty), while the absence of any repeat over the $\approx$\qty{8}{yr} ZTF baseline sets a lower limit of $\approx$\qty{8}{yr}.

\item \textit{ZTF~J0436+5728: a rock at the tidal disruption radius:} ZTF~J0436+5728 differs from the other five: it shows the sudden onset of a \qty{5.910}{h} periodic signal, most likely its orbital period (Fig.~\ref{fig:ZTFJ0436}), coherent over hundreds of orbits. It is the seventh white dwarf with orbiting debris for which an orbital period is measured, and the only one showing strictly periodic variability alone. This is well explained by a differentiated rocky body at the tidal disruption radius that intermittently sheds low-density outer layers while remaining largely intact: its average density (assuming no tensile strength and a circular orbit) is $\bar{\rho}\gtrsim$\qty{4.4}{g.cm^{-3}}, higher than most stony meteorites and pointing to iron-rich material (Fig.~\ref{fig:period_density}).

\item \textit{Occurrence fractions and rates:} A fraction \qty{2E-4}{} of the white dwarfs monitored by ZTF show debris transits detectable in our search, a factor of $\approx$40 below the fraction inferred from the more precise \textit{TESS} photometry \citep{robert2024}; ZTF sees only the tip of the iceberg. The events correspond to occurrence rates of $\mathcal{R}_\mathrm{G<20,depth>50\%}=$\qty{2.6e-5}{yr^{-1}} and $\mathcal{R}_\mathrm{G<19.5,depth>25\%}=$\qty{8e-5}{yr^{-1}} per white dwarf. Considering only the confirmed objects, we find no correlation between occurrence rate and white dwarf temperature, and therefore age; the three candidates are all hot, however, and if they do host debris they would imply a slight excess around young white dwarfs (Fig.~\ref{fig:agehist}).

\item \textit{Nature of the irregularly variable white dwarfs:} In addition to the six white dwarfs with single dimming events, we identified three white dwarfs that show long-term irregular variability which may also be caused by orbiting debris (Figs.~\ref{fig:ZTFJ0115}--\ref{fig:ZTFJ1046}). We speculate that this is driven by strong UV evaporation of debris near the tidal disruption radius (Sect.~\ref{sec:hotWDnature}).

\item \textit{Alerts as a discovery tool and future surveys:} This is the first systematic use of a negative alert stream to identify fading stars, and because alerts arrive within minutes the same search can be run in real time; about 10 white dwarfs per month pass our alert criteria. Given the estimated event rate, ZTF should detect $\approx$1 further event per year, and Rubin, reaching a limiting magnitude of $\approx$24, a few per year. ZTF~J2327+0019 already triggered Rubin alerts while this manuscript was being prepared.

\end{enumerate}

Several questions remain open. How do these events fit into the wider picture of how planetary systems around white dwarfs are destroyed? How massive are the disrupted bodies, and what is their origin: are they the remnants of rocky planets, or asteroids? The masses are the least certain of these: our estimates span nearly seven orders of magnitude depending on the method and the assumed geometry. Answering these questions requires measuring the occurrence rate, the recurrence time, and the orbital separation or period of the debris, as well as the connection with metal enrichment and the relation to infrared excess and flares. That in turn needs continued monitoring of the known systems, and real-time searches for new events. Rubin, but also future optical surveys such as the Argus Array \citep{law2022} and GOTTA/SiTian \citep{liu2021a}, are promising. Especially promising is combining such optical monitoring with alerts from the \textit{Roman} Space Telescope \citep{akeson2019}, which will restore the time-domain infrared capability lost when \textit{NEOWISE} ended in 2024: together they can find more systems that show optical dips coincident with infrared brightening. Immediate follow-up of events with ground-based and space-based telescopes will then allow us to measure the orbital period from high-speed photometry, the dust temperature, grain sizes, and composition with JWST, and possibly the metal enrichment of the white dwarf in real time. What our six systems do show is that these disruptions can be caught in the act: for the first time we can watch planetesimals being ground down and accreted, one collision at a time.

\section*{Data availability}
The combined optical light curves of the nine white dwarfs presented in this
paper (ZTF, ATLAS, CRTS, PTF/iPTF, and LINEAR), the \textit{WISE/NEOWISE}
light curves, the reduced high-speed photometry obtained with CHIMERA,
ULTRACAM, and PRISM, and the reduced follow-up spectra obtained with LRIS,
DBSP, and DeVeny will be made available on Zenodo.

The archival data underlying this work are available from their respective
public archives: ZTF and PTF/iPTF light curves and \textit{WISE/NEOWISE}
images from IRSA, ATLAS photometry from the ATLAS forced-photometry service,
and the SDSS, SDSS-V~DR19, DESI~DR1, and LAMOST spectra from the corresponding
survey archives.

\section*{Use of AI-assisted technologies}
Large language model assistants (Claude, Gemini) were used in preparation of this manuscript to make the text more concise, consistent, and precise, and to do copy editing. The authors take full responsibility for the content of this paper.

\begin{acknowledgements}

We thank Siyi Xu, Laura Rogers, and Jay Farihi for useful discussions.\\

This publication is part of the project ``The life and death of white dwarf binary stars'' (with project number VI.Veni.212.201) of the research programme NWO Talent Programme Veni Science domain 2021 which is financed by the Dutch Research Council (NWO). This material is based upon work supported by the National Aeronautics and Space Administration under Grant No. 80NSSC23K1068 issued through the Science Mission Directorate.\\

Based on observations obtained with the Samuel Oschin Telescope 48-inch and the 60-inch Telescope at the Palomar Observatory as part of the Zwicky Transient Facility project. ZTF is supported by the National Science Foundation under Grants No. AST-1440341 (Phase 1) and AST-2034437 (Phase 2) and a collaboration including Caltech, IPAC, the Weizmann Institute for Science, the Oskar Klein Center at Stockholm University, the University of Maryland, Deutsches Elektronen-Synchrotron and Humboldt University, the TANGO Consortium of Taiwan, the University of Wisconsin at Milwaukee, Trinity College Dublin, Lawrence Livermore National Laboratories, and IN2P3, France. Operations are conducted by COO, IPAC, and UW.

The ZTF forced-photometry service was funded under the Heising-Simons Foundation grant \#12540303 (PI: Graham).

This work used the Palomar Observatory 5.0m Hale telescope, which is owned and operated by the Caltech Optical Observatories.

Some of the data presented herein were obtained at the W. M. Keck Observatory, which is operated as a scientific partnership among the California Institute of Technology, the University of California and the National Aeronautics and Space Administration. The Observatory was made possible by the generous financial support of the W. M. Keck Foundation.

This research is based on observations obtained with the Perkins Telescope at the Perkins Telescope Observatory (PTO), which is owned and operated by Boston University. The PRISM instrument was built at Boston University with funding from the National Science Foundation, Boston University, and Lowell Observatory.

This publication makes use of data products from the Wide-field Infrared Survey Explorer, which is a joint project of the University of California, Los Angeles, and the Jet Propulsion Laboratory/California Institute of Technology, and NEOWISE, which is a project of the Jet Propulsion Laboratory/California Institute of Technology. WISE and NEOWISE are funded by the National Aeronautics and Space Administration.

This work made use of data from the Pan-STARRS1 Surveys (PS1) and the PS1 public science archive.

Funding for the Sloan Digital Sky Survey (SDSS) and SDSS-V has been provided by the Alfred P. Sloan Foundation and the Participating Institutions. The SDSS website is \url{https://www.sdss.org}.

This research used data obtained with the Dark Energy Spectroscopic Instrument (DESI). DESI construction and operations is managed by the Lawrence Berkeley National Laboratory. This material is based upon work supported by the U.S. Department of Energy, Office of Science, Office of High-Energy Physics, under Contract No. DE-AC02-05CH11231, and by the National Energy Research Scientific Computing Center, a DOE Office of Science User Facility under the same contract. Additional support for DESI was provided by the U.S. National Science Foundation (NSF), Division of Astronomical Sciences under Contract No. AST-0950945 to the NSF's National Optical-Infrared Astronomy Research Laboratory; the Science and Technology Facilities Council of the United Kingdom; the Gordon and Betty Moore Foundation; the Heising-Simons Foundation; the French Alternative Energies and Atomic Energy Commission (CEA); the National Council of Humanities, Science and Technology of Mexico (CONAHCYT); the Ministry of Science and Innovation of Spain (MICINN); and by the DESI Member Institutions: \url{https://www.desi.lbl.gov/collaborating-institutions}. Any opinions, findings, and conclusions or recommendations expressed in this material are those of the authors and do not necessarily reflect the views of the U.S. National Science Foundation, the U.S. Department of Energy, or any of the listed funding agencies.

The authors are honored to be permitted to conduct scientific research on Iolkam Du'ag (Kitt Peak), a mountain with particular significance to the Tohono O'odham Nation.

This work has made use of data from the Guoshoujing Telescope (LAMOST), a National Major Scientific Project built by the Chinese Academy of Sciences. Funding for the project has been provided by the National Development and Reform Commission. LAMOST is operated and managed by the National Astronomical Observatories, Chinese Academy of Sciences.

This research has made use of the SIMBAD database and VizieR catalogue access tool, CDS, Strasbourg Astronomical Observatory, France (DOI : 10.26093/cds/vizier).

\end{acknowledgements}

\bibliographystyle{aa}
\bibliography{references} 

@article{adams2013,
  title   = {Evolution of {Planetary} {Orbits} with {Stellar} {Mass} {Loss} and {Tidal} {Dissipation}},
  volume  = {777},
  doi     = {10.1088/2041-8205/777/2/L30},
  journal = {\apj},
  author  = {Adams, Fred C. and Bloch, Anthony M.},
  month   = nov,
  year    = {2013},
  pages   = {L30},
}

@article{allegre1995,
  title   = {The chemical composition of the {Earth}},
  volume  = {134},
  doi     = {10.1016/0012-821X(95)00123-T},
  journal = {Earth and Planetary Science Letters},
  author  = {Allègre, Claude J. and Poirier, Jean-Paul and Humler, Eric and Hofmann, Albrecht W.},
  month   = sep,
  year    = {1995},
  pages   = {515--526},
}

@article{antoniadou2019,
  title   = {Driving white dwarf metal pollution through unstable eccentric periodic orbits},
  volume  = {629},
  doi     = {10.1051/0004-6361/201935996},
  journal = {\aap},
  author  = {Antoniadou, Kyriaki I. and Veras, Dimitri},
  month   = sep,
  year    = {2019},
  pages   = {A126},
}

@article{aungwerojwit2024,
  author  = {{Aungwerojwit}, Amornrat and {G{\"a}nsicke}, Boris T. and {Dhillon}, Vikram S. and {Drake}, Andrew and {Inight}, Keith and {Kaye}, Thomas G. and {Marsh}, T.~R. and {Mullen}, Ed and {Pelisoli}, Ingrid and {Swan}, Andrew},
  title   = "{Long-term variability in debris transiting white dwarfs}",
  journal = {\mnras},
  year    = 2024,
  month   = may,
  volume  = {530},
  number  = {1},
  pages   = {117-128},
  doi     = {10.1093/mnras/stae750},
}

@article{bailer-jones2021,
  title   = {Estimating {Distances} from {Parallaxes}. {V}. {Geometric} and {Photogeometric} {Distances} to 1.47 {Billion} {Stars} in {Gaia} {Early} {Data} {Release} 3},
  volume  = {161},
  doi     = {10.3847/1538-3881/abd806},
  journal = {\aj},
  author  = {Bailer-Jones, C. A. L. and Rybizki, J. and Fouesneau, M. and Demleitner, M. and Andrae, R.},
  month   = mar,
  year    = {2021},
  pages   = {147},
}

@article{ballering2022,
  title     = {The {Geometry} of the {G29}-38 {White} {Dwarf} {Dust} {Disk} from {Radiative} {Transfer} {Modeling}},
  volume    = {939},
  doi       = {10.3847/1538-4357/ac9a4a},
  journal   = {\apj},
  publisher = {IOP},
  author    = {Ballering, Nicholas P. and Levens, Colette I. and Su, Kate Y. L. and Cleeves, L. Ilsedore},
  month     = nov,
  year      = {2022},
  pages     = {108},
}

@article{barstow2003,
  title   = {Heavy-element abundance patterns in hot {DA} white dwarfs},
  volume  = {341},
  doi     = {10.1046/j.1365-8711.2003.06462.x},
  journal = {\mnras},
  author  = {Barstow, M. A. and Good, S. A. and Holberg, J. B. and Hubeny, I. and Bannister, N. P. and Bruhweiler, F. C. and Burleigh, M. R. and Napiwotzki, R.},
  month   = may,
  year    = {2003},
  pages   = {870--890},
}

@article{bauer2019,
  title   = {Polluted {White} {Dwarfs}: {Mixing} {Regions} and {Diffusion} {Timescales}},
  volume  = {872},
  doi     = {10.3847/1538-4357/ab0028},
  journal = {\apj},
  author  = {Bauer, Evan B. and Bildsten, Lars},
  month   = feb,
  year    = {2019},
  pages   = {96},
}

@article{bellm2019,
  title   = {The {Zwicky} {Transient} {Facility}: {Surveys} and {Scheduler}},
  journal = {\pasp},
  volume  = {131},
  doi     = {10.1088/1538-3873/ab0c2a},
  author  = {Bellm, Eric C. and Kulkarni, Shrinivas R. and Barlow, Tom and others},
  month   = jun,
  year    = {2019},
  pages   = {068003},
}

@misc{bhattacharjee2025,
  title  = {A {ZTF} {Search} for {Circumstellar} {Debris} {Transits} in {White} {Dwarfs}: {Six} {New} {Candidates}, one with {Gas} {Disk} {Emission}, identified in a {Novel} {Metric} {Space}},
  doi    = {10.48550/arXiv.2502.05502},
  author = {Bhattacharjee, Soumyadeep and Vanderbosch, Zachary P. and Hollands, Mark A. and Tremblay, Pier-Emmanuel and Xu, Siyi and Guidry, Joseph A. and Hermes, J. J. and Caiazzo, Ilaria and Rodriguez, Antonio C. and van Roestel, Jan and Roulston, Benjamin R. and Riddle, Reed and Rusholme, Ben and Groom, Steven L. and Smith, Roger and Toloza, Odette},
  month  = feb,
  year   = {2025},
}

@misc{bhattacharjee2025a,
  title     = {Thick {Disks} around {White} {Dwarfs} viewed '{Edge}-off': {Effects} on {Transit} {Properties} and {Infrared} {Excess}},
  doi       = {10.48550/arXiv.2507.20594},
  publisher = {arXiv},
  author    = {Bhattacharjee, Soumyadeep},
  month     = jul,
  year      = {2025},
}

@ARTICLE{bianchi2017,
       author = {{Bianchi}, Luciana and {Shiao}, Bernie and {Thilker}, David},
        title = "{Revised Catalog of GALEX Ultraviolet Sources. I. The All-Sky Survey: GUVcat\_AIS}",
      journal = {\apjs},
         year = 2017,
        month = jun,
       volume = {230},
       number = {2},
          eid = {24},
        pages = {24},
          doi = {10.3847/1538-4365/aa7053},
archivePrefix = {arXiv},
       eprint = {1704.05903},
 primaryClass = {astro-ph.GA},
       adsurl = {https://ui.adsabs.harvard.edu/abs/2017ApJS..230...24B}
}

@article{bonsor2012,
  title   = {The scattering of small bodies in planetary systems: constraints on the possible orbits of cometary material},
  volume  = {420},
  doi     = {10.1111/j.1365-2966.2011.20156.x},
  journal = {\mnras},
  author  = {Bonsor, A. and Wyatt, M. C.},
  month   = mar,
  year    = {2012},
  pages   = {2990--3002},
}

@article{bonsor2017,
  title   = {Infrared observations of white dwarfs and the implications for the accretion of dusty planetary material},
  volume  = {468},
  doi     = {10.1093/mnras/stx425},
  journal = {\mnras},
  author  = {Bonsor, Amy and Farihi, Jay and Wyatt, Mark C. and van Lieshout, Rik},
  month   = jun,
  year    = {2017},
  pages   = {154--164},
}

@article{brouwers2022,
  title     = {A road-map to white dwarf pollution: tidal disruption, eccentric grind-down, and dust accretion},
  volume    = {509},
  doi       = {10.1093/mnras/stab3009},
  journal   = {\mnras},
  publisher = {OUP},
  author    = {Brouwers, Marc G. and Bonsor, Amy and Malamud, Uri},
  month     = jan,
  year      = {2022},
  pages     = {2404--2422},
}

@article{brown2017,
  title   = {Discovery of a {Detached}, {Eclipsing} 40 {Minute} {Period} {Double} {White} {Dwarf} {Binary} and a {Friend}: {Implications} for {He}+{CO} {White} {Dwarf} {Mergers}},
  volume  = {847},
  doi     = {10.3847/1538-4357/aa8724},
  journal = {\apj},
  author  = {Brown, Warren R. and Kilic, Mukremin and Kosakowski, Alekzander and Gianninas, A.},
  month   = sep,
  year    = {2017},
  pages   = {10},
}

@article{bruhweiler1981,
  title   = {Ionized species observed in the spectrum of the nearby white dwarf {G} 191-{B2B}.},
  volume  = {248},
  doi     = {10.1086/183639},
  journal = {\apj},
  author  = {Bruhweiler, F. C. and Kondo, Y.},
  month   = sep,
  year    = {1981},
  pages   = {L123--L127},
}

@article{buchan2025,
  author  = {{Buchan}, Andrew M. and {Tremblay}, Pier-Emmanuel and {B{\'e}dard}, Antoine and {Bauer}, Evan B. and {Cunningham}, Tim},
  title   = "{Exogeological inferences from white dwarf pollutants: the impact of stellar physics}",
  journal = {\mnras},
  year    = 2025,
  month   = dec,
  volume  = {544},
  number  = {2},
  pages   = {2098-2119},
  doi     = {10.1093/mnras/staf1832},
}

@article{caiazzo2017,
  title   = {Polluting white dwarfs with perturbed exo-comets},
  volume  = {469},
  doi     = {10.1093/mnras/stx1036},
  journal = {\mnras},
  author  = {Caiazzo, Ilaria and Heyl, Jeremy S.},
  month   = aug,
  year    = {2017},
  pages   = {2750--2759},
}

@article{carrasco-davis2021,
  title   = {Alert {Classification} for the {ALeRCE} {Broker} {System}: {The} {Real}-time {Stamp} {Classifier}},
  volume  = {162},
  doi     = {10.3847/1538-3881/ac0ef1},
  journal = {\aj},
  author  = {Carrasco-Davis, R. and Reyes, E. and Valenzuela, C. and Förster, F. and Estévez, P. A. and Pignata, G. and Bauer, F. E. and Reyes, I. and Sánchez-Sáez, P. and Cabrera-Vives, G. and Eyheramendy, S. and Catelan, M. and Arredondo, J. and Castillo-Navarrete, E. and Rodríguez-Mancini, D. and Ruz-Mieres, D. and Moya, A. and Sabatini-Gacitúa, L. and Sepúlveda-Cobo, C. and Mahabal, A. A. and Silva-Farfán, J. and Camacho-Iñiguez, E. and Galbany, L.},
  month   = dec,
  year    = {2021},
  pages   = {231},
}

@ARTICLE{chamandy2024,
       author = {{Chamandy}, Luke and {Nordhaus}, Jason and {Blackman}, Eric G. and {Wilson}, Emily},
        title = "{Second-generation planet formation after tidal disruption from common envelope evolution}",
      journal = {\pasa},
         year = 2025,
        month = feb,
       volume = {42},
          eid = {e027},
        pages = {e027},
          doi = {10.1017/pasa.2025.4},
archivePrefix = {arXiv},
       eprint = {2407.14190},
 primaryClass = {astro-ph.EP},
       adsurl = {https://ui.adsabs.harvard.edu/abs/2025PASA...42...27C}
}

@ARTICLE{chambers2016,
       author = {{Chambers}, K.~C. and {Magnier}, E.~A. and {Metcalfe}, N. and {Flewelling}, H.~A. and {Huber}, M.~E. and {Waters}, C.~Z. and {Denneau}, L. and {Draper}, P.~W. and {Farrow}, D. and {Finkbeiner}, D.~P. and {Holmberg}, C. and {Koppenhoefer}, J. and {Price}, P.~A. and {Rest}, A. and {Saglia}, R.~P. and {Schlafly}, E.~F. and {Smartt}, S.~J. and {Sweeney}, W. and {Wainscoat}, R.~J. and {Burgett}, W.~S. and {Chastel}, S. and {Grav}, T. and {Heasley}, J.~N. and {Hodapp}, K.~W. and {Jedicke}, R. and {Kaiser}, N. and {Kudritzki}, R.-P. and {Luppino}, G.~A. and {Lupton}, R.~H. and {Monet}, D.~G. and {Morgan}, J.~S. and {Onaka}, P.~M. and {Shiao}, B. and {Stubbs}, C.~W. and {Tonry}, J.~L. and {White}, R. and {Ba{\~n}ados}, E. and {Bell}, E.~F. and {Bender}, R. and {Bernard}, E.~J. and {Boegner}, M. and {Boffi}, F. and {Botticella}, M.~T. and {Calamida}, A. and {Casertano}, S. and {Chen}, W.-P. and {Chen}, X. and {Cole}, S. and {Deacon}, N. and {Frenk}, C. and {Fitzsimmons}, A. and {Gezari}, S. and {Gibbs}, V. and {Goessl}, C. and {Goggia}, T. and {Gourgue}, R. and {Goldman}, B. and {Grant}, P. and {Grebel}, E.~K. and {Hambly}, N.~C. and {Hasinger}, G. and {Heavens}, A.~F. and {Heckman}, T.~M. and {Henderson}, R. and {Henning}, T. and {Holman}, M. and {Hopp}, U. and {Ip}, W.-H. and {Isani}, S. and {Jackson}, M. and {Keyes}, C.~D. and {Koekemoer}, A.~M. and {Kotak}, R. and {Le}, D. and {Liska}, D. and {Long}, K.~S. and {Lucey}, J.~R. and {Liu}, M. and {Martin}, N.~F. and {Masci}, G. and {McLean}, B. and {Mindel}, E. and {Misra}, P. and {Morganson}, E. and {Murphy}, D.~N.~A. and {Obaika}, A. and {Narayan}, G. and {Nieto-Santisteban}, M.~A. and {Norberg}, P. and {Peacock}, J.~A. and {Pier}, E.~A. and {Postman}, M. and {Primak}, N. and {Rae}, C. and {Rai}, A. and {Riess}, A. and {Riffeser}, A. and {Rix}, H.~W. and {R{\"o}ser}, S. and {Russel}, R. and {Rutz}, L. and {Schilbach}, E. and {Schultz}, A.~S.~B. and {Scolnic}, D. and {Strolger}, L. and {Szalay}, A. and {Seitz}, S. and {Small}, E. and {Smith}, K.~W. and {Soderblom}, D.~R. and {Taylor}, P. and {Thomson}, R. and {Taylor}, A.~N. and {Thakar}, A.~R. and {Thiel}, J. and {Thilker}, D. and {Unger}, D. and {Urata}, Y. and {Valenti}, J. and {Wagner}, J. and {Walder}, T. and {Walter}, F. and {Watters}, S.~P. and {Werner}, S. and {Wood-Vasey}, W.~M. and {Wyse}, R.},
        title = "{The Pan-STARRS1 Surveys}",
      journal = {arXiv e-prints},
         year = 2016,
        month = dec,
          eid = {arXiv:1612.05560},
        pages = {arXiv:1612.05560},
          doi = {10.48550/arXiv.1612.05560},
archivePrefix = {arXiv},
       eprint = {1612.05560},
 primaryClass = {astro-ph.IM},
       adsurl = {https://ui.adsabs.harvard.edu/abs/2016arXiv161205560C}
}

@article{chen2001,
  title   = {A {Possible} {Massive} {Asteroid} {Belt} around ζ {Leporis}},
  volume  = {560},
  doi     = {10.1086/324057},
  journal = {\apj},
  author  = {Chen, C. H. and Jura, M.},
  month   = oct,
  year    = {2001},
  pages   = {L171--L174},
}

@article{chiang1997,
  title     = {Spectral {Energy} {Distributions} of {T} {Tauri} {Stars} with {Passive} {Circumstellar} {Disks}},
  volume    = {490},
  doi       = {10.1086/304869},
  journal   = {\apj},
  publisher = {IOP},
  author    = {Chiang, E. I. and Goldreich, P.},
  month     = nov,
  year      = {1997},
  pages     = {368--376},
}

@article{coughlin2023,
  title   = {A {Data} {Science} {Platform} to {Enable} {Time}-domain {Astronomy}},
  volume  = {267},
  doi     = {10.3847/1538-4365/acdee1},
  journal = {\apjs},
  author  = {Coughlin, Michael W. and Bloom, Joshua S. and Nir, Guy and Antier, Sarah and du Laz, Theophile Jegou and van der Walt, Stéfan and Crellin-Quick, Arien and Culino, Thomas and Duev, Dmitry A. and Goldstein, Daniel A. and Healy, Brian F. and Karambelkar, Viraj and Lilleboe, Jada and Shin, Kyung Min and Singer, Leo P. and Ahumada, Tomás and Anand, Shreya and Bellm, Eric C. and Dekany, Richard and Graham, Matthew J. and Kasliwal, Mansi M. and Kostadinova, Ivona and Kiendrebeogo, R. Weizmann and Kulkarni, Shrinivas R. and Jenkins, Sydney and LeBaron, Natalie and Mahabal, Ashish A. and Neill, James D. and Parazin, B. and Peloton, Julien and Perley, Daniel A. and Riddle, Reed and Rusholme, Ben and van Santen, Jakob and Sollerman, Jesper and Stein, Robert and Turpin, D. and Wold, Avery and Amat, Carla and Bonnefon, Adrien and Bonnefoy, Adrien and Flament, Manon and Kerkow, Frank and Kishore, Sulekha and Jani, Shloke and Mahanty, Stephen K. and Liu, Céline and Llinares, Laura and Makarison, Jolyane and Olliéric, Alix and Perez, Inès and Pont, Lydie and Sharma, Vyom},
  month   = aug,
  year    = {2023},
  pages   = {31},
}

@article{coutu2019,
  author  = {{Coutu}, S. and {Dufour}, P. and {Bergeron}, P. and {Blouin}, S. and {Loranger}, E. and {Allard}, N.~F. and {Dunlap}, B.~H.},
  title   = "{Analysis of Helium-rich White Dwarfs Polluted by Heavy Elements in the Gaia Era}",
  journal = {\apj},
  year    = 2019,
  month   = nov,
  volume  = {885},
  number  = {1},
  eid     = {74},
  pages   = {74},
  doi     = {10.3847/1538-4357/ab46b9},
}

@article{cui2012,
  title   = {The {Large} {Sky} {Area} {Multi}-{Object} {Fiber} {Spectroscopic} {Telescope} ({LAMOST})},
  volume  = {12},
  doi     = {10.1088/1674-4527/12/9/003},
  journal = {Research in Astronomy and Astrophysics},
  author  = {Cui, Xiang-Qun and Zhao, Yong-Heng and Chu, Yao-Quan and Li, Guo-Ping and Li, Qi and Zhang, Li-Ping and Su, Hong-Jun and Yao, Zheng-Qiu and Wang, Ya-Nan and Xing, Xiao-Zheng and Li, Xin-Nan and Zhu, Yong-Tian and Wang, Gang and Gu, Bo-Zhong and Luo, A. -Li and Xu, Xin-Qi and Zhang, Zhen-Chao and Liu, Gen-Rong and Zhang, Hao-Tong and Yang, De-Hua and Cao, Shu-Yun and Chen, Hai-Yuan and Chen, Jian-Jun and Chen, Kun-Xin and Chen, Ying and Chu, Jia-Ru and Feng, Lei and Gong, Xue-Fei and Hou, Yong-Hui and Hu, Hong-Zhuan and Hu, Ning-Sheng and Hu, Zhong-Wen and Jia, Lei and Jiang, Fang-Hua and Jiang, Xiang and Jiang, Zi-Bo and Jin, Ge and Li, Ai-Hua and Li, Yan and Li, Ye-Ping and Liu, Guan-Qun and Liu, Zhi-Gang and Lu, Wen-Zhi and Mao, Yin-Dun and Men, Li and Qi, Yong-Jun and Qi, Zhao-Xiang and Shi, Huo-Ming and Tang, Zheng-Hong and Tao, Qing-Sheng and Wang, Da-Qi and Wang, Dan and Wang, Guo-Min and Wang, Hai and Wang, Jia-Ning and Wang, Jian and Wang, Jian-Ling and Wang, Jian-Ping and Wang, Lei and Wang, Shu-Qing and Wang, You and Wang, Yue-Fei and Xu, Ling-Zhe and Xu, Yan and Yang, Shi-Hai and Yu, Yong and Yuan, Hui and Yuan, Xiang-Yan and Zhai, Chao and Zhang, Jing and Zhang, Yan-Xia and Zhang, Yong and Zhao, Ming and Zhou, Fang and Zhou, Guo-Hua and Zhu, Jie and Zou, Si-Cheng},
  month   = sep,
  year    = {2012},
  pages   = {1197--1242},
}

@article{cunningham2019,
  author  = {{Cunningham}, Tim and {Tremblay}, Pier-Emmanuel and {Freytag}, Bernd and {Ludwig}, Hans-G{\"u}nter and {Koester}, Detlev},
  title   = "{Convective overshoot and macroscopic diffusion in pure-hydrogen-atmosphere white dwarfs}",
  journal = {\mnras},
  year    = 2019,
  month   = sep,
  volume  = {488},
  number  = {2},
  pages   = {2503-2522},
  doi     = {10.1093/mnras/stz1759},
}

@article{cunningham2021,
  title   = {Horizontal spreading of planetary debris accreted by white dwarfs},
  volume  = {503},
  doi     = {10.1093/mnras/stab553},
  journal = {\mnras},
  author  = {Cunningham, Tim and Tremblay, Pier-Emmanuel and Bauer, Evan B. and Toloza, Odette and Cukanovaite, Elena and Koester, Detlev and Farihi, Jay and Freytag, Bernd and Gänsicke, Boris T. and Ludwig, Hans-Günter and Veras, Dimitri},
  month   = may,
  year    = {2021},
  pages   = {1646--1667},
}

@article{cunningham2022,
  title   = {A white dwarf accreting planetary material determined from {X}-ray observations},
  volume  = {602},
  doi     = {10.1038/s41586-021-04300-w},
  journal = {\nat},
  author  = {Cunningham, Tim and Wheatley, Peter J. and Tremblay, Pier-Emmanuel and Gänsicke, Boris T. and King, George W. and Toloza, Odette and Veras, Dimitri},
  month   = feb,
  year    = {2022},
  pages   = {219--222},
}

@article{cunningham2025,
  author  = {{Cunningham}, Tim and {Tremblay}, Pier-Emmanuel and {O'Brien}, Mairi and {Bauer}, Evan B. and {Hollands}, Mark A. and {Koester}, Detlev and {Kenyon}, Scott J. and {Charbonneau}, David and {Veras}, Dimitri and {Yusaf}, Muhammad Furqaan},
  title   = "{The dearth of high-mass hydrogen-atmosphere metal-polluted white dwarfs within 40 pc}",
  journal = {\mnras},
  year    = 2025,
  month   = may,
  volume  = {539},
  number  = {3},
  pages   = {2021-2038},
  doi     = {10.1093/mnras/staf428},
}

@article{cutri2003,
  title   = {{VizieR} {Online} {Data} {Catalog}: {2MASS} {All}-{Sky} {Catalog} of {Point} {Sources} ({Cutri}+ 2003)},
  journal = {VizieR Online Data Catalog},
  author  = {Cutri, R. M. and Skrutskie, M. F. and van Dyk, S. and Beichman, C. A. and Carpenter, J. M. and Chester, T. and Cambresy, L. and Evans, T. and Fowler, J. and Gizis, J. and Howard, E. and Huchra, J. and Jarrett, T. and Kopan, E. L. and Kirkpatrick, J. D. and Light, R. M. and Marsh, K. A. and McCallon, H. and Schneider, S. and Stiening, R. and Sykes, M. and Weinberg, M. and Wheaton, W. A. and Wheelock, S. and Zacarias, N.},
  month   = jun,
  year    = {2003},
  pages   = {II/246},
}

@article{de2020,
  title   = {The {Zwicky} {Transient} {Facility} {Census} of the {Local} {Universe}. {I}. {Systematic} {Search} for {Calcium}-rich {Gap} {Transients} {Reveals} {Three} {Related} {Spectroscopic} {Subclasses}},
  volume  = {905},
  doi     = {10.3847/1538-4357/abb45c},
  journal = {\apj},
  author  = {De, Kishalay and Kasliwal, Mansi M. and Tzanidakis, Anastasios and Fremling, U. Christoffer and Adams, Scott and Aloisi, Robert and Andreoni, Igor and Bagdasaryan, Ashot and Bellm, Eric C. and Bildsten, Lars and Cannella, Christopher and Cook, David O. and Delacroix, Alexandre and Drake, Andrew and Duev, Dmitry and Dugas, Alison and Frederick, Sara and Gal-Yam, Avishay and Goldstein, Daniel and Golkhou, V. Zach and Graham, Matthew J. and Hale, David and Hankins, Matthew and Helou, George and Ho, Anna Y. Q. and Irani, Ido and Jencson, Jacob E. and Kaplan, David L. and Kaye, Stephen and Kulkarni, S. R. and Kupfer, Thomas and Laher, Russ R. and Leadbeater, Robin and Lunnan, Ragnhild and Masci, Frank J. and Miller, Adam A. and Neill, James D. and Ofek, Eran O. and Perley, Daniel A. and Polin, Abigail and Prince, Thomas A. and Quataert, Eliot and Reiley, Dan and Riddle, Reed L. and Rusholme, Ben and Sharma, Yashvi and Shupe, David L. and Sollerman, Jesper and Tartaglia, Leonardo and Walters, Richard and Yan, Lin and Yao, Yuhan},
  month   = dec,
  year    = {2020},
  pages   = {58},
}

@article{debes2002,
  title   = {Are {There} {Unstable} {Planetary} {Systems} around {White} {Dwarfs}?},
  volume  = {572},
  doi     = {10.1086/340291},
  journal = {\apj},
  author  = {Debes, John H. and Sigurdsson, Steinn},
  month   = jun,
  year    = {2002},
  pages   = {556--565},
}

@article{debes2012,
  title   = {The {Link} {Between} {Planetary} {Systems}, {Dusty} {White} {Dwarfs}, and {Metal} {Polluted} {White} {Dwarfs}},
  volume  = {747},
  doi     = {10.1088/0004-637X/747/2/148},
  number  = {2},
  journal = {\apj},
  author  = {Debes, J. and Walsh, K. and Stark, C.},
  month   = mar,
  year    = {2012},
  pages   = {148},
}

@article{debes2025,
  title     = {Metal-polluted {White} {Dwarfs} with 21 μm {IR} {Excesses} from {JWST}/{MIRI}: {Planets} or {Dust}?},
  volume    = {170},
  doi       = {10.3847/1538-3881/ade68c},
  journal   = {\aj},
  publisher = {IOP},
  author    = {Debes, John H. and Poulsen, Sabrina and Messier, Ashley and Mullally, Susan E. and Thibault, Katherine and Albert, Loïc and Cracraft, Misty and Le Bourdais, Érika and Dufour, Patrick and Barclay, Tom and Hermes, J. J. and Kilic, Mukremin and Lafrenière, David and Mullally, Fergal and Reach, William and Quintana, Elisa},
  month     = aug,
  year      = {2025},
  pages     = {123},
}

@article{dekany2020,
  title   = {The {Zwicky} {Transient} {Facility}: {Observing} {System}},
  volume  = {132},
  doi     = {10.1088/1538-3873/ab4ca2},
  journal = {\pasp},
  author  = {Dekany, Richard and Smith, Roger M. and Riddle, Reed and Feeney, Michael and Porter, Michael and Hale, David and Zolkower, Jeffry and Belicki, Justin and Kaye, Stephen and Henning, John and Walters, Richard and Cromer, John and Delacroix, Alex and Rodriguez, Hector and Reiley, Daniel J. and Mao, Peter and Hover, David and Murphy, Patrick and Burruss, Rick and Baker, John and Kowalski, Marek and Reif, Klaus and Mueller, Phillip and Bellm, Eric and Graham, Matthew and Kulkarni, Shrinivas R.},
  month   = mar,
  year    = {2020},
  pages   = {038001},
}

@article{dennihy2018,
  title   = {Rapid {Evolution} of the {Gaseous} {Exoplanetary} {Debris} around the {White} {Dwarf} {Star} {HE} 1349-2305},
  volume  = {854},
  doi     = {10.3847/1538-4357/aaa89b},
  journal = {\apj},
  author  = {Dennihy, E. and Clemens, J. C. and Dunlap, B. H. and Fanale, S. M. and Fuchs, J. T. and Hermes, J. J.},
  month   = feb,
  year    = {2018},
  pages   = {40},
}

@ARTICLE{desicollaboration2025,
       author = {{DESI Collaboration} and {Abdul Karim}, M. and {Adame}, A.~G. and {Aguado}, D. and {Aguilar}, J. and {Ahlen}, S. and {Alam}, S. and {Aldering}, G. and {Alexander}, D.~M. and {Alfarsy}, R. and {Allen}, L. and {Allende Prieto}, C. and {Alves}, O. and {Anand}, A. and {Andrade}, U. and {Armengaud}, E. and {Avila}, S. and {Aviles}, A. and {Awan}, H. and {Bailey}, S. and {Baleato Lizancos}, A. and {Ballester}, O. and {Bault}, A. and {Bautista}, J. and {Bean}, R. and {Behera}, J. and {BenZvi}, S. and {Beraldo e Silva}, L. and {Bermejo-Climent}, J.~R. and {Beutler}, F. and {Bianchi}, D. and {Blake}, C. and {Blum}, R. and {Bolton}, A.~S. and {Bonici}, M. and {Brieden}, S. and {Brodzeller}, A. and {Brooks}, D. and {Buckley-Geer}, E. and {Burtin}, E. and {Bystr{\"o}m}, A. and {Canning}, R. and {Carnero Rosell}, A. and {Carr}, A. and {Carrilho}, P. and {Casas}, L. and {Castander}, F.~J. and {Cereskaite}, R. and {Cervantes-Cota}, J.~L. and {Chaussidon}, E. and {Chaves-Montero}, J. and {Chen}, S. and {Chen}, X. and {Circosta}, C. and {Claybaugh}, T. and {Cole}, S. and {Cooper}, A.~P. and {Cousinou}, M.-C. and {Cuceu}, A. and {Davis}, T.~M. and {Dawson}, K.~S. and {de Belsunce}, R. and {de la Cruz}, R. and {de la Macorra}, A. and {de Mattia}, A. and {Deiosso}, N. and {Della Costa}, J. and {Demina}, R. and {Demirbozan}, U. and {DeRose}, J. and {Dey}, A. and {Dey}, B. and {Ding}, J. and {Ding}, Z. and {Doel}, P. and {Douglass}, K. and {Dowicz}, M. and {Ebina}, H. and {Edelstein}, J. and {Eisenstein}, D.~J. and {Elbers}, W. and {Emas}, N. and {Escoffier}, S. and {Fagrelius}, P. and {Fan}, X. and {Fanning}, K. and {Favole}, G. and {Fawcett}, V.~A. and {Fern{\'a}ndez-Garc{\'\i}a}, E. and {Ferraro}, S. and {Findlay}, N. and {Font-Ribera}, A. and {Forero-Romero}, J.~E. and {Forero-S{\'a}nchez}, D. and {Frenk}, C.~S. and {G{\"a}nsicke}, B.~T. and {Galbany}, L. and {Garc{\'\i}a-Bellido}, J. and {Garcia-Quintero}, C. and {Garrison}, L.~H. and {Gazta{\~n}aga}, E. and {Gil-Mar{\'\i}n}, H. and {Gloudemans}, A. and {Gnedin}, O.~Y. and {Gontcho A Gontcho}, S. and {Gonzalez}, D. and {Gonzalez-Morales}, A.~X. and {Gonzalez-Perez}, V. and {Gordon}, C. and {Graur}, O. and {Green}, D. and {Gruen}, D. and {Gsponer}, R. and {Guandalin}, C. and {Gutierrez}, G. and {Guy}, J. and {Hahn}, C. and {Han}, J.~J. and {Han}, J. and {He}, S. and {Herrera-Alcantar}, H.~K. and {Heydenreich}, S. and {Honscheid}, K. and {Hou}, J. and {Howlett}, C. and {Huterer}, D. and {Ir{\v{s}}i{\v{c}}}, V. and {Ishak}, M. and {Jacques}, A. and {Jiang}, L. and {Jimenez}, J. and {Jing}, Y.~P. and {Joachimi}, B. and {Joudaki}, S. and {Joyce}, R. and {Jullo}, E. and {Juneau}, S. and {Kara{\c{c}}ayl{\i}}, N.~G. and {Karim}, T. and {Kehoe}, R. and {Kent}, S. and {Khederlarian}, A. and {Kirkby}, D. and {Kisner}, T. and {Kitaura}, F.-S. and {Kizhuprakkat}, N. and {Kong}, H. and {Koposov}, S.~E. and {Kremin}, A. and {Krolewski}, A. and {Lahav}, O. and {Lai}, Y. and {Lamman}, C. and {Lan}, T.-W. and {Landriau}, M. and {Lang}, D. and {Lange}, J.~U. and {Lasker}, J. and {Le Goff}, J.~M. and {Le Guillou}, L. and {Leauthaud}, A. and {Levi}, M.~E. and {Li}, S. and {Li}, T.~S. and {Liu}, W. and {Lodha}, K. and {Lokken}, M. and {Luo}, Y. and {Magneville}, C. and {Manera}, M. and {Manser}, C.~J. and {Margala}, D. and {Martini}, P. and {Maus}, M. and {McCullough}, J. and {McDonald}, P. and {Medina}, G.~E. and {Medina-Varela}, L. and {Meisner}, A. and {Mena-Fern{\'a}ndez}, J. and {Menegas}, A. and {Meneses-Rizo}, J. and {Mezcua}, M. and {Miquel}, R. and {Montero-Camacho}, P. and {Moon}, J. and {Moustakas}, J. and {Mu{\~n}oz-Guti{\'e}rrez}, A. and {Mu noz-Santos}, D. and {Myers}, A.~D. and {Myles}, J. and {Nadathur}, S. and {Najita}, J. and {Napolitano}, L. and {Newman}, J.~A. and {Nikakhtar}, F. and {Nikutta}, R. and {Niz}, G. and {Noriega}, H.~E. and {Nugent}, P.},
        title = "{Data Release 1 of the Dark Energy Spectroscopic Instrument}",
      journal = {\aj},
         year = 2026,
        month = may,
       volume = {171},
       number = {5},
          eid = {285},
        pages = {285},
          doi = {10.3847/1538-3881/ae4c43},
archivePrefix = {arXiv},
       eprint = {2503.14745},
 primaryClass = {astro-ph.CO},
       adsurl = {https://ui.adsabs.harvard.edu/abs/2026AJ....171..285D}
}

@article{dhillon2007,
  title   = {{ULTRACAM}: an ultrafast, triple-beam {CCD} camera for high-speed astrophysics},
  volume  = {378},
  doi     = {10.1111/j.1365-2966.2007.11881.x},
  journal = {\mnras},
  author  = {Dhillon, V. S. and Marsh, T. R. and Stevenson, M. J. and Atkinson, D. C. and Kerry, P. and Peacocke, P. T. and Vick, A. J. A. and Beard, S. M. and Ives, D. J. and Lunney, D. W. and McLay, S. A. and Tierney, C. J. and Kelly, J. and Littlefair, S. P. and Nicholson, R. and Pashley, R. and Harlaftis, E. T. and O'Brien, K.},
  month   = jul,
  year    = {2007},
  pages   = {825--840},
}

@article{drake2009,
  title     = {{FIRST} {RESULTS} {FROM} {THE} {CATALINA} {REAL}-{TIME} {TRANSIENT} {SURVEY}},
  volume    = {696},
  doi       = {10.1088/0004-637X/696/1/870},
  number    = {1},
  journal   = {\apj},
  publisher = {American Astronomical Society},
  author    = {Drake, A. J. and Djorgovski, S. G. and Mahabal, A. and Beshore, E. and Larson, S. and Graham, M. J. and Williams, R. and Christensen, E. and Catelan, M. and Boattini, A. and Gibbs, A. and Hill, R. and Kowalski, R.},
  month     = apr,
  year      = {2009},
  pages     = {870--884},
}

@ARTICLE{duev2019,
       author = {{Duev}, Dmitry A. and {Mahabal}, Ashish and {Masci}, Frank J. and {Graham}, Matthew J. and {Rusholme}, Ben and {Walters}, Richard and {Karmarkar}, Ishani and {Frederick}, Sara and {Kasliwal}, Mansi M. and {Rebbapragada}, Umaa and {Ward}, Charlotte},
        title = "{Real-bogus classification for the Zwicky Transient Facility using deep learning}",
      journal = {\mnras},
         year = 2019,
        month = nov,
       volume = {489},
       number = {3},
        pages = {3582-3590},
          doi = {10.1093/mnras/stz2357},
archivePrefix = {arXiv},
       eprint = {1907.11259},
 primaryClass = {astro-ph.IM},
       adsurl = {https://ui.adsabs.harvard.edu/abs/2019MNRAS.489.3582D}
}

@misc{duev2021,
  title  = {Phenomenological classification of the {Zwicky} {Transient} {Facility} astronomical event alerts},
  doi    = {10.48550/arXiv.2111.12142},
  author = {Duev, Dmitry A. and van der Walt, Stéfan J.},
  month  = nov,
  year   = {2021},
}

@misc{dufour2017,
  address   = {eprint: arXiv:1610.00986},
  title     = {The {Montreal} {White} {Dwarf} {Database}: {A} {Tool} for the {Community}},
  doi       = {10.48550/arXiv.1610.00986},
  publisher = {arXiv},
  author    = {Dufour, P. and Blouin, S. and Coutu, S. and Fortin-Archambault, M. and Thibeault, C. and Bergeron, P. and Fontaine, G.},
  month     = mar,
  year      = {2017},
}

@article{dupuis1992,
  author  = {{Dupuis}, J. and {Fontaine}, G. and {Pelletier}, C. and {Wesemael}, F.},
  title   = "{A Study of Metal Abundance Patterns in Cool White Dwarfs. I. Time-dependent Calculations of Gravitational Settling}",
  journal = {\apjs},
  year    = 1992,
  month   = oct,
  volume  = {82},
  pages   = {505},
  doi     = {10.1086/191728},
}

@article{eisenhardt2020,
  title   = {The {CatWISE} {Preliminary} {Catalog}: {Motions} from {WISE} and {NEOWISE} {Data}},
  volume  = {247},
  doi     = {10.3847/1538-4365/ab7f2a},
  journal = {\apjs},
  author  = {Eisenhardt, Peter R. M. and Marocco, Federico and Fowler, John W. and Meisner, Aaron M. and Kirkpatrick, J. Davy and Garcia, Nelson and Jarrett, Thomas H. and Koontz, Renata and Marchese, Elijah J. and Stanford, S. Adam and Caselden, Dan and Cushing, Michael C. and Cutri, Roc M. and Faherty, Jacqueline K. and Gelino, Christopher R. and Gonzalez, Anthony H. and Mainzer, Amanda and Mobasher, Bahram and Schlegel, David J. and Stern, Daniel and Teplitz, Harry I. and Wright, Edward L.},
  month   = apr,
  year    = {2020},
  pages   = {69},
}

@article{fantin2020,
  title     = {White {Dwarfs} in the {Era} of the {LSST} and {Its} {Synergies} with {Space}-based {Missions}},
  volume    = {900},
  doi       = {10.3847/1538-4357/aba270},
  number    = {2},
  journal   = {\apj},
  publisher = {The American Astronomical Society},
  author    = {Fantin, Nicholas J. and Côté, Patrick and McConnachie, Alan W.},
  month     = sep,
  year      = {2020},
  pages     = {139},
}

@article{farihi2022,
  title   = {Relentless and complex transits from a planetesimal debris disc},
  volume  = {511},
  doi     = {10.1093/mnras/stab3475},
  journal = {\mnras},
  author  = {Farihi, J. and Hermes, J. J. and Marsh, T. R. and Mustill, A. J. and Wyatt, M. C. and Guidry, J. A. and Wilson, T. G. and Redfield, S. and Izquierdo, P. and Toloza, O. and Gänsicke, B. T. and Aungwerojwit, A. and Kaewmanee, C. and Dhillon, V. S. and Swan, A.},
  month   = apr,
  year    = {2022},
  pages   = {1647--1666},
}

@article{farihi2026,
  title     = {Accretion {Rate} {Changes} {Detected} in a {Polluted} {White} {Dwarf}},
  doi       = {10.1093/mnras/stag176},
  journal   = {\mnras},
  publisher = {OUP},
  author    = {Farihi, Jay and Noor, Hiba Tu and Melis, Carl and Klein, Beth L. and Sahu, Snehalata and Gänsicke, Boris T. and Wyatt, Mark C. and Redfield, Seth and Johnson, Ted M.},
  month     = jan,
  year      = {2026},
}

@article{foreman-mackey2013,
  title   = {emcee: {The} {MCMC} {Hammer}},
  volume  = {125},
  doi     = {10.1086/670067},
  number  = {925},
  journal = {\pasp},
  author  = {Foreman-Mackey, Daniel and Hogg, David W. and Lang, Dustin and Goodman, Jonathan},
  month   = mar,
  year    = {2013},
  pages   = {306},
}

@article{foreman-mackey2017,
  title   = {Fast and {Scalable} {Gaussian} {Process} {Modeling} with {Applications} to {Astronomical} {Time} {Series}},
  volume  = {154},
  doi     = {10.3847/1538-3881/aa9332},
  journal = {\aj},
  author  = {Foreman-Mackey, Daniel and Agol, Eric and Ambikasaran, Sivaram and Angus, Ruth},
  month   = dec,
  year    = {2017},
  pages   = {220},
}

@article{fortin-archambault2020,
  title   = {Modeling of the {Variable} {Circumstellar} {Absorption} {Features} of {WD} 1145+017},
  volume  = {888},
  doi     = {10.3847/1538-4357/ab585a},
  journal = {\apj},
  author  = {Fortin-Archambault, M. and Dufour, P. and Xu, S.},
  month   = jan,
  year    = {2020},
  pages   = {47},
}

@article{frewen2014,
  title   = {Eccentric planets and stellar evolution as a cause of polluted white dwarfs},
  volume  = {439},
  doi     = {10.1093/mnras/stu097},
  journal = {\mnras},
  author  = {Frewen, S. F. N. and Hansen, B. M. S.},
  month   = apr,
  year    = {2014},
  pages   = {2442--2458},
}

@techreport{friedman1984,
  title       = {A {Variable} {Span} {Smoother}},
  institution = {Laboratory for Computational Statistics, Stanford University},
  number      = {LCS 005},
  url         = {https://purl.stanford.edu/fy579rh2058},
  author      = {Friedman, J. H.},
  year        = {1984},
}

@article{gaiacollaboration2021,
  title   = {Gaia {Early} {Data} {Release} 3. {Summary} of the contents and survey properties},
  volume  = {649},
  doi     = {10.1051/0004-6361/202039657},
  journal = {\aap},
  author  = {{Gaia Collaboration} and Brown, A. G. A. and Vallenari, A. and Prusti, T. and de Bruijne, J. H. J. and Babusiaux, C. and Biermann, M. and Creevey, O. L. and Evans, D. W. and Eyer, L. and Hutton, A. and Jansen, F. and Jordi, C. and Klioner, S. A. and Lammers, U. and Lindegren, L. and Luri, X. and Mignard, F. and Panem, C. and Pourbaix, D. and Randich, S. and Sartoretti, P. and Soubiran, C. and Walton, N. A. and Arenou, F. and Bailer-Jones, C. A. L. and Bastian, U. and Cropper, M. and Drimmel, R. and Katz, D. and Lattanzi, M. G. and van Leeuwen, F. and Bakker, J. and Cacciari, C. and Castañeda, J. and De Angeli, F. and Ducourant, C. and Fabricius, C. and Fouesneau, M. and Frémat, Y. and Guerra, R. and Guerrier, A. and Guiraud, J. and Jean-Antoine Piccolo, A. and Masana, E. and Messineo, R. and Mowlavi, N. and Nicolas, C. and Nienartowicz, K. and Pailler, F. and Panuzzo, P. and Riclet, F. and Roux, W. and Seabroke, G. M. and Sordo, R. and Tanga, P. and Thévenin, F. and Gracia-Abril, G. and Portell, J. and Teyssier, D. and Altmann, M. and Andrae, R. and Bellas-Velidis, I. and Benson, K. and Berthier, J. and Blomme, R. and Brugaletta, E. and Burgess, P. W. and Busso, G. and Carry, B. and Cellino, A. and Cheek, N. and Clementini, G. and Damerdji, Y. and Davidson, M. and Delchambre, L. and Dell'Oro, A. and Fernández-Hernández, J. and Galluccio, L. and García-Lario, P. and Garcia-Reinaldos, M. and González-Núñez, J. and Gosset, E. and Haigron, R. and Halbwachs, J. -L. and Hambly, N. C. and Harrison, D. L. and Hatzidimitriou, D. and Heiter, U. and Hernández, J. and Hestroffer, D. and Hodgkin, S. T. and Holl, B. and Janßen, K. and Jevardat de Fombelle, G. and Jordan, S. and Krone-Martins, A. and Lanzafame, A. C. and Löffler, W. and Lorca, A. and Manteiga, M. and Marchal, O. and Marrese, P. M. and Moitinho, A. and Mora, A. and Muinonen, K. and Osborne, P. and Pancino, E. and Pauwels, T. and Petit, J. -M. and Recio-Blanco, A. and Richards, P. J. and Riello, M. and Rimoldini, L. and Robin, A. C. and Roegiers, T. and Rybizki, J. and Sarro, L. M. and Siopis, C. and Smith, M. and Sozzetti, A. and Ulla, A. and Utrilla, E. and van Leeuwen, M. and van Reeven, W. and Abbas, U. and Abreu Aramburu, A. and Accart, S. and Aerts, C. and Aguado, J. J. and Ajaj, M. and Altavilla, G. and Álvarez, M. A. and Álvarez Cid-Fuentes, J. and Alves, J. and Anderson, R. I. and Anglada Varela, E. and Antoja, T. and Audard, M. and Baines, D. and Baker, S. G. and Balaguer-Núñez, L. and Balbinot, E. and Balog, Z. and Barache, C. and Barbato, D. and Barros, M. and Barstow, M. A. and Bartolomé, S. and Bassilana, J. -L. and Bauchet, N. and Baudesson-Stella, A. and Becciani, U. and Bellazzini, M. and Bernet, M. and Bertone, S. and Bianchi, L. and Blanco-Cuaresma, S. and Boch, T. and Bombrun, A. and Bossini, D. and Bouquillon, S. and Bragaglia, A. and Bramante, L. and Breedt, E. and Bressan, A. and Brouillet, N. and Bucciarelli, B. and Burlacu, A. and Busonero, D. and Butkevich, A. G. and Buzzi, R. and Caffau, E. and Cancelliere, R. and Cánovas, H. and Cantat-Gaudin, T. and Carballo, R. and Carlucci, T. and Carnerero, M. I. and Carrasco, J. M. and Casamiquela, L. and Castellani, M. and Castro-Ginard, A. and Castro Sampol, P. and Chaoul, L. and Charlot, P. and Chemin, L. and Chiavassa, A. and Cioni, M. -R. L. and Comoretto, G. and Cooper, W. J. and Cornez, T. and Cowell, S. and Crifo, F. and Crosta, M. and Crowley, C. and Dafonte, C. and Dapergolas, A. and David, M. and David, P. and de Laverny, P. and De Luise, F. and De March, R. and De Ridder, J. and de Souza, R. and de Teodoro, P. and de Torres, A. and del Peloso, E. F. and del Pozo, E. and Delbo, M. and Delgado, A. and Delgado, H. E. and Delisle, J. -B. and Di Matteo, P. and Diakite, S. and Diener, C. and Distefano, E. and Dolding, C. and Eappachen, D. and Edvardsson, B. and Enke, H. and Esquej, P. and Fabre, C. and Fabrizio, M. and Faigler, S. and Fedorets, G. and Fernique, P. and Fienga, A. and Figueras, F. and Fouron, C. and Fragkoudi, F. and Fraile, E. and Franke, F. and Gai, M. and Garabato, D. and Garcia-Gutierrez, A. and García-Torres, M. and Garofalo, A. and Gavras, P. and Gerlach, E. and Geyer, R. and Giacobbe, P. and Gilmore, G. and Girona, S. and Giuffrida, G. and Gomel, R. and Gomez, A. and Gonzalez-Santamaria, I. and González-Vidal, J. J. and Granvik, M. and Gutiérrez-Sánchez, R. and Guy, L. P. and Hauser, M. and Haywood, M. and Helmi, A. and Hidalgo, S. L. and Hilger, T. and Hładczuk, N. and Hobbs, D. and Holland, G. and Huckle, H. E. and Jasniewicz, G. and Jonker, P. G. and Juaristi Campillo, J. and Julbe, F. and Karbevska, L. and Kervella, P. and Khanna, S. and Kochoska, A. and Kontizas, M. and Kordopatis, G. and Korn, A. J. and Kostrzewa-Rutkowska, Z. and Kruszyńska, K. and Lambert, S. and Lanza, A. F. and Lasne, Y. and Le Campion, J. -F. and Le Fustec, Y. and Lebreton, Y. and Lebzelter, T. and Leccia, S. and Leclerc, N. and Lecoeur-Taibi, I. and Liao, S. and Licata, E. and Lindstrøm, E. P. and Lister, T. A. and Livanou, E. and Lobel, A. and Madrero Pardo, P. and Managau, S. and Mann, R. G. and Marchant, J. M. and Marconi, M. and Marcos Santos, M. M. S. and Marinoni, S. and Marocco, F. and Marshall, D. J. and Martin Polo, L. and Martín-Fleitas, J. M. and Masip, A. and Massari, D. and Mastrobuono-Battisti, A. and Mazeh, T. and McMillan, P. J. and Messina, S. and Michalik, D. and Millar, N. R. and Mints, A. and Molina, D. and Molinaro, R. and Molnár, L. and Montegriffo, P. and Mor, R. and Morbidelli, R. and Morel, T. and Morris, D. and Mulone, A. F. and Munoz, D. and Muraveva, T. and Murphy, C. P. and Musella, I. and Noval, L. and Ordénovic, C. and Orrù, G. and Osinde, J. and Pagani, C. and Pagano, I. and Palaversa, L. and Palicio, P. A. and Panahi, A. and Pawlak, M. and Peñalosa Esteller, X. and Penttilä, A. and Piersimoni, A. M. and Pineau, F. -X. and Plachy, E. and Plum, G. and Poggio, E. and Poretti, E. and Poujoulet, E. and Prša, A. and Pulone, L. and Racero, E. and Ragaini, S. and Rainer, M. and Raiteri, C. M. and Rambaux, N. and Ramos, P. and Ramos-Lerate, M. and Re Fiorentin, P. and Regibo, S. and Reylé, C. and Ripepi, V. and Riva, A. and Rixon, G. and Robichon, N. and Robin, C. and Roelens, M. and Rohrbasser, L. and Romero-Gómez, M. and Rowell, N. and Royer, F. and Rybicki, K. A. and Sadowski, G. and Sagristà Sellés, A. and Sahlmann, J. and Salgado, J. and Salguero, E. and Samaras, N. and Sanchez Gimenez, V. and Sanna, N. and Santoveña, R. and Sarasso, M. and Schultheis, M. and Sciacca, E. and Segol, M. and Segovia, J. C. and Ségransan, D. and Semeux, D. and Shahaf, S. and Siddiqui, H. I. and Siebert, A. and Siltala, L. and Slezak, E. and Smart, R. L. and Solano, E. and Solitro, F. and Souami, D. and Souchay, J. and Spagna, A. and Spoto, F. and Steele, I. A. and Steidelmüller, H. and Stephenson, C. A. and Süveges, M. and Szabados, L. and Szegedi-Elek, E. and Taris, F. and Tauran, G. and Taylor, M. B. and Teixeira, R. and Thuillot, W. and Tonello, N. and Torra, F. and Torra, J. and Turon, C. and Unger, N. and Vaillant, M. and van Dillen, E. and Vanel, O. and Vecchiato, A. and Viala, Y. and Vicente, D. and Voutsinas, S. and Weiler, M. and Wevers, T. and Wyrzykowski, Ł. and Yoldas, A. and Yvard, P. and Zhao, H. and Zorec, J. and Zucker, S. and Zurbach, C. and Zwitter, T.},
  month   = may,
  year    = {2021},
  pages   = {A1},
}

@article{gallay2025,
  title   = {Technosignature {Searches} with {Real}-time {Alert} {Brokers}},
  volume  = {170},
  doi     = {10.3847/1538-3881/ade4bb},
  journal = {\aj},
  author  = {Gallay, Eleanor M. and Davenport, James R. A. and Croft, Steve},
  month   = aug,
  year    = {2025},
  pages   = {95},
}

@article{gansicke2006,
  title   = {A {Gaseous} {Metal} {Disk} {Around} a {White} {Dwarf}},
  volume  = {314},
  doi     = {10.1126/science.1135033},
  journal = {Science},
  author  = {Gänsicke, B. T. and Marsh, T. R. and Southworth, J. and Rebassa-Mansergas, A.},
  month   = dec,
  year    = {2006},
  pages   = {1908},
}

@article{gansicke2016,
  title   = {High-speed {Photometry} of the {Disintegrating} {Planetesimals} at {WD1145}+017: {Evidence} for {Rapid} {Dynamical} {Evolution}},
  volume  = {818},
  doi     = {10.3847/2041-8205/818/1/L7},
  journal = {\apj},
  author  = {Gänsicke, B. T. and Aungwerojwit, A. and Marsh, T. R. and Dhillon, V. S. and Sahman, D. I. and Veras, Dimitri and Farihi, J. and Chote, P. and Ashley, R. and Arjyotha, S. and Rattanasoon, S. and Littlefair, S. P. and Pollacco, D. and Burleigh, M. R.},
  month   = feb,
  year    = {2016},
  pages   = {L7},
}

@article{gansicke2019,
  title   = {Accretion of a giant planet onto a white dwarf star},
  volume  = {576},
  doi     = {10.1038/s41586-019-1789-8},
  journal = {\nat},
  author  = {Gänsicke, Boris T. and Schreiber, Matthias R. and Toloza, Odette and Fusillo, Nicola P. Gentile and Koester, Detlev and Manser, Christopher J.},
  month   = dec,
  year    = {2019},
  pages   = {61--64},
}

@article{gary2017,
  title     = {{WD} 1145+017 photometric observations during eight months of high activity},
  volume    = {465},
  doi       = {10.1093/mnras/stw2921},
  journal   = {\mnras},
  publisher = {OUP},
  author    = {Gary, B. L. and Rappaport, S. and Kaye, T. G. and Alonso, R. and Hambschs, F. -J.},
  month     = mar,
  year      = {2017},
  pages     = {3267--3280},
}

@article{gentilefusillo2019,
  title   = {A {Gaia} {Data} {Release} 2 catalogue of white dwarfs and a comparison with {SDSS}},
  volume  = {482},
  doi     = {10.1093/mnras/sty3016},
  number  = {4},
  journal = {\mnras},
  author  = {Gentile Fusillo, Nicola Pietro and Tremblay, Pier-Emmanuel and Gänsicke, Boris T. and Manser, Christopher J. and Cunningham, Tim and Cukanovaite, Elena and Hollands, Mark and Marsh, Thomas and Raddi, Roberto and Jordan, Stefan and Toonen, Silvia and Geier, Stephan and Barstow, Martin and Cummings, Jeffrey D.},
  month   = feb,
  year    = {2019},
  pages   = {4570--4591},
}

@article{gentilefusillo2021,
  title   = {White dwarfs with planetary remnants in the era of {Gaia} - {I}. {Six} emission line systems},
  volume  = {504},
  doi     = {10.1093/mnras/stab992},
  journal = {\mnras},
  author  = {Gentile Fusillo, N. P. and Manser, C. J. and Gänsicke, Boris T. and Toloza, O. and Koester, D. and Dennihy, E. and Brown, W. R. and Farihi, J. and Hollands, M. A. and Hoskin, M. J. and Izquierdo, P. and Kinnear, T. and Marsh, T. R. and Santamaría-Miranda, A. and Pala, A. F. and Redfield, S. and Rodríguez-Gil, P. and Schreiber, M. R. and Veras, Dimitri and Wilson, D. J.},
  month   = jun,
  year    = {2021},
  pages   = {2707--2726},
}

@article{gentilefusillo2021a,
  title     = {A Catalogue of White Dwarfs in {{Gaia EDR3}}},
  author    = {Gentile Fusillo, N. P. and Tremblay, P.-E. and Cukanovaite, E. and Vorontseva, A. and Lallement, R. and Hollands, M. and G{\"a}nsicke, B. T. and Burdge, K. B. and McCleery, J. and Jordan, S.},
  year      = 2021,
  month     = dec,
  journal   = {\mnras},
  volume    = {508},
  pages     = {3877--3896},
  publisher = {OUP},
  doi       = {10.1093/mnras/stab2672},
}

@article{girven2012,
  title   = {Constraints on the {Lifetimes} of {Disks} {Resulting} from {Tidally} {Destroyed} {Rocky} {Planetary} {Bodies}},
  volume  = {749},
  doi     = {10.1088/0004-637X/749/2/154},
  journal = {\apj},
  author  = {Girven, J. and Brinkworth, C. S. and Farihi, J. and Gänsicke, B. T. and Hoard, D. W. and Marsh, T. R. and Koester, D.},
  month   = apr,
  year    = {2012},
  pages   = {154},
}

@article{goldstein2015,
  title   = {Automated {Transient} {Identification} in the {Dark} {Energy} {Survey}},
  volume  = {150},
  doi     = {10.1088/0004-6256/150/3/82},
  journal = {\aj},
  author  = {Goldstein, D. A. and D'Andrea, C. B. and Fischer, J. A. and Foley, R. J. and Gupta, R. R. and Kessler, R. and Kim, A. G. and Nichol, R. C. and Nugent, P. E. and Papadopoulos, A. and Sako, M. and Smith, M. and Sullivan, M. and Thomas, R. C. and Wester, W. and Wolf, R. C. and Abdalla, F. B. and Banerji, M. and Benoit-Lévy, A. and Bertin, E. and Brooks, D. and Carnero Rosell, A. and Castander, F. J. and da Costa, L. N. and Covarrubias, R. and DePoy, D. L. and Desai, S. and Diehl, H. T. and Doel, P. and Eifler, T. F. and Fausti Neto, A. and Finley, D. A. and Flaugher, B. and Fosalba, P. and Frieman, J. and Gerdes, D. and Gruen, D. and Gruendl, R. A. and James, D. and Kuehn, K. and Kuropatkin, N. and Lahav, O. and Li, T. S. and Maia, M. A. G. and Makler, M. and March, M. and Marshall, J. L. and Martini, P. and Merritt, K. W. and Miquel, R. and Nord, B. and Ogando, R. and Plazas, A. A. and Romer, A. K. and Roodman, A. and Sanchez, E. and Scarpine, V. and Schubnell, M. and Sevilla-Noarbe, I. and Smith, R. C. and Soares-Santos, M. and Sobreira, F. and Suchyta, E. and Swanson, M. E. C. and Tarle, G. and Thaler, J. and Walker, A. R.},
  month   = sep,
  year    = {2015},
  pages   = {82},
}

@article{graham2019,
  title   = {The {Zwicky} {Transient} {Facility}: {Science} {Objectives}},
  volume  = {131},
  doi     = {10.1088/1538-3873/ab006c},
  number  = {1001},
  journal = {\pasp},
  author  = {Graham, Matthew J. and Kulkarni, S. R. and Bellm, Eric C. and Adams, Scott M. and Barbarino, Cristina and Blagorodnova, Nadejda and Bodewits, Dennis and Bolin, Bryce and Brady, Patrick R. and Cenko, S. Bradley and Chang, Chan-Kao and Coughlin, Michael W. and De, Kishalay and Eadie, Gwendolyn and Farnham, Tony L. and Feindt, Ulrich and Franckowiak, Anna and Fremling, Christoffer and Gezari, Suvi and Ghosh, Shaon and Goldstein, Daniel A. and Golkhou, V. Zach and Goobar, Ariel and Ho, Anna Y. Q. and Huppenkothen, Daniela and Ivezić, Željko and Jones, R. Lynne and Juric, Mario and Kaplan, David L. and Kasliwal, Mansi M. and Kelley, Michael S. P. and Kupfer, Thomas and Lee, Chien-De and Lin, Hsing Wen and Lunnan, Ragnhild and Mahabal, Ashish A. and Miller, Adam A. and Ngeow, Chow-Choong and Nugent, Peter and Ofek, Eran O. and Prince, Thomas A. and Rauch, Ludwig and van Roestel, Jan and Schulze, Steve and Singer, Leo P. and Sollerman, Jesper and Taddia, Francesco and Yan, Lin and Ye, Quan-Zhi and Yu, Po-Chieh and Barlow, Tom and Bauer, James and Beck, Ron and Belicki, Justin and Biswas, Rahul and Brinnel, Valery and Brooke, Tim and Bue, Brian and Bulla, Mattia and Burruss, Rick and Connolly, Andrew and Cromer, John and Cunningham, Virginia and Dekany, Richard and Delacroix, Alex and Desai, Vandana and Duev, Dmitry A. and Feeney, Michael and Flynn, David and Frederick, Sara and Gal-Yam, Avishay and Giomi, Matteo and Groom, Steven and Hacopians, Eugean and Hale, David and Helou, George and Henning, John and Hover, David and Hillenbrand, Lynne A. and Howell, Justin and Hung, Tiara and Imel, David and Ip, Wing-Huen and Jackson, Edward and Kaspi, Shai and Kaye, Stephen and Kowalski, Marek and Kramer, Emily and Kuhn, Michael and Landry, Walter and Laher, Russ R. and Mao, Peter and Masci, Frank J. and Monkewitz, Serge and Murphy, Patrick and Nordin, Jakob and Patterson, Maria T. and Penprase, Bryan and Porter, Michael and Rebbapragada, Umaa and Reiley, Dan and Riddle, Reed and Rigault, Mickael and Rodriguez, Hector and Rusholme, Ben and van Santen, Jakob and Shupe, David L. and Smith, Roger M. and Soumagnac, Maayane T. and Stein, Robert and Surace, Jason and Szkody, Paula and Terek, Scott and Van Sistine, Angela and van Velzen, Sjoert and Vestrand, W. Thomas and Walters, Richard and Ward, Charlotte and Zhang, Chaoran and Zolkower, Jeffry},
  month   = jul,
  year    = {2019},
  pages   = {078001},
}

@article{green2019a,
  title   = {A {3D} {Dust} {Map} {Based} on {Gaia}, {Pan}-{STARRS} 1, and {2MASS}},
  volume  = {887},
  doi     = {10.3847/1538-4357/ab5362},
  journal = {\apj},
  author  = {Green, Gregory M. and Schlafly, Edward and Zucker, Catherine and Speagle, Joshua S. and Finkbeiner, Douglas},
  month   = dec,
  year    = {2019},
  pages   = {93},
}

@article{groot2024,
  title   = {The {BlackGEM} {Telescope} {Array}. {I}. {Overview}},
  volume  = {136},
  doi     = {10.1088/1538-3873/ad8b6a},
  journal = {\pasp},
  author  = {Groot, P. J. and Bloemen, S. and Vreeswijk, P. M. and van Roestel, J. C. J. and Jonker, P. G. and Nelemans, G. and Klein-Wolt, M. and Lepoole, R. and Pieterse, D. L. A. and Rodenhuis, M. and Boland, W. and Haverkorn, M. and Aerts, C. and Bakker, R. and Balster, H. and Bekema, M. and Dijkstra, E. and Dolron, P. and Elswijk, E. and van Elteren, A. and Engels, A. and Fokker, M. and de Haan, M. and Hahn, F. and ter Horst, R. and Lesman, D. and Kragt, J. and Morren, J. and Nillissen, H. and Pessemier, W. and Raskin, G. and de Rijke, A. and Scheers, L. H. A. and Schuil, M. and Timmer, S. T. and Antunes Amaral, L. and Arancibia-Rojas, E. and Arcavi, I. and Blagorodnova, N. and Biswas, S. and Breton, R. P. and Dawson, H. and Dayal, P. and De Wet, S. and Duffy, C. and Faris, S. and Fausnaugh, M. and Gal-Yam, A. and Geier, S. and Horesh, A. and Johnston, C. and Katusiime, G. and Kelley, C. and Kosakowski, A. and Kupfer, T. and Leloudas, G. and Levan, A. and Modiano, D. and Mogawana, O. and Munday, J. and Paice, J. and Patat, F. and Pelisoli, I. and Ramsay, G. and Ranaivomanana, P. T. and Ruiz-Carmona, R. and Schaffenroth, V. and Scaringi, S. and Stoppa, F. and Street, R. and Tranin, H. and Uzundag, M. and Valenti, S. and Veresvarska, M. and Vu\u{c}kovi\' {c}, M. and Wichern, H. C. I. and Wijers, R. A. M. J. and Wijnands, R. A. D. and Zimmerman, E.},
  month   = nov,
  year    = {2024},
  pages   = {115003},
}

@article{guidry2021,
  title   = {I {Spy} {Transits} and {Pulsations}: {Empirical} {Variability} in {White} {Dwarfs} {Using} {Gaia} and the {Zwicky} {Transient} {Facility}},
  volume  = {912},
  doi     = {10.3847/1538-4357/abee68},
  journal = {\apj},
  author  = {Guidry, Joseph A. and Vanderbosch, Zachary P. and Hermes, J. J. and Barlow, Brad N. and Lopez, Isaac D. and Boudreaux, Thomas M. and Corcoran, Kyle A. and Bell, Keaton J. and Montgomery, M. H. and Heintz, Tyler M. and Castanheira, Barbara G. and Reding, Joshua S. and Dunlap, Bart H. and Winget, D. E. and Winget, Karen I. and Kuehne, J. W.},
  month   = may,
  year    = {2021},
  pages   = {125},
}

@article{guidry2024,
  title   = {Using 3.4 μm {Variability} toward {White} {Dwarfs} as a {Signpost} of {Remnant} {Planetary} {Systems}},
  volume  = {972},
  doi     = {10.3847/1538-4357/ad5be7},
  journal = {\apj},
  author  = {Guidry, Joseph A. and Hermes, J. J. and De, Kishalay and Ould Rouis, Lou Baya and Ewing, Brison B. and Kaiser, B. C.},
  month   = sep,
  year    = {2024},
  pages   = {126},
}

@article{guidry2025,
  title     = {Transiting {Planetary} {Debris} near the {Roche} {Limit} of a {White} {Dwarf} on a 4.97 hr {Orbit}—and its {Vanishing}},
  volume    = {992},
  doi       = {10.3847/1538-4357/adfecb},
  journal   = {\apj},
  publisher = {IOP},
  author    = {Guidry, Joseph A. and Vanderbosch, Zachary P. and Hermes, J. J. and Veras, Dimitri and Hollands, Mark A. and Bhattacharjee, Soumyadeep and Caiazzo, Ilaria and El-Badry, Kareem and Kao, Malia L. and Ould Rouis, Lou Baya and Rodriguez, Antonio C. and van Roestel, Jan},
  month     = oct,
  year      = {2025},
  pages     = {167},
}

@article{gunn2006,
  title   = {The 2.5 m {Telescope} of the {Sloan} {Digital} {Sky} {Survey}},
  volume  = {131},
  doi     = {10.1086/500975},
  journal = {\aj},
  author  = {Gunn, James E. and Siegmund, Walter A. and Mannery, Edward J. and Owen, Russell E. and Hull, Charles L. and Leger, R. French and Carey, Larry N. and Knapp, Gillian R. and York, Donald G. and Boroski, William N. and Kent, Stephen M. and Lupton, Robert H. and Rockosi, Constance M. and Evans, Michael L. and Waddell, Patrick and Anderson, John E. and Annis, James and Barentine, John C. and Bartoszek, Larry M. and Bastian, Steven and Bracker, Stephen B. and Brewington, Howard J. and Briegel, Charles I. and Brinkmann, Jon and Brown, Yorke J. and Carr, Michael A. and Czarapata, Paul C. and Drennan, Craig C. and Dombeck, Thomas and Federwitz, Glenn R. and Gillespie, Bruce A. and Gonzales, Carlos and Hansen, Sten U. and Harvanek, Michael and Hayes, Jeffrey and Jordan, Wendell and Kinney, Ellyne and Klaene, Mark and Kleinman, S. J. and Kron, Richard G. and Kresinski, Jurek and Lee, Glenn and Limmongkol, Siriluk and Lindenmeyer, Carl W. and Long, Daniel C. and Loomis, Craig L. and McGehee, Peregrine M. and Mantsch, Paul M. and Neilsen, Jr., Eric H. and Neswold, Richard M. and Newman, Peter R. and Nitta, Atsuko and Peoples, Jr., John and Pier, Jeffrey R. and Prieto, Peter S. and Prosapio, Angela and Rivetta, Claudio and Schneider, Donald P. and Snedden, Stephanie and Wang, Shu-i.},
  month   = apr,
  year    = {2006},
  pages   = {2332--2359},
}

@article{harding2016,
  title   = {{CHIMERA}: a wide-field, multi-colour, high-speed photometer at the prime focus of the {Hale} telescope},
  volume  = {457},
  doi     = {10.1093/mnras/stw094},
  number  = {3},
  journal = {\mnras},
  author  = {Harding, L. K. and Hallinan, G. and Milburn, J. and Gardner, P. and Konidaris, N. and Singh, N. and Shao, M. and Sandhu, J. and Kyne, G. and Schlichting, H. E.},
  month   = apr,
  year    = {2016},
  pages   = {3036},
}

@ARTICLE{hermes2025,
       author = {{Hermes}, J.~J. and {Guidry}, Joseph A. and {Vanderbosch}, Zachary P. and {Badenas-Agusti}, Mariona and {Xu}, Siyi and {Kao}, Malia L. and {Rodriguez}, Antonio C. and {Hawkins}, Keith},
        title = "{Sporadic Dips from Extended Debris Transiting the Metal-rich White Dwarf SBSS 1232+563}",
      journal = {\apj},
         year = 2025,
        month = feb,
       volume = {980},
       number = {1},
          eid = {56},
        pages = {56},
          doi = {10.3847/1538-4357/ada5fd},
archivePrefix = {arXiv},
       eprint = {2501.02050},
 primaryClass = {astro-ph.SR},
       adsurl = {https://ui.adsabs.harvard.edu/abs/2025ApJ...980...56H}
}

@article{hillenbrand2022,
  title   = {A {Zwicky} {Transient} {Facility} {Look} at {Optical} {Variability} of {Young} {Stellar} {Objects} in the {North} {America} and {Pelican} {Nebulae} {Complex}},
  volume  = {163},
  doi     = {10.3847/1538-3881/ac62d8},
  journal = {\aj},
  author  = {Hillenbrand, Lynne A. and Kiker, Thaddaeus J. and Gee, Miles and Lester, Owen and Braunfeld, Noah L. and Rebull, Luisa M. and Kuhn, Michael A.},
  month   = jun,
  year    = {2022},
  pages   = {263},
}

@misc{hogg2010,
  title     = {Data analysis recipes: {Fitting} a model to data},
  doi       = {10.48550/arXiv.1008.4686},
  publisher = {arXiv},
  author    = {Hogg, David W. and Bovy, Jo and Lang, Dustin},
  month     = aug,
  year      = {2010},
}

@article{hollands2017,
  title   = {Cool {DZ} white dwarfs - {I}. {Identification} and spectral analysis},
  volume  = {467},
  doi     = {10.1093/mnras/stx250},
  journal = {\mnras},
  author  = {Hollands, M. A. and Koester, D. and Alekseev, V. and Herbert, E. L. and Gänsicke, B. T.},
  month   = jun,
  year    = {2017},
  pages   = {4970--5000},
}

@article{hollands2018,
  author  = {{Hollands}, M.~A. and {G{\"a}nsicke}, B.~T. and {Koester}, D.},
  title   = "{Cool DZ white dwarfs II: compositions and evolution of old remnant planetary systems}",
  journal = {\mnras},
  year    = 2018,
  month   = jun,
  volume  = {477},
  number  = {1},
  pages   = {93-111},
  doi     = {10.1093/mnras/sty592},
}

@article{ivezic2019,
  title   = {{LSST}: {From} {Science} {Drivers} to {Reference} {Design} and {Anticipated} {Data} {Products}},
  volume  = {873},
  doi     = {10.3847/1538-4357/ab042c},
  number  = {2},
  journal = {\apj},
  author  = {Ivezić, Zeljko and Kahn, Steven M. and Tyson, J. Anthony and Abel, Bob and Acosta, Emily and Allsman, Robyn and Alonso, David and AlSayyad, Yusra and Anderson, Scott F. and Andrew, John and Angel, James Roger P. and Angeli, George Z. and Ansari, Reza and Antilogus, Pierre and Araujo, Constanza and Armstrong, Robert and Arndt, Kirk T. and Astier, Pierre and Aubourg, Éric and Auza, Nicole and Axelrod, Tim S. and Bard, Deborah J. and Barr, Jeff D. and Barrau, Aurelian and Bartlett, James G. and Bauer, Amanda E. and Bauman, Brian J. and Baumont, Sylvain and Bechtol, Ellen and Bechtol, Keith and Becker, Andrew C. and Becla, Jacek and Beldica, Cristina and Bellavia, Steve and Bianco, Federica B. and Biswas, Rahul and Blanc, Guillaume and Blazek, Jonathan and Blandford, Roger D. and Bloom, Josh S. and Bogart, Joanne and Bond, Tim W. and Booth, Michael T. and Borgland, Anders W. and Borne, Kirk and Bosch, James F. and Boutigny, Dominique and Brackett, Craig A. and Bradshaw, Andrew and Brandt, William Nielsen and Brown, Michael E. and Bullock, James S. and Burchat, Patricia and Burke, David L. and Cagnoli, Gianpietro and Calabrese, Daniel and Callahan, Shawn and Callen, Alice L. and Carlin, Jeffrey L. and Carlson, Erin L. and Chandrasekharan, Srinivasan and Charles-Emerson, Glenaver and Chesley, Steve and Cheu, Elliott C. and Chiang, Hsin-Fang and Chiang, James and Chirino, Carol and Chow, Derek and Ciardi, David R. and Claver, Charles F. and Cohen-Tanugi, Johann and Cockrum, Joseph J. and Coles, Rebecca and Connolly, Andrew J. and Cook, Kem H. and Cooray, Asantha and Covey, Kevin R. and Cribbs, Chris and Cui, Wei and Cutri, Roc and Daly, Philip N. and Daniel, Scott F. and Daruich, Felipe and Daubard, Guillaume and Daues, Greg and Dawson, William and Delgado, Francisco and Dellapenna, Alfred and Peyster, Robert de and Val-Borro, Miguel de and Digel, Seth W. and Doherty, Peter and Dubois, Richard and Dubois-Felsmann, Gregory P. and Durech, Josef and Economou, Frossie and Eifler, Tim and Eracleous, Michael and Emmons, Benjamin L. and Neto, Angelo Fausti and Ferguson, Henry and Figueroa, Enrique and Fisher-Levine, Merlin and Focke, Warren and Foss, Michael D. and Frank, James and Freemon, Michael D. and Gangler, Emmanuel and Gawiser, Eric and Geary, John C. and Gee, Perry and Geha, Marla and Gessner, Charles J. B. and Gibson, Robert R. and Gilmore, D. Kirk and Glanzman, Thomas and Glick, William and Goldina, Tatiana and Goldstein, Daniel A. and Goodenow, Iain and Graham, Melissa L. and Gressler, William J. and Gris, Philippe and Guy, Leanne P. and Guyonnet, Augustin and Haller, Gunther and Harris, Ron and Hascall, Patrick A. and Haupt, Justine and Hernandez, Fabio and Herrmann, Sven and Hileman, Edward and Hoblitt, Joshua and Hodgson, John A. and Hogan, Craig and Howard, James D. and Huang, Dajun and Huffer, Michael E. and Ingraham, Patrick and Innes, Walter R. and Jacoby, Suzanne H. and Jain, Bhuvnesh and Jammes, Fabrice and Jee, M. James and Jenness, Tim and Jernigan, Garrett and Jevremović, Darko and Johns, Kenneth and Johnson, Anthony S. and Johnson, Margaret W. G. and Jones, R. Lynne and Juramy-Gilles, Claire and Jurić, Mario and Kalirai, Jason S. and Kallivayalil, Nitya J. and Kalmbach, Bryce and Kantor, Jeffrey P. and Karst, Pierre and Kasliwal, Mansi M. and Kelly, Heather and Kessler, Richard and Kinnison, Veronica and Kirkby, David and Knox, Lloyd and Kotov, Ivan V. and Krabbendam, Victor L. and Krughoff, K. Simon and Kubánek, Petr and Kuczewski, John and Kulkarni, Shri and Ku, John and Kurita, Nadine R. and Lage, Craig S. and Lambert, Ron and Lange, Travis and Langton, J. Brian and Guillou, Laurent Le and Levine, Deborah and Liang, Ming and Lim, Kian-Tat and Lintott, Chris J. and Long, Kevin E. and Lopez, Margaux and Lotz, Paul J. and Lupton, Robert H. and Lust, Nate B. and MacArthur, Lauren A. and Mahabal, Ashish and Mandelbaum, Rachel and Markiewicz, Thomas W. and Marsh, Darren S. and Marshall, Philip J. and Marshall, Stuart and May, Morgan and McKercher, Robert and McQueen, Michelle and Meyers, Joshua and Migliore, Myriam and Miller, Michelle and Mills, David J. and Miraval, Connor and Moeyens, Joachim and Moolekamp, Fred E. and Monet, David G. and Moniez, Marc and Monkewitz, Serge and Montgomery, Christopher and Morrison, Christopher B. and Mueller, Fritz and Muller, Gary P. and Arancibia, Freddy Muñoz and Neill, Douglas R. and Newbry, Scott P. and Nief, Jean-Yves and Nomerotski, Andrei and Nordby, Martin and O’Connor, Paul and Oliver, John and Olivier, Scot S. and Olsen, Knut and O’Mullane, William and Ortiz, Sandra and Osier, Shawn and Owen, Russell E. and Pain, Reynald and Palecek, Paul E. and Parejko, John K. and Parsons, James B. and Pease, Nathan M. and Peterson, J. Matt and Peterson, John R. and Petravick, Donald L. and Petrick, M. E. Libby and Petry, Cathy E. and Pierfederici, Francesco and Pietrowicz, Stephen and Pike, Rob and Pinto, Philip A. and Plante, Raymond and Plate, Stephen and Plutchak, Joel P. and Price, Paul A. and Prouza, Michael and Radeka, Veljko and Rajagopal, Jayadev and Rasmussen, Andrew P. and Regnault, Nicolas and Reil, Kevin A. and Reiss, David J. and Reuter, Michael A. and Ridgway, Stephen T. and Riot, Vincent J. and Ritz, Steve and Robinson, Sean and Roby, William and Roodman, Aaron and Rosing, Wayne and Roucelle, Cecille and Rumore, Matthew R. and Russo, Stefano and Saha, Abhijit and Sassolas, Benoit and Schalk, Terry L. and Schellart, Pim and Schindler, Rafe H. and Schmidt, Samuel and Schneider, Donald P. and Schneider, Michael D. and Schoening, William and Schumacher, German and Schwamb, Megan E. and Sebag, Jacques and Selvy, Brian and Sembroski, Glenn H. and Seppala, Lynn G. and Serio, Andrew and Serrano, Eduardo and Shaw, Richard A. and Shipsey, Ian and Sick, Jonathan and Silvestri, Nicole and Slater, Colin T. and Smith, J. Allyn and Smith, R. Chris and Sobhani, Shahram and Soldahl, Christine and Storrie-Lombardi, Lisa and Stover, Edward and Strauss, Michael A. and Street, Rachel A. and Stubbs, Christopher W. and Sullivan, Ian S. and Sweeney, Donald and Swinbank, John D. and Szalay, Alexander and Takacs, Peter and Tether, Stephen A. and Thaler, Jon J. and Thayer, John Gregg and Thomas, Sandrine and Thornton, Adam J. and Thukral, Vaikunth and Tice, Jeffrey and Trilling, David E. and Turri, Max and Berg, Richard Van and Berk, Daniel Vanden and Vetter, Kurt and Virieux, Francoise and Vucina, Tomislav and Wahl, William and Walkowicz, Lucianne and Walsh, Brian and Walter, Christopher W. and Wang, Daniel L. and Wang, Shin-Yawn and Warner, Michael and Wiecha, Oliver and Willman, Beth and Winters, Scott E. and Wittman, David and Wolff, Sidney C. and Wood-Vasey, W. Michael and Wu, Xiuqin and Xin, Bo and Yoachim, Peter and Zhan, Hu},
  month   = mar,
  year    = {2019},
  pages   = {111},
}

@article{izquierdo2021,
  title   = {{GD} 424 - a helium-atmosphere white dwarf with a large amount of trace hydrogen in the process of digesting a rocky planetesimal},
  volume  = {501},
  doi     = {10.1093/mnras/staa3987},
  journal = {\mnras},
  author  = {Izquierdo, Paula and Toloza, Odette and Gänsicke, Boris T. and Rodríguez-Gil, Pablo and Farihi, Jay and Koester, Detlev and Guo, Jincheng and Redfield, Seth},
  month   = mar,
  year    = {2021},
  pages   = {4276--4288},
}

@article{jewitt2012,
  title     = {The {Active} {Asteroids}},
  volume    = {143},
  doi       = {10.1088/0004-6256/143/3/66},
  journal   = {\aj},
  publisher = {IOP},
  author    = {Jewitt, David},
  month     = mar,
  year      = {2012},
  pages     = {66},
}

@article{jura2003,
  title   = {A {Tidally} {Disrupted} {Asteroid} around the {White} {Dwarf} {G29}-38},
  volume  = {584},
  doi     = {10.1086/374036},
  journal = {\apj},
  author  = {Jura, M.},
  month   = feb,
  year    = {2003},
  pages   = {L91--L94},
}

@article{jura2008,
  title   = {Pollution of {Single} {White} {Dwarfs} by {Accretion} of {Many} {Small} {Asteroids}},
  volume  = {135},
  doi     = {10.1088/0004-6256/135/5/1785},
  journal = {\aj},
  author  = {Jura, M.},
  month   = may,
  year    = {2008},
  pages   = {1785--1792},
}

@article{kenworthy2015,
  title   = {Modeling {Giant} {Extrasolar} {Ring} {Systems} in {Eclipse} and the {Case} of {J1407b}: {Sculpting} by {Exomoons}?},
  volume  = {800},
  doi     = {10.1088/0004-637X/800/2/126},
  journal = {\apj},
  author  = {Kenworthy, M. A. and Mamajek, E. E.},
  month   = feb,
  year    = {2015},
  pages   = {126},
}

@article{kenworthy2023,
  title     = {A planetary collision afterglow and transit of the resultant debris cloud},
  volume    = {622},
  doi       = {10.1038/s41586-023-06573-9},
  number    = {7982},
  journal   = {\nat},
  publisher = {Nature Publishing Group},
  author    = {Kenworthy, Matthew and Lock, Simon and Kennedy, Grant and van Capelleveen, Richelle and Mamajek, Eric and Carone, Ludmila and Hambsch, Franz-Josef and Masiero, Joseph and Mainzer, Amy and Kirkpatrick, J. Davy and Gomez, Edward and Leinhardt, Zoë and Dou, Jingyao and Tanna, Pavan and Sainio, Arttu and Barker, Hamish and Charbonnel, Stéphane and Garde, Olivier and Le Dû, Pascal and Mulato, Lionel and Petit, Thomas and Rizzo Smith, Michael},
  month     = oct,
  year      = {2023},
  pages     = {251--254},
}

@article{kenyon2017,
  title   = {Numerical {Simulations} of {Gaseous} {Disks} {Generated} from {Collisional} {Cascades} at the {Roche} {Limits} of {White} {Dwarf} {Stars}},
  volume  = {850},
  doi     = {10.3847/1538-4357/aa9570},
  journal = {\apj},
  author  = {Kenyon, Scott J. and Bromley, Benjamin C.},
  month   = nov,
  year    = {2017},
  pages   = {50},
}

@article{kenyon2017a,
  title   = {Numerical {Simulations} of {Collisional} {Cascades} at the {Roche} {Limits} of {White} {Dwarf} {Stars}},
  volume  = {844},
  doi     = {10.3847/1538-4357/aa7b85},
  journal = {\apj},
  author  = {Kenyon, Scott J. and Bromley, Benjamin C.},
  month   = aug,
  year    = {2017},
  pages   = {116},
}

@article{knight2010,
  title   = {Photometric {Study} of the {Kreutz} {Comets} {Observed} by {SOHO} from 1996 to 2005},
  volume  = {139},
  doi     = {10.1088/0004-6256/139/3/926},
  journal = {\aj},
  author  = {Knight, Matthew M. and A'Hearn, Michael F. and Biesecker, Douglas A. and Faury, Guillaume and Hamilton, Douglas P. and Lamy, Philippe and Llebaria, Antoine},
  month   = mar,
  year    = {2010},
  pages   = {926--949},
}

@article{kochanek2008,
  title   = {A {Survey} {About} {Nothing}: {Monitoring} a {Million} {Supergiants} for {Failed} {Supernovae}},
  volume  = {684},
  doi     = {10.1086/590053},
  journal = {\apj},
  author  = {Kochanek, Christopher S. and Beacom, John F. and Kistler, Matthew D. and Prieto, José L. and Stanek, Krzysztof Z. and Thompson, Todd A. and Yüksel, Hasan},
  month   = sep,
  year    = {2008},
  pages   = {1336--1342},
}

@article{koester2009,
  title   = {Accretion and diffusion in white dwarfs: {New} diffusion timescales and applications to {GD} 362 and {G} 29-38},
  volume  = {498},
  doi     = {10.1051/0004-6361/200811468},
  number  = {2},
  journal = {\aap},
  author  = {Koester, D.},
  month   = may,
  year    = {2009},
  pages   = {517--525},
}

@article{koester2014,
  title   = {The frequency of planetary debris around young white dwarfs},
  volume  = {566},
  doi     = {10.1051/0004-6361/201423691},
  journal = {\aap},
  author  = {Koester, D. and Gänsicke, B. T. and Farihi, J.},
  month   = jun,
  year    = {2014},
  pages   = {A34},
}

@article{koester2020,
  author  = {{Koester}, D. and {Kepler}, S.~O. and {Irwin}, A.~W.},
  title   = "{New white dwarf envelope models and diffusion. Application to DQ white dwarfs}",
  journal = {\aap},
  year    = 2020,
  month   = mar,
  volume  = {635},
  eid     = {A103},
  pages   = {A103},
  doi     = {10.1051/0004-6361/202037530},
}

@article{kovacs2002,
  title   = {A box-fitting algorithm in the search for periodic transits},
  volume  = {391},
  doi     = {10.1051/0004-6361:20020802},
  journal = {\aap},
  author  = {Kovács, G. and Zucker, S. and Mazeh, T.},
  month   = aug,
  year    = {2002},
  pages   = {369--377},
}

@ARTICLE{kupfer2021,
       author = {{Kupfer}, Thomas and {Prince}, Thomas A. and {van Roestel}, Jan and {Bellm}, Eric C. and {Bildsten}, Lars and {Coughlin}, Michael W. and {Drake}, Andrew J. and {Graham}, Matthew J. and {Klein}, Courtney and {Kulkarni}, Shrinivas R. and {Masci}, Frank J. and {Walters}, Richard and {Andreoni}, Igor and {Biswas}, Rahul and {Bradshaw}, Corey and {Duev}, Dmitry A. and {Dekany}, Richard and {Guidry}, Joseph A. and {Hermes}, J.~J. and {Laher}, Russ R. and {Riddle}, Reed},
        title = "{Year 1 of the ZTF high-cadence Galactic plane survey: strategy, goals, and early results on new single-mode hot subdwarf B-star pulsators}",
      journal = {\mnras},
         year = 2021,
        month = jul,
       volume = {505},
       number = {1},
        pages = {1254-1267},
          doi = {10.1093/mnras/stab1344},
archivePrefix = {arXiv},
       eprint = {2105.02758},
 primaryClass = {astro-ph.SR},
       adsurl = {https://ui.adsabs.harvard.edu/abs/2021MNRAS.505.1254K}
}

@article{laor1993,
  title   = {Spectroscopic constraints on the properties of dust in active galactic nuclei},
  journal = {\apj},
  volume  = {402},
  doi    = {10.1086/172149},
  author = {Laor, Ari and Draine, Bruce T.},
  month  = jan,
  year   = {1993},
  pages  = {441},
}

@article{law2009,
  title   = {The {Palomar} {Transient} {Factory}: {System} {Overview}, {Performance}, and {First} {Results}},
  volume  = {121},
  doi     = {10.1086/648598},
  journal = {\pasp},
  author  = {Law, Nicholas M. and Kulkarni, Shrinivas R. and Dekany, Richard G. and Ofek, Eran O. and Quimby, Robert M. and Nugent, Peter E. and Surace, Jason and Grillmair, Carl C. and Bloom, Joshua S. and Kasliwal, Mansi M. and Bildsten, Lars and Brown, Tim and Cenko, S. Bradley and Ciardi, David and Croner, Ernest and Djorgovski, S. George and van Eyken, Julian and Filippenko, Alexei V. and Fox, Derek B. and Gal-Yam, Avishay and Hale, David and Hamam, Nouhad and Helou, George and Henning, John and Howell, D. Andrew and Jacobsen, Janet and Laher, Russ and Mattingly, Sean and McKenna, Dan and Pickles, Andrew and Poznanski, Dovi and Rahmer, Gustavo and Rau, Arne and Rosing, Wayne and Shara, Michael and Smith, Roger and Starr, Dan and Sullivan, Mark and Velur, Viswa and Walters, Richard and Zolkower, Jeff},
  month   = dec,
  year    = {2009},
  pages   = {1395},
}

@article{law2022,
  title     = {Low-cost {Access} to the {Deep}, {High}-cadence {Sky}: the {Argus} {Optical} {Array}},
  volume    = {134},
  doi       = {10.1088/1538-3873/ac4811},
  journal   = {\pasp},
  publisher = {IOP},
  author    = {Law, Nicholas M. and Corbett, Hank and Galliher, Nathan W. and Gonzalez, Ramses and Vasquez, Alan and Walters, Glenn and Machia, Lawrence and Ratzloff, Jeff and Ackley, Kendall and Bizon, Chris and Clemens, Christopher and Cox, Steven and Eikenberry, Steven and Howard, Ward S. and Glazier, Amy and Mann, Andrew W. and Quimby, Robert and Reichart, Daniel and Trilling, David},
  month     = mar,
  year      = {2022},
  pages     = {035003},
}

@article{lebourdais2025,
  title   = {Tracing {Planetary} {Accretion} in a 3 {Gyr} old {Hydrogen}-rich {White} {Dwarf}: {The} {Extremely} {Polluted} {Atmosphere} of {LSPM} {J0207}+3331},
  volume  = {993},
  doi     = {10.3847/1538-4357/ae0ace},
  journal = {\apj},
  author  = {Le Bourdais, Erika and Dufour, Patrick and Melis, Carl and Klein, Beth L. and Rogers, Laura K. and Bédard, Antoine and Debes, John and Messier, Ashley and Weinberger, Alycia J. and Xu, Siyi},
  month   = nov,
  year    = {2025},
  pages   = {8},
}

@article{limbach2024,
  title   = {The {MIRI} {Exoplanets} {Orbiting} {White} dwarfs ({MEOW}) {Survey}: {Mid}-infrared {Excess} {Reveals} a {Giant} {Planet} {Candidate} around a {Nearby} {White} {Dwarf}},
  volume  = {973},
  doi     = {10.3847/2041-8213/ad74ed},
  journal = {\apj},
  author  = {Limbach, Mary Anne and Vanderburg, Andrew and Venner, Alexander and Blouin, Simon and Stevenson, Kevin B. and MacDonald, Ryan J. and Jenkins, Sydney and Bowens-Rubin, Rachel and Soares-Furtado, Melinda and Morley, Caroline and Janson, Markus and Debes, John and Xu, Siyi and Kleisioti, Evangelia and Kenworthy, Matthew and Butler, Paul and Crane, Jeffrey D. and Osip, Dave and Shectman, Stephen and Teske, Johanna},
  month   = sep,
  year    = {2024},
  pages   = {L11},
}

@article{limbach2025,
  title     = {Thermal {Emission} and {Confirmation} of the {Frigid} {White} {Dwarf} {Exoplanet} {WD} 1856+534 b},
  volume    = {984},
  doi       = {10.3847/2041-8213/adc9ad},
  journal   = {\apj},
  publisher = {IOP},
  author    = {Limbach, Mary Anne and Vanderburg, Andrew and MacDonald, Ryan J. and Stevenson, Kevin B. and Jenkins, Sydney and Blouin, Simon and Rauscher, Emily and Bowens-Rubin, Rachel and Gallo, Elena and Mang, James and Morley, Caroline V. and Sing, David K. and O'Connor, Christopher and Venner, Alexander and Xu, Siyi},
  month     = may,
  year      = {2025},
  pages     = {L28},
}

@article{lomb1976,
  title     = {Least-{Squares} {Frequency} {Analysis} of {Unequally} {Spaced} {Data}},
  volume    = {39},
  doi       = {10.1007/BF00648343},
  journal   = {\apss},
  publisher = {Springer},
  author    = {Lomb, N. R.},
  month     = feb,
  year      = {1976},
  pages     = {447--462},
}

@article{lopez-sanjuan2024,
  title   = {J-{PLUS}: {The} fraction of calcium white dwarfs along the cooling sequence},
  volume  = {691},
  doi     = {10.1051/0004-6361/202451226},
  journal = {\aap},
  author  = {López-Sanjuan, C. and Tremblay, P. -E. and O'Brien, M. W. and Spinoso, D. and Ederoclite, A. and Vázquez Ramió, H. and Cenarro, A. J. and Marín-Franch, A. and Civera, T. and Carrasco, J. M. and Gänsicke, B. T. and Gentile Fusillo, N. P. and Hernán-Caballero, A. and Hollands, M. A. and del Pino, A. and Domínguez Sánchez, H. and Fernández-Ontiveros, J. A. and Jiménez-Esteban, F. M. and Rebassa-Mansergas, A. and Schmidtobreick, L. and Angulo, R. E. and Cristóbal-Hornillos, D. and Dupke, R. A. and Hernández-Monteagudo, C. and Moles, M. and Sodré, L. and Varela, J.},
  month   = nov,
  year    = {2024},
  pages   = {A211},
}

@article{mahabal2019,
  title   = {Machine {Learning} for the {Zwicky} {Transient} {Facility}},
  volume  = {131},
  doi     = {10.1088/1538-3873/aaf3fa},
  journal = {\pasp},
  author  = {Mahabal, Ashish and Rebbapragada, Umaa and Walters, Richard and Masci, Frank J. and Blagorodnova, Nadejda and van Roestel, Jan and Ye, Quan-Zhi and Biswas, Rahul and Burdge, Kevin and Chang, Chan-Kao and Duev, Dmitry A. and Golkhou, V. Zach and Miller, Adam A. and Nordin, Jakob and Ward, Charlotte and Adams, Scott and Bellm, Eric C. and Branton, Doug and Bue, Brian and Cannella, Chris and Connolly, Andrew and Dekany, Richard and Feindt, Ulrich and Hung, Tiara and Fortson, Lucy and Frederick, Sara and Fremling, C. and Gezari, Suvi and Graham, Matthew and Groom, Steven and Kasliwal, Mansi M. and Kulkarni, Shrinivas and Kupfer, Thomas and Lin, Hsing Wen and Lintott, Chris and Lunnan, Ragnhild and Parejko, John and Prince, Thomas A. and Riddle, Reed and Rusholme, Ben and Saunders, Nicholas and Sedaghat, Nima and Shupe, David L. and Singer, Leo P. and Soumagnac, Maayane T. and Szkody, Paula and Tachibana, Yutaro and Tirumala, Kushal and van Velzen, Sjoert and Wright, Darryl},
  month   = mar,
  year    = {2019},
  pages   = {038002},
}

@article{mainzer2011,
  title   = {{NEOWISE} {Observations} of {Near}-{Earth} {Objects}: {Preliminary} {Results}},
  volume  = {743},
  doi     = {10.1088/0004-637X/743/2/156},
  journal = {\apj},
  author  = {Mainzer, A. and Grav, T. and Bauer, J. and Masiero, J. and McMillan, R. S. and Cutri, R. M. and Walker, R. and Wright, E. and Eisenhardt, P. and Tholen, D. J. and Spahr, T. and Jedicke, R. and Denneau, L. and DeBaun, E. and Elsbury, D. and Gautier, T. and Gomillion, S. and Hand, E. and Mo, W. and Watkins, J. and Wilkins, A. and Bryngelson, G. L. and Del Pino Molina, A. and Desai, S. and Gómez Camus, M. and Hidalgo, S. L. and Konstantopoulos, I. and Larsen, J. A. and Maleszewski, C. and Malkan, M. A. and Mauduit, J. -C. and Mullan, B. L. and Olszewski, E. W. and Pforr, J. and Saro, A. and Scotti, J. V. and Wasserman, L. H.},
  month   = dec,
  year    = {2011},
  pages   = {156},
}

@article{malamud2020a,
  title   = {Tidal disruption of planetary bodies by white dwarfs {I}: a hybrid {SPH}-analytical approach},
  volume  = {492},
  doi     = {10.1093/mnras/staa142},
  number  = {4},
  journal = {\mnras},
  author  = {Malamud, Uri and Perets, Hagai B.},
  month   = mar,
  year    = {2020},
  pages   = {5561},
}

@article{maldonado2021,
  title   = {Do instabilities in high-multiplicity systems explain the existence of close-in white dwarf planets?},
  volume  = {501},
  doi     = {10.1093/mnrasl/slaa193},
  journal = {\mnras},
  author  = {Maldonado, R. F. and Villaver, E. and Mustill, A. J. and Chávez, M. and Bertone, E.},
  month   = jan,
  year    = {2021},
  pages   = {L43--L48},
}

@article{manser2016,
  title     = {Another one grinds the dust: variability of the planetary debris disc at the white dwarf {SDSS} {J104341}.53+085558.2},
  volume    = {462},
  doi       = {10.1093/mnras/stw1760},
  journal   = {\mnras},
  publisher = {OUP},
  author    = {Manser, Christopher J. and Gänsicke, Boris T. and Koester, Detlev and Marsh, Thomas R. and Southworth, John},
  month     = oct,
  year      = {2016},
  pages     = {1461--1469},
}

@article{manser2019,
  title   = {A planetesimal orbiting within the debris disc around a white dwarf star},
  journal = {Science},
  volume  = {364},
  doi     = {10.1126/science.aat5330},
  author  = {Manser, Christopher J. and G\"ansicke, Boris T. and Eggl, Siegfried and others},
  month   = apr,
  year    = {2019},
  pages   = {66},
}

@article{manser2020,
  title   = {The frequency of gaseous debris discs around white dwarfs},
  volume  = {493},
  doi     = {10.1093/mnras/staa359},
  journal = {\mnras},
  author  = {Manser, Christopher J. and Gänsicke, Boris T. and Gentile Fusillo, Nicola Pietro and Ashley, Richard and Breedt, Elmé and Hollands, Mark and Izquierdo, Paula and Pelisoli, Ingrid},
  month   = apr,
  year    = {2020},
  pages   = {2127--2139},
}

@article{marocco2021,
  title   = {The {CatWISE2020} {Catalog}},
  volume  = {253},
  doi     = {10.3847/1538-4365/abd805},
  journal = {\apjs},
  author  = {Marocco, Federico and Eisenhardt, Peter R. M. and Fowler, John W. and Kirkpatrick, J. Davy and Meisner, Aaron M. and Schlafly, Edward F. and Stanford, S. A. and Garcia, Nelson and Caselden, Dan and Cushing, Michael C. and Cutri, Roc M. and Faherty, Jacqueline K. and Gelino, Christopher R. and Gonzalez, Anthony H. and Jarrett, Thomas H. and Koontz, Renata and Mainzer, Amanda and Marchese, Elijah J. and Mobasher, Bahram and Schlegel, David J. and Stern, Daniel and Teplitz, Harry I. and Wright, Edward L.},
  month   = mar,
  year    = {2021},
  pages   = {8},
}

@article{martin2020,
  title   = {Asteroid belt survival through stellar evolution: dependence on the stellar mass},
  volume  = {494},
  doi     = {10.1093/mnrasl/slaa030},
  journal = {\mnras},
  author  = {Martin, Rebecca G. and Livio, Mario and Smallwood, Jeremy L. and Chen, Cheng},
  month   = may,
  year    = {2020},
  pages   = {L17--L21},
}

@article{masci2019,
  title   = {The {Zwicky} {Transient} {Facility}: {Data} {Processing}, {Products}, and {Archive}},
  volume  = {131},
  doi     = {10.1088/1538-3873/aae8ac},
  number  = {995},
  journal = {\pasp},
  author  = {Masci, Frank J. and Laher, Russ R. and Rusholme, Ben and Shupe, David L. and Groom, Steven and Surace, Jason and Jackson, Edward and Monkewitz, Serge and Beck, Ron and Flynn, David and Terek, Scott and Landry, Walter and Hacopians, Eugean and Desai, Vandana and Howell, Justin and Brooke, Tim and Imel, David and Wachter, Stefanie and Ye, Quan-Zhi and Lin, Hsing-Wen and Cenko, S. Bradley and Cunningham, Virginia and Rebbapragada, Umaa and Bue, Brian and Miller, Adam A. and Mahabal, Ashish and Bellm, Eric C. and Patterson, Maria T. and Jurić, Mario and Golkhou, V. Zach and Ofek, Eran O. and Walters, Richard and Graham, Matthew and Kasliwal, Mansi M. and Dekany, Richard G. and Kupfer, Thomas and Burdge, Kevin and Cannella, Christopher B. and Barlow, Tom and Van Sistine, Angela and Giomi, Matteo and Fremling, Christoffer and Blagorodnova, Nadejda and Levitan, David and Riddle, Reed and Smith, Roger M. and Helou, George and Prince, Thomas A. and Kulkarni, Shrinivas R.},
  month   = jan,
  year    = {2019},
  pages   = {018003},
}

@misc{masci2023,
  title  = {A {New} {Forced} {Photometry} {Service} for the {Zwicky} {Transient} {Facility}},
  doi    = {10.48550/arXiv.2305.16279},
  author = {Masci, Frank J. and Laher, Russ R. and Rusholme, Benjamin and Shupe, David and Paladini, Roberta and Groom, Steve and Wold, Avery and Miller, Adam A. and Drake, Andrew},
  month  = may,
  year   = {2023},
}

@inproceedings{mccarthy1998,
  title     = {Blue channel of the {Keck} low-resolution imaging spectrometer},
  volume    = {3355},
  doi       = {10.1117/12.316831},
  booktitle = {Optical astronomical instrumentation},
  publisher = {SPIE},
  author    = {McCarthy, James K. and Cohen, Judith G. and Butcher, Brad and Cromer, John and Croner, Ernest and Jr., William R. Douglas and Goeden, Richard M. and Grewal, Tony and Lu, Barry and Petrie, Harold L. and Weng, Tianxiang and Weber, Bob and Koch, Donald G. and Rodgers, J. Michael},
  editor    = {D'Odorico, Sandro},
  year      = {1998},
  pages     = {81 -- 92},
}

@article{metzger2012,
  title   = {Global models of runaway accretion in white dwarf debris discs},
  volume  = {423},
  doi     = {10.1111/j.1365-2966.2012.20895.x},
  journal = {\mnras},
  author  = {Metzger, Brian D. and Rafikov, Roman R. and Bochkarev, Konstantin V.},
  month   = jun,
  year    = {2012},
  pages   = {505--528},
}

@misc{miller2025,
  title     = {The {La} {Silla} {Schmidt} {Southern} {Survey}},
  doi       = {10.48550/arXiv.2503.14579},
  publisher = {arXiv},
  author    = {Miller, Adam A. and Abrams, Natasha S. and Aldering, Greg and Anand, Shreya and Angus, Charlotte R. and Arcavi, Iair and Baltay, Charles and Bauer, Franz E. and Brethauer, Daniel and Bloom, Joshua S. and Bommireddy, Hemanth and Catelan, Marcio and Chornock, Ryan and Clark, Peter and Collett, Thomas E. and Dimitriadis, Georgios and Faris, Sara and Forster, Francisco and Franckowiak, Anna and Frohmaier, Christopher and Galbany, Lluıs and Galleguillos, Renato B. and Goobar, Ariel and Gutierrez, Claudia P. and Hall, Saarah and Hammerstein, Erica and Herner, Kenneth R. and Hook, Isobel M. and Huston, Macy J. and Johansson, Joel and Kilpatrick, Charles D. and Kim, Alex G. and Knop, Robert A. and Kowalski, Marek P. and Kwok, Lindsey A. and LeBaron, Natalie and Lin, Kenneth W. and Liu, Chang and Lu, Jessica R. and Lu, Wenbin and Lunnan, Ragnhild and Maguire, Kate and Makrygianni, Lydia and Margutti, Raffaella and Maoz, Dan and Milan Veres, Patrik and Moore, Thomas and Nayana, A. J. and Nicholl, Matt and Nordin, Jakob and Pignata, Giuliano and Polin, Abigail and Poznanski, Dovi and Prieto, Jose L. and Rabinowitz, David L. and Rehemtulla, Nabeel and Rigault, Mickael and Ryczanowski, Dan and Sarin, Nikhil and Schulze, Steve and Shah, Ved G. and Sheng, Xinyue and Shilling, Samuel P. R. and Simmons, Brooke D. and Singh, Avinash and Smith, Graham P. and Smith, Mathew and Sollerman, Jesper and Soumagnac, Maayane T. and Stubbs, Christopher W. and Sullivan, Mark and Suresh, Aswin and Trakhtenbrot, Benny and Ward, Charlotte and Wiston, Eli and Xiong, Helen and Yao, Yuhan and Nugent, Peter E.},
  month     = mar,
  year      = {2025},
}

@article{miranda2018,
  title   = {Fast and {Slow} {Precession} of {Gaseous} {Debris} {Disks} around {Planet}-accreting {White} {Dwarfs}},
  volume  = {857},
  doi     = {10.3847/1538-4357/aab9a2},
  journal = {\apj},
  author  = {Miranda, Ryan and Rafikov, Roman R.},
  month   = apr,
  year    = {2018},
  pages   = {135},
}

@article{mullally2024,
  title   = {{JWST} {Directly} {Images} {Giant} {Planet} {Candidates} {Around} {Two} {Metal}-polluted {White} {Dwarf} {Stars}},
  volume  = {962},
  doi     = {10.3847/2041-8213/ad2348},
  journal = {\apj},
  author  = {Mullally, Susan E. and Debes, John and Cracraft, Misty and Mullally, Fergal and Poulsen, Sabrina and Albert, Loic and Thibault, Katherine and Reach, William T. and Hermes, J. J. and Barclay, Thomas and Kilic, Mukremin and Quintana, Elisa V.},
  month   = feb,
  year    = {2024},
  pages   = {L32},
}

@article{mustill2012,
  title   = {Foretellings of {Ragnarök}: {World}-engulfing {Asymptotic} {Giants} and the {Inheritance} of {White} {Dwarfs}},
  volume  = {761},
  doi     = {10.1088/0004-637X/761/2/121},
  journal = {\apj},
  author  = {Mustill, Alexander J. and Villaver, Eva},
  month   = dec,
  year    = {2012},
  pages   = {121},
}

@article{mustill2018,
  author  = {{Mustill}, Alexander J. and {Villaver}, Eva and {Veras}, Dimitri and {G{\"a}nsicke}, Boris T. and {Bonsor}, Amy},
  title   = "{Unstable low-mass planetary systems as drivers of white dwarf pollution}",
  journal = {\mnras},
  year    = 2018,
  month   = may,
  volume  = {476},
  number  = {3},
  pages   = {3939-3955},
  doi     = {10.1093/mnras/sty446},
}

@article{noor2025,
  author  = {{Noor}, Hiba Tu and {Farihi}, Jay and {Kenyon}, Scott J. and {Rafikov}, Roman R. and {Wyatt}, Mark C. and {Su}, Kate Y.~L. and {Melis}, Carl and {Swan}, Andrew and {Wilson}, Thomas G. and {G{\"a}nsicke}, Boris T. and {Bonsor}, Amy and {Rogers}, Laura K. and {Redfield}, Seth and {Kilic}, Mukremin},
  title   = "{Activity in white dwarf debris discs I: Spitzer legacy reveals variability incompatible with the canonical model}",
  journal = {\mnras},
  year    = 2025,
  month   = oct,
  volume  = {543},
  number  = {2},
  pages   = {1602-1623},
  doi     = {10.1093/mnras/staf1380},
}

@article{obrien2024,
  author  = {{O'Brien}, Mairi W. and {Tremblay}, P.-E. and {Klein}, B.~L. and {Koester}, D. and {Melis}, C. and {B{\'e}dard}, A. and {Cukanovaite}, E. and {Cunningham}, T. and {Doyle}, A.~E. and {G{\"a}nsicke}, B.~T. and {Gentile Fusillo}, N.~P. and {Hollands}, M.~A. and {McCleery}, J. and {Pelisoli}, I. and {Toonen}, S. and {Weinberger}, A.~J. and {Zuckerman}, B.},
  title   = "{The 40 pc sample of white dwarfs from Gaia}",
  journal = {\mnras},
  year    = 2024,
  month   = jan,
  volume  = {527},
  number  = {3},
  pages   = {8687-8705},
  doi     = {10.1093/mnras/stad3773},
}

@article{oconnor2023,
  title   = {On the pollution of white dwarfs by exo-{Oort} cloud comets},
  volume  = {524},
  doi     = {10.1093/mnras/stad2281},
  number  = {4},
  journal = {\mnras},
  author  = {O’Connor, Christopher E and Lai, Dong and Seligman, Darryl Z},
  month   = oct,
  year    = {2023},
  pages   = {6181--6197},
}

@article{oke1982,
  title   = {An {Efficient} {Low} {Resolution} and {Moderate} {Resolution} {Spectrograph} for the {Hale} {Telescope}},
  volume  = {94},
  doi     = {10.1086/131027},
  journal = {\pasp},
  author  = {Oke, J. B. and Gunn, J. E.},
  month   = jun,
  year    = {1982},
  pages   = {586},
}

@article{oke1995,
  title   = {The {Keck} {Low}-{Resolution} {Imaging} {Spectrometer}},
  volume  = {107},
  doi     = {10.1086/133562},
  journal = {\pasp},
  author  = {Oke, J. B. and Cohen, J. G. and Carr, M. and Cromer, J. and Dingizian, A. and Harris, F. H. and Labrecque, S. and Lucinio, R. and Schaal, W. and Epps, H. and Miller, J.},
  month   = apr,
  year    = {1995},
  pages   = {375},
}

@article{ossenkopf1992,
  title   = {Constraints on cosmic silicates.},
  volume  = {261},
  journal = {\aap},
  author  = {Ossenkopf, V. and Henning, Th. and Mathis, J. S.},
  month   = aug,
  year    = {1992},
  pages   = {567--578},
}

@article{ouldrouis2024,
  title   = {Constraints on {Remnant} {Planetary} {Systems} as a {Function} of {Main}-sequence {Mass} with {HST}/{COS}},
  volume  = {976},
  doi     = {10.3847/1538-4357/ad86bb},
  journal = {\apj},
  author  = {Ould Rouis, Lou Baya and Hermes, J. J. and Gänsicke, Boris T. and Sahu, Snehalata and Koester, Detlev and Tremblay, P. -E. and Veras, Dimitri and Farihi, Jay and Heintz, Tyler M. and Gentile Fusillo, Nicola Pietro and Redfield, Seth},
  month   = dec,
  year    = {2024},
  pages   = {156},
}

@misc{palaversa2015,
  title     = {Highlights of the {LINEAR} survey},
  doi       = {10.48550/arXiv.1505.02082},
  publisher = {arXiv},
  author    = {Palaversa, Lovro},
  month     = apr,
  year      = {2015},
}

@article{patterson2019,
  title   = {The {Zwicky} {Transient} {Facility} {Alert} {Distribution} {System}},
  volume  = {131},
  doi     = {10.1088/1538-3873/aae904},
  journal = {\pasp},
  author  = {Patterson, Maria T. and Bellm, Eric C. and Rusholme, Ben and Masci, Frank J. and Juric, Mario and Krughoff, K. Simon and Golkhou, V. Zach and Graham, Matthew J. and Kulkarni, Shrinivas R. and Helou, George and {Zwicky Transient Facility Collaboration}},
  month   = jan,
  year    = {2019},
  pages   = {018001},
}

@article{payne2016,
  title   = {Liberating exomoons in white dwarf planetary systems},
  volume  = {457},
  doi     = {10.1093/mnras/stv2966},
  journal = {\mnras},
  author  = {Payne, Matthew J. and Veras, Dimitri and Holman, Matthew J. and Gänsicke, Boris T.},
  month   = mar,
  year    = {2016},
  pages   = {217--231},
}

@article{perley2019,
  series  = {{PASP}},
  title   = {Fully {Automated} {Reduction} of {Longslit} {Spectroscopy} with the {Low} {Resolution} {Imaging} {Spectrometer} at the {Keck} {Observatory}},
  volume  = {131},
  doi     = {10.1088/1538-3873/ab215d},
  number  = {8},
  journal = {\pasp},
  author  = {Perley, D. A.},
  month   = aug,
  year    = {2019},
  pages   = {084503},
}

@article{pham2024,
  author  = {{Pham}, Dang and {Rein}, Hanno},
  title   = "{Polluting white dwarfs with Oort cloud comets}",
  journal = {\mnras},
  year    = 2024,
  month   = may,
  volume  = {530},
  number  = {3},
  pages   = {2526-2547},
  doi     = {10.1093/mnras/stae986},
}

@article{preval2019,
  title     = {A far-{UV} survey of three hot, metal-polluted white dwarf stars: {WD0455}-282, {WD0621}-376, and {WD2211}-495},
  volume    = {487},
  doi       = {10.1093/mnras/stz1506},
  journal   = {\mnras},
  publisher = {OUP},
  author    = {Preval, Simon P. and Barstow, Martin A. and Bainbridge, Matthew and Reindl, Nicole and Ayres, Thomas and Holberg, Jay B. and Barrow, John D. and Lee, Chung-Chi and Webb, John K. and Hu, Jiting},
  month     = aug,
  year      = {2019},
  pages     = {3470--3487},
}

@article{prochaska2020,
  title   = {{PypeIt}: {The} {Python} {Spectroscopic} {Data} {Reduction} {Pipeline}},
  volume  = {5},
  doi     = {10.21105/joss.02308},
  number  = {56},
  journal = {Journal of Open Source Software},
  author  = {Prochaska, J. and Hennawi, Joseph and Westfall, Kyle and Cooke, Ryan and Wang, Feige and Hsyu, Tiffany and Davies, Frederick and Farina, Emanuele and Pelliccia, Debora},
  month   = dec,
  year    = {2020},
  pages   = {2308},
}

@article{rappaport2012,
  title   = {Possible {Disintegrating} {Short}-period {Super}-{Mercury} {Orbiting} {KIC} 12557548},
  volume  = {752},
  doi     = {10.1088/0004-637X/752/1/1},
  journal = {\apj},
  author  = {Rappaport, S. and Levine, A. and Chiang, E. and El Mellah, I. and Jenkins, J. and Kalomeni, B. and Kite, E. S. and Kotson, M. and Nelson, L. and Rousseau-Nepton, L. and Tran, K.},
  month   = jun,
  year    = {2012},
  pages   = {1},
}

@article{rappaport2013,
  title   = {The {Roche} {Limit} for {Close}-orbiting {Planets}: {Minimum} {Density}, {Composition} {Constraints}, and {Application} to the 4.2 hr {Planet} {KOI} 1843.03},
  volume  = {773},
  doi     = {10.1088/2041-8205/773/1/L15},
  journal = {\apjl},
  author  = {Rappaport, Saul and Sanchis-Ojeda, Roberto and Rogers, Leslie A. and Levine, Alan and Winn, Joshua N.},
  month   = aug,
  year    = {2013},
  pages   = {L15},
}

@article{rappaport2016,
  title     = {Drifting asteroid fragments around {WD} 1145+017},
  volume    = {458},
  doi       = {10.1093/mnras/stw612},
  journal   = {\mnras},
  publisher = {OUP},
  author    = {Rappaport, S. and Gary, B. L. and Kaye, T. and Vanderburg, A. and Croll, B. and Benni, P. and Foote, J.},
  month     = jun,
  year      = {2016},
  pages     = {3904--3917},
}

@article{rappaport2018,
  title     = {Likely transiting exocomets detected by {Kepler}},
  volume    = {474},
  doi       = {10.1093/mnras/stx2735},
  journal   = {\mnras},
  publisher = {OUP},
  author    = {Rappaport, S. and Vanderburg, A. and Jacobs, T. and LaCourse, D. and Jenkins, J. and Kraus, A. and Rizzuto, A. and Latham, D. W. and Bieryla, A. and Lazarevic, M. and Schmitt, A.},
  month     = feb,
  year      = {2018},
  pages     = {1453--1468},
}

@ARTICLE{rappaport2021,
       author = {{Rappaport}, S. and {Vanderburg}, A. and {Schwab}, J. and {Nelson}, L.},
        title = "{Minimum Orbital Periods of H-rich Bodies}",
      journal = {\apj},
         year = 2021,
        month = jun,
       volume = {913},
       number = {2},
          eid = {118},
        pages = {118},
          doi = {10.3847/1538-4357/abf7b0},
archivePrefix = {arXiv},
       eprint = {2104.12083},
 primaryClass = {astro-ph.SR},
       adsurl = {https://ui.adsabs.harvard.edu/abs/2021ApJ...913..118R}
}

@article{rau2009,
  title   = {Exploring the {Optical} {Transient} {Sky} with the {Palomar} {Transient} {Factory}},
  volume  = {121},
  doi     = {10.1086/605911},
  journal = {\pasp},
  author  = {Rau, Arne and Kulkarni, Shrinivas R. and Law, Nicholas M. and Bloom, Joshua S. and Ciardi, David and Djorgovski, George S. and Fox, Derek B. and Gal-Yam, Avishay and Grillmair, Carl C. and Kasliwal, Mansi M. and Nugent, Peter E. and Ofek, Eran O. and Quimby, Robert M. and Reach, William T. and Shara, Michael and Bildsten, Lars and Cenko, S. Bradley and Drake, Andrew J. and Filippenko, Alexei V. and Helfand, David J. and Helou, George and Howell, D. Andrew and Poznanski, Dovi and Sullivan, Mark},
  month   = dec,
  year    = {2009},
  pages   = {1334},
}

@article{robert2024,
  title   = {The frequency of transiting planetary systems around polluted white dwarfs},
  volume  = {533},
  doi     = {10.1093/mnras/stae1859},
  journal = {\mnras},
  author  = {Robert, Akshay and Farihi, Jay and Van Eylen, Vincent and Aungwerojwit, Amornrat and Gänsicke, Boris T. and Redfield, Seth and Dhillon, Vikram S. and Marsh, Thomas R. and Swan, Andrew},
  month   = sep,
  year    = {2024},
  pages   = {1756--1765},
}

@article{rogers2025,
  title     = {Simultaneous emission from dust and gas in the planetary debris orbiting a white dwarf},
  volume    = {537},
  doi       = {10.1093/mnrasl/slae117},
  journal   = {\mnras},
  publisher = {OUP},
  author    = {Rogers, Laura K. and Manser, Christopher J. and Bonsor, Amy and Dennihy, Erik and Hodgkin, Simon and Kissler-Patig, Markus and Lai, Samuel and Melis, Carl and Xu, Siyi and Gentile Fusillo, Nicola and Gänsicke, Boris and Swan, Andrew and Toloza, Odette and Veras, Dimitri},
  month     = feb,
  year      = {2025},
  pages     = {L72--L79},
}

@article{sackmann1993,
  title   = {Our {Sun}. {III}. {Present} and {Future}},
  volume  = {418},
  doi     = {10.1086/173407},
  journal = {\apj},
  author  = {Sackmann, I. -Juliana and Boothroyd, Arnold I. and Kraemer, Kathleen E.},
  month   = nov,
  year    = {1993},
  pages   = {457},
}

@article{scargle1982,
  title   = {Studies in astronomical time series analysis. {II}. {Statistical} aspects of spectral analysis of unevenly spaced data.},
  volume  = {263},
  doi     = {10.1086/160554},
  journal = {\apj},
  author  = {Scargle, J. D.},
  month   = dec,
  year    = {1982},
  pages   = {835--853},
}

@article{schreiber2019,
  title   = {Cold {Giant} {Planets} {Evaporated} by {Hot} {White} {Dwarfs}},
  volume  = {887},
  doi     = {10.3847/2041-8213/ab42e2},
  journal = {\apj},
  author  = {Schreiber, Matthias R. and Gänsicke, Boris T. and Toloza, Odette and Hernandez, Mercedes-S. and Lagos, Felipe},
  month   = dec,
  year    = {2019},
  pages   = {L4},
}

@article{schroder2008,
  title   = {Distant future of the {Sun} and {Earth} revisited},
  volume  = {386},
  doi     = {10.1111/j.1365-2966.2008.13022.x},
  journal = {\mnras},
  author  = {Schröder, K. -P. and Smith, Robert Connon},
  month   = may,
  year    = {2008},
  pages   = {155--163},
}

@ARTICLE{sdsscollaboration2025,
       author = {{SDSS Collaboration} and {Aghakhanloo}, Mojgan and {Aird}, James and {Almeida}, Andr{\'e}s and {Amrita}, Singh and {Anders}, Friedrich and {Anderson}, Scott F. and {Arseneau}, Stefan and {Gonz{\'a}lez {\'A}vila}, Consuelo and {Aviram}, Shir and {Aydar}, Catarina and {Badenes}, Carles and {Barrera-Ballesteros}, Jorge K. and {Bauer}, Franz E. and {Behmard}, Aida and {Berg}, Michelle and {Besser}, F. and {Moni Bidin}, Christian and {Bizyaev}, Dmitry and {Blanc}, Guillermo and {Blanton}, Michael R. and {Bovy}, Jo and {Brandt}, William Nielsen and {Brownstein}, Joel R. and {Buchner}, Johannes and {Bulbul}, Esra and {Burchett}, Joseph N. and {Carigi}, Leticia and {Carlberg}, Joleen K. and {Casey}, Andrew R. and {Chakraborty}, Priyanka and {Chanam{\'e}}, Julio and {Chandra}, Vedant and {Chiappini}, Cristina and {Chilingarian}, Igor and {Comparat}, Johan and {Covey}, Kevin and {Crumpler}, Nicole and {Cunha}, Katia and {D'Onghia}, Elena and {Dai}, Xinyu and {Darling}, Jeremy and {Davis}, Megan and {De Lee}, Nathan and {Deacon}, Niall and {M{\'e}ndez Delgado}, Jos{\'e} Eduardo and {Demasi}, Sebastian and {Demianenko}, Mariia and {Demke}, Delvin and {Donor}, John and {Drory}, Niv and {Villa Durango}, Monica Alejandra and {Dwelly}, Tom and {Egorov}, Oleg and {Egorova}, Evgeniya and {El-Badry}, Kareem and {Eracleous}, Mike and {Fan}, Xiaohui and {Farr}, Emily and {Finkbeiner}, Douglas P. and {Fries}, Logan and {Frinchaboy}, Peter and {Gentile Fusillo}, Nicola Pietro and {Serrano F{\'e}lix}, Luis Daniel and {G{\"a}nsicke}, Boris T. and {Galligan}, Emma and {Garc{\'\i}a}, Pablo and {Gelfand}, Joseph and {Grabowski}, Katie and {Grebel}, Eva and {Green}, Paul J. and {Greve}, Hannah and {Grier}, Catherine and {Griffith}, Emily J. and {Guetzoyan}, Paloma and {Gupta}, Pramod and {Hackshaw}, Zoe and {Hall}, Patrick B. and {Hawkins}, Keith and {Heged{\H{u}}s}, Viola and {Hekker}, Saskia and {Herbst}, T.~M. and {Hermes}, J.~J. and {Hern{\'a}ndez-Garc{\'\i}a}, Lorena and {Hiremath}, Pranavi and {Hogg}, David W. and {Holtzman}, Jon and {Horne}, Keith and {Horta}, Danny and {Huang}, Yang and {Hutchinson}, Brian and {H{\"a}berle}, Maximilian and {Ibarra-Medel}, Hector Javier and {Ji}, Alexander P. and {Jofre}, Paula and {Johnson}, James W. and {Johnson}, Jennifer and {Johnston}, Evelyn J. and {Kaldor}, Mary and {Katkov}, Ivan and {Khalatyan}, Arman and {Khoperskov}, Sergey and {Klessen}, Ralf and {Kluge}, Matthias and {Koekemoer}, Anton M. and {Kollmeier}, Juna A. and {Kounkel}, Marina and {Kreckel}, Kathryn and {Krishnarao}, Dhanesh and {Krumpe}, Mirko and {Lacerna}, Ivan and {Laporte}, Chervin and {L{\'e}pine}, S{\'e}bastien and {Li}, Jing and {Liang}, Fu-Heng and {Limberg}, Guilherme and {Liu}, Xin and {Loebman}, Sarah and {Long}, Knox and {Lu}, Yuxi(Lucy) and {Lucey}, Madeline and {Lugo-Aranda}, Alejandra Z. and {Mart{\'\i}nez-Aldama}, Mary Loli and {McKinnon}, Kevin A. and {Medan}, Ilija and {Merloni}, Andrea and {Morrison}, Sean and {Myers}, Natalie and {M{\'e}sz{\'a}ros}, Szabolcs and {M{\"u}ller-Horn}, Johanna and {Nepal}, Samir and {Ness}, Melissa and {Nidever}, David and {Nitschelm}, Christian and {Oravetz}, Audrey and {Otto}, Jonah and {Adamane Pallathadka}, Gautham and {Pan}, Kaike and {P{\'e}rez Paolino}, Facundo and {Negrete Pe{\~n}aloza}, Castalia Alenka and {Pinsonneault}, Marc and {Taghizadeh-Popp}, Manuchehr and {Price-Whelan}, Adrian and {Pulatova}, Nadiia and {Queiroz}, A.~B.~A. and {Raddick}, Jordan and {Rankine}, Amy L. and {Rix}, Hans-Walter and {Rom{\'a}n-Z{\'u}{\~n}iga}, Carlos G. and {Fern{\'a}ndez Rosso}, Daniela and {Runnoe}, Jessie and {Saad}, Serat Mahmud and {Salvato}, Mara and {Sanchez}, Sebastian F. and {Sattler}, Natascha and {Saydjari}, Andrew K. and {Sayres}, Conor and {Schlaufman}, Kevin C. and {Schneider}, Donald P. and {Schwope}, Axel and {Seaton}, Lucas M. and {Seeburger}, Rhys and {Serna}, Javier and {Sharma}, Sanjib and {Shen}, Yue and {Sinha}, Amaya and {Sizemore}, Logan and {Sniegowska}, Marzena and {Song}, Ying-Yi and {Souto}, Diogo and {Stassun}, Keivan G. and {Steinmetz}, Matthias and {Stone}, Zachary and {Stone-Martinez}, Alexander and {Stringfellow}, Guy S. and {Mata S{\'a}nchez}, Aurora and {S{\'a}nchez-Gallego}, Jos{\'e} and {Tan}, Jonathan C. and {Tayar}, Jamie and {Thai}, Riley and {Thakar}, Ani and {Thibodeaux}, Pierre and {Ting}, Yuan-Sen and {Tkachenko}, Andrew and {Trakhtenbrot}, Benny and {Fern{\'a}ndez-Trincado}, Jos{\'e} G. and {Troup}, Nicholas and {Trump}, Jonathan R. and {Ulloa}, Natalie and {Van Der Marel}, Roeland P. and {Vera}, Pablo and {Villanova}, Sandro and {Villase{\~n}or}, Jaime I. and {Wang}, Ji and {Way}, Zachary and {Weijmans}, Anne-Marie and {Wheeler}, Adam and {Wilson}, John C. and {Wofford}, Aida and {Wong}, Tony},
        title = "{The Nineteenth Data Release of the Sloan Digital Sky Survey}",
      journal = {\apjs},
         year = 2026,
        month = jul,
       volume = {285},
       number = {1},
          eid = {9},
        pages = {9},
          doi = {10.3847/1538-4365/ae4697},
archivePrefix = {arXiv},
       eprint = {2507.07093},
 primaryClass = {astro-ph.GA},
       adsurl = {https://ui.adsabs.harvard.edu/abs/2026ApJS..285....9S}
}

@article{shingles2021,
  title   = {Release of the {ATLAS} {Forced} {Photometry} server for public use},
  volume  = {7},
  journal = {Transient Name Server AstroNote},
  author  = {Shingles, L. and Smith, K. W. and Young, D. R. and Smartt, S. J. and Tonry, J. and Denneau, L. and Heinze, A. and Weiland, H. and Flewelling, H. and Stalder, B. and Clocchiatti, A. and Förster, F. and Pignata, G. and Rest, A. and Anderson, J. and Stubbs, C. and Erasmus, N.},
  month   = jan,
  year    = {2021},
  pages   = {1--7},
}

@article{smallwood2018,
  title   = {White dwarf pollution by asteroids from secular resonances},
  volume  = {480},
  doi     = {10.1093/mnras/sty1819},
  journal = {\mnras},
  author  = {Smallwood, Jeremy L. and Martin, Rebecca G. and Livio, Mario and Lubow, Stephen H.},
  month   = oct,
  year    = {2018},
  pages   = {57--67},
}

@article{solano2022,
  title   = {Discovering vanishing objects in {POSS} {I} red images using the {Virtual} {Observatory}},
  volume  = {515},
  doi     = {10.1093/mnras/stac1552},
  number  = {1},
  journal = {\mnras},
  author  = {Solano, Enrique and Villarroel, B and Rodrigo, C},
  month   = sep,
  year    = {2022},
  pages   = {1380--1391},
}

@article{steeghs2022,
  title   = {The {Gravitational}-wave {Optical} {Transient} {Observer} ({GOTO}): prototype performance and prospects for transient science},
  volume  = {511},
  doi     = {10.1093/mnras/stac013},
  journal = {\mnras},
  author  = {Steeghs, D. and Galloway, D. K. and Ackley, K. and Dyer, M. J. and Lyman, J. and Ulaczyk, K. and Cutter, R. and Mong, Y. -L. and Dhillon, V. and O'Brien, P. and Ramsay, G. and Poshyachinda, S. and Kotak, R. and Nuttall, L. K. and Pallé, E. and Breton, R. P. and Pollacco, D. and Thrane, E. and Aukkaravittayapun, S. and Awiphan, S. and Burhanudin, U. and Chote, P. and Chrimes, A. and Daw, E. and Duffy, C. and Eyles-Ferris, R. and Gompertz, B. and Heikkilä, T. and Irawati, P. and Kennedy, M. R. and Killestein, T. and Kuncarayakti, H. and Levan, A. J. and Littlefair, S. and Makrygianni, L. and Marsh, T. and Mata-Sanchez, D. and Mattila, S. and Maund, J. and McCormac, J. and Mkrtichian, D. and Mullaney, J. and Noysena, K. and Patel, M. and Rol, E. and Sawangwit, U. and Stanway, E. R. and Starling, R. and Strøm, P. and Tooke, S. and West, R. and White, D. J. and Wiersema, K.},
  month   = apr,
  year    = {2022},
  pages   = {2405--2422},
}

@article{stokes2000,
  title   = {Lincoln {Near}-{Earth} {Asteroid} {Program} ({LINEAR})},
  volume  = {148},
  doi     = {10.1006/icar.2000.6493},
  journal = {\icarus},
  author  = {Stokes, Grant H. and Evans, Jenifer B. and Viggh, Herbert E. M. and Shelly, Frank C. and Pearce, Eric C.},
  month   = nov,
  year    = {2000},
  pages   = {21--28},
}

@article{swan2019,
  title   = {Most white dwarfs with detectable dust discs show infrared variability},
  volume  = {484},
  doi     = {10.1093/mnrasl/slz014},
  journal = {\mnras},
  author  = {Swan, Andrew and Farihi, Jay and Wilson, Thomas G.},
  month   = mar,
  year    = {2019},
  pages   = {L109--L113},
}

@article{swan2020,
  title   = {The dust never settles: collisional production of gas and dust in evolved planetary systems},
  volume  = {496},
  doi     = {10.1093/mnras/staa1688},
  journal = {\mnras},
  author  = {Swan, Andrew and Farihi, Jay and Wilson, Thomas G. and Parsons, Steven G.},
  month   = aug,
  year    = {2020},
  pages   = {5233--5242},
}

@article{szkody2020,
  title   = {Cataclysmic {Variables} in the {First} {Year} of the {Zwicky} {Transient} {Facility}},
  volume  = {159},
  doi     = {10.3847/1538-3881/ab7cce},
  journal = {\aj},
  author  = {Szkody, Paula and Dicenzo, Brooke and Ho, Anna Y. Q. and Hillenbrand, Lynne A. and van Roestel, Jan and Ridder, Margaret and DeJesus Lima, Isabel and Graham, Melissa L. and Bellm, Eric C. and Burdge, Kevin and Kupfer, Thomas and Prince, Thomas A. and Masci, Frank J. and Mróz, Przemyslaw J. and Golkhou, V. Zach and Coughlin, Michael and Cunningham, Virginia A. and Dekany, Richard and Graham, Matthew J. and Hale, David and Kaplan, David and Kasliwal, Mansi M. and Miller, Adam A. and Neill, James D. and Patterson, Maria T. and Riddle, Reed and Smith, Roger and Soumagnac, Maayane T.},
  month   = may,
  year    = {2020},
  pages   = {198},
}

@ARTICLE{szkody2021,
       author = {{Szkody}, Paula and {Olde Loohuis}, Claire and {Koplitz}, Brad and {van Roestel}, Jan and {Dicenzo}, Brooke and {Ho}, Anna Y.~Q. and {Hillenbrand}, Lynne A. and {Bellm}, Eric C. and {Dekany}, Richard and {Drake}, Andrew J. and {Duev}, Dmitry A. and {Graham}, Matthew J. and {Kasliwal}, Mansi M. and {Mahabal}, Ashish A. and {Masci}, Frank J. and {Neill}, James D. and {Riddle}, Reed and {Rusholme}, Benjamin and {Sollerman}, Jesper and {Walters}, Richard},
        title = "{Cataclysmic Variables in the Second Year of the Zwicky Transient Facility}",
      journal = {\aj},
         year = 2021,
        month = sep,
       volume = {162},
       number = {3},
          eid = {94},
        pages = {94},
          doi = {10.3847/1538-3881/ac0efb},
archivePrefix = {arXiv},
       eprint = {2107.07051},
 primaryClass = {astro-ph.SR},
       adsurl = {https://ui.adsabs.harvard.edu/abs/2021AJ....162...94S}
}

@article{tonry2018,
  title   = {{ATLAS}: {A} {High}-cadence {All}-sky {Survey} {System}},
  volume  = {130},
  doi     = {10.1088/1538-3873/aabadf},
  journal = {\pasp},
  author  = {Tonry, J. L. and Denneau, L. and Heinze, A. N. and Stalder, B. and Smith, K. W. and Smartt, S. J. and Stubbs, C. W. and Weiland, H. J. and Rest, A.},
  month   = jun,
  year    = {2018},
  pages   = {064505},
}

@article{torres2025,
  author  = {{Torres}, Santiago},
  title   = "{Implications for the formation of Oort cloud-like structures and interstellar comets in dense environments}",
  journal = {arXiv e-prints},
  year    = 2025,
  month   = oct,
  eid     = {arXiv:2510.23653},
  pages   = {arXiv:2510.23653},
  doi     = {10.48550/arXiv.2510.23653},
}

@article{tzanidakis2025,
  title   = {A {Systematic} {Search} for {Main}-sequence {Dipper} {Stars} {Using} the {Zwicky} {Transient} {Facility}},
  volume  = {991},
  doi     = {10.3847/1538-4357/adf5bb},
  journal = {\apj},
  author  = {Tzanidakis, Anastasios and Davenport, James R. A. and Caplar, Neven and Bellm, Eric C. and Beebe, Wilson and Branton, Doug and Campos, Sandro and Connolly, Andrew J. and DeLucchi, Melissa and Malanchev, Konstantin and McGuire, Sean},
  month   = sep,
  year    = {2025},
  pages   = {118},
}

@INPROCEEDINGS{uomoto1999,
       author = {{Uomoto}, A. and {Smee}, S. and {Rockosi}, C. and {Burles}, S. and {Pope}, A. and {Friedman}, S. and {Brinkmann}, J. and {Gunn}, J. and {Nichol}, R. and {SDSS Collaboration}},
        title = "{The Sloan Digital Sky Survey Spectrographs}",
    booktitle = {American Astronomical Society Meeting Abstracts},
         year = 1999,
       series = {American Astronomical Society Meeting Abstracts},
       volume = {195},
        month = dec,
          eid = {87.01},
        pages = {87.01},
       adsurl = {https://ui.adsabs.harvard.edu/abs/1999AAS...195.8701U}
}

@article{vanderbosch2020,
  title   = {A {White} {Dwarf} with {Transiting} {Circumstellar} {Material} {Far} outside the {Roche} {Limit}},
  volume  = {897},
  doi     = {10.3847/1538-4357/ab9649},
  journal = {\apj},
  author  = {Vanderbosch, Z. and Hermes, J. J. and Dennihy, E. and Dunlap, B. H. and Izquierdo, P. and Tremblay, P.-E. and Cho, P. B. and Gänsicke, B. T. and Toloza, O. and Bell, K. J. and Montgomery, M. H. and Winget, D. E.},
  month   = jul,
  year    = {2020},
  pages   = {171},
}

@article{vanderbosch2021,
  title   = {Recurring {Planetary} {Debris} {Transits} and {Circumstellar} {Gas} around {White} {Dwarf} {ZTF} {J0328}-1219},
  volume  = {917},
  doi     = {10.3847/1538-4357/ac0822},
  journal = {\apj},
  author  = {Vanderbosch, Zachary P. and Rappaport, Saul and Guidry, Joseph A. and Gary, Bruce L. and Blouin, Simon and Kaye, Thomas G. and Weinberger, Alycia J. and Melis, Carl and Klein, Beth L. and Zuckerman, B. and Vanderburg, Andrew and Hermes, J. J. and Hegedus, Ryan J. and Burleigh, Matthew. R. and Sefako, Ramotholo and Worters, Hannah L. and Heintz, Tyler M.},
  month   = aug,
  year    = {2021},
  pages   = {41},
}

@article{vanderburg2015,
  title   = {A disintegrating minor planet transiting a white dwarf},
  volume  = {526},
  doi     = {10.1038/nature15527},
  journal = {\nat},
  author  = {Vanderburg, Andrew and Johnson, John Asher and Rappaport, Saul and Bieryla, Allyson and Irwin, Jonathan and Lewis, John Arban and Kipping, David and Brown, Warren R. and Dufour, Patrick and Ciardi, David R. and Angus, Ruth and Schaefer, Laura and Latham, David W. and Charbonneau, David and Beichman, Charles and Eastman, Jason and McCrady, Nate and Wittenmyer, Robert A. and Wright, Jason T.},
  month   = oct,
  year    = {2015},
  pages   = {546--549},
}

@article{vanderburg2020,
  title   = {A {Giant} {Planet} {Candidate} {Transiting} a {White} {Dwarf}},
  volume  = {585},
  doi     = {10.1038/s41586-020-2713-y},
  number  = {7825},
  journal = {\nat},
  author  = {Vanderburg, Andrew and Rappaport, Saul A. and Xu, Siyi and Crossfield, Ian and Becker, Juliette C. and Gary, Bruce and Murgas, Felipe and Blouin, Simon and Kaye, Thomas G. and Palle, Enric and Melis, Carl and Morris, Brett and Kreidberg, Laura and Gorjian, Varoujan and Morley, Caroline V. and Mann, Andrew W. and Parviainen, Hannu and Pearce, Logan A. and Newton, Elisabeth R. and Carrillo, Andreia and Zuckerman, Ben and Nelson, Lorne and Zeimann, Greg and Brown, Warren R. and Tronsgaard, René and Klein, Beth and Ricker, George R. and Vanderspek, Roland K. and Latham, David W. and Seager, Sara and Winn, Joshua N. and Jenkins, Jon M. and Adams, Fred C. and Benneke, Björn and Berardo, David and Buchhave, Lars A. and Caldwell, Douglas A. and Christiansen, Jessie L. and Collins, Karen A. and Colón, Knicole D. and Daylan, Tansu and Doty, John and Doyle, Alexandra E. and Dragomir, Diana and Dressing, Courtney and Dufour, Patrick and Fukui, Akihiko and Glidden, Ana and Guerrero, Natalia M. and Guo, Xueying and Heng, Kevin and Henriksen, Andreea I. and Huang, Chelsea X. and Kaltenegger, Lisa and Kane, Stephen R. and Lewis, John A. and Lissauer, Jack J. and Morales, Farisa and Narita, Norio and Pepper, Joshua and Rose, Mark E. and Smith, Jeffrey C. and Stassun, Keivan G. and Yu, Liang},
  month   = sep,
  year    = {2020},
  pages   = {363--367},
}

@article{vanderwalt2019,
  title   = {{SkyPortal}: {An} {Astronomical} {Data} {Platform}},
  volume  = {4},
  doi     = {10.21105/joss.01247},
  number  = {37},
  journal = {Journal of Open Source Software},
  author  = {Van Der Walt, Stéfan and Crellin-Quick, Arien and Bloom, Joshua},
  month   = may,
  year    = {2019},
  pages   = {1247},
}

@article{veras2014,
  title   = {Formation of planetary debris discs around white dwarfs - {I}. {Tidal} disruption of an extremely eccentric asteroid},
  volume  = {445},
  doi     = {10.1093/mnras/stu1871},
  journal = {\mnras},
  author  = {Veras, Dimitri and Leinhardt, Zoë M. and Bonsor, Amy and Gänsicke, Boris T.},
  month   = dec,
  year    = {2014},
  pages   = {2244--2255},
}

@article{veras2015,
  title   = {Formation of planetary debris discs around white dwarfs - {II}. {Shrinking} extremely eccentric collisionless rings},
  volume  = {451},
  doi     = {10.1093/mnras/stv1195},
  journal = {\mnras},
  author  = {Veras, Dimitri and Leinhardt, Zoë M. and Eggl, Siegfried and Gänsicke, Boris T.},
  month   = aug,
  year    = {2015},
  pages   = {3453--3459},
}

@article{veras2016,
  title   = {Post-main-sequence planetary system evolution},
  volume  = {3},
  doi     = {10.1098/rsos.150571},
  number  = {2},
  journal = {Royal Society Open Science},
  author  = {Veras, Dimitri},
  month   = feb,
  year    = {2016},
  pages   = {150571},
}

@article{veras2017,
  title     = {Explaining the variability of {WD} 1145+017 with simulations of asteroid tidal disruption},
  volume    = {465},
  doi       = {10.1093/mnras/stw2748},
  journal   = {\mnras},
  publisher = {OUP},
  author    = {Veras, Dimitri and Carter, Philip J. and Leinhardt, Zoë M. and Gänsicke, Boris T.},
  month     = feb,
  year      = {2017},
  pages     = {1008--1022},
}

@INCOLLECTION{veras2021,
       author = {{Veras}, Dimitri},
        title = "{Planetary Systems Around White Dwarfs}",
    booktitle = {Oxford Research Encyclopedia of Planetary Science},
    publisher = {Oxford University Press},
         year = 2021,
          eid = {1},
        pages = {1},
          doi = {10.1093/acrefore/9780190647926.013.238},
       adsurl = {https://ui.adsabs.harvard.edu/abs/2021orel.bookE...1V}
}

@article{veras2022,
  title   = {Orbit decay of 2-100 au planetary remnants around white dwarfs with no gravitational assistance from planets},
  volume  = {510},
  doi     = {10.1093/mnras/stab3490},
  journal = {\mnras},
  author  = {Veras, Dimitri and Birader, Yusuf and Zaman, Uwais},
  month   = mar,
  year    = {2022},
  pages   = {3379--3388},
}

@article{veras2024a,
  title   = {The {Evolution} and {Delivery} of {Rocky} {Extra}-{Solar} {Materials} to {White} {Dwarfs}},
  volume  = {90},
  doi     = {10.2138/rmg.2024.90.05},
  journal = {Reviews in Mineralogy and Geochemistry},
  author  = {Veras, Dimitri and Mustill, Alexander J. and Bonsor, Amy},
  month   = jul,
  year    = {2024},
  pages   = {141--170},
}

@article{veras2025,
  author  = {{Veras}, Dimitri},
  title   = "{Resupplying planetary debris to old white dwarfs with supernova blast waves}",
  journal = {\mnras},
  year    = 2025,
  month   = aug,
  volume  = {541},
  number  = {3},
  pages   = {2119-2133},
  doi     = {10.1093/mnras/staf1051},
}

@article{verbunt1988,
  title   = {Mass transfer instabilities due to angular momentum flows in close binaries},
  volume  = {332},
  doi     = {10.1086/166645},
  journal = {\apj},
  author  = {Verbunt, Frank and Rappaport, Saul},
  month   = sep,
  year    = {1988},
  pages   = {193--198},
}

@article{villarroel2020,
  title     = {The {Vanishing} and {Appearing} {Sources} during a {Century} of {Observations} {Project}. {I}. {USNO} {Objects} {Missing} in {Modern} {Sky} {Surveys} and {Follow}-up {Observations} of a “{Missing} {Star}”},
  volume    = {159},
  doi       = {10.3847/1538-3881/ab570f},
  journal   = {\aj},
  publisher = {IOP},
  author    = {Villarroel, Beatriz and Soodla, Johan and Comerón, Sébastien and Mattsson, Lars and Pelckmans, Kristiaan and López-Corredoira, Martín and Krisciunas, Kevin and Guerras, Eduardo and Kochukhov, Oleg and Bergstedt, Josefine and Buelens, Bart and Bär, Rudolf E. and Cubo, Rubén and Enriquez, J. Emilio and Gupta, Alok C. and Imaz, Iñigo and Karlsson, Torgny and Prieto, M. Almudena and Shlyapnikov, Aleksey A. and de Souza, Rafael S. and Vavilova, Irina B. and Ward, Martin J.},
  month     = jan,
  year      = {2020},
  pages     = {8},
}

@article{virtanen2020,
  title   = {{SciPy} 1.0: fundamental algorithms for scientific computing in {Python}},
  volume  = {17},
  doi     = {10.1038/s41592-019-0686-2},
  journal = {Nature Methods},
  author  = {Virtanen, Pauli and Gommers, Ralf and Oliphant, Travis E. and Haberland, Matt and Reddy, Tyler and Cournapeau, David and Burovski, Evgeni and Peterson, Pearu and Weckesser, Warren and Bright, Jonathan and van der Walt, Stéfan J. and Brett, Matthew and Wilson, Joshua and Millman, K. Jarrod and Mayorov, Nikolay and Nelson, Andrew R. J. and Jones, Eric and Kern, Robert and Larson, Eric and Carey, C. J. and Polat, İlhan and Feng, Yu and Moore, Eric W. and VanderPlas, Jake and Laxalde, Denis and Perktold, Josef and Cimrman, Robert and Henriksen, Ian and Quintero, E. A. and Harris, Charles R. and Archibald, Anne M. and Ribeiro, Antônio H. and Pedregosa, Fabian and van Mulbregt, Paul and {SciPy 1. 0 Contributors}},
  month   = feb,
  year    = {2020},
  pages   = {261--272},
}

@misc{wang2019,
  title     = {An {On}-going {Mid}-infrared {Outburst} in the {White} {Dwarf} 0145+234: {Catching} in {Action} of {Tidal} {Disruption} of an {Exoasteroid}?},
  doi       = {10.48550/arXiv.1910.04314},
  publisher = {arXiv},
  author    = {Wang, Ting-Gui and Jiang, Ning and Ge, Jian and Cutri, Roc M. and Jiang, Peng and Sheng, Zhengfeng and Zhou, Hongyan and Bauer, James and Mainzer, Amy and Wright, Edward L.},
  month     = oct,
  year      = {2019},
}

@article{williams2024,
  title   = {{PEWDD}: {A} database of white dwarfs enriched by exo-planetary material},
  volume  = {691},
  doi     = {10.1051/0004-6361/202450509},
  journal = {\aap},
  author  = {Williams, J. T. and Gänsicke, B. T. and Swan, A. and O’Brien, M. W. and Izquierdo, P. and Cutolo, A.-M. and Cunningham, T.},
  month   = nov,
  year    = {2024},
  pages   = {A352},
}

@article{wilson2014,
  title   = {Variable emission from a gaseous disc around a metal-polluted white dwarf},
  volume  = {445},
  doi     = {10.1093/mnras/stu1876},
  journal = {\mnras},
  author  = {Wilson, D. J. and Gänsicke, B. T. and Koester, D. and Raddi, R. and Breedt, E. and Southworth, J. and Parsons, S. G.},
  month   = dec,
  year    = {2014},
  pages   = {1878--1884},
}

@article{wilson2019,
  title   = {The unbiased frequency of planetary signatures around single and binary white dwarfs using {Spitzer} and {Hubble}},
  volume  = {487},
  doi     = {10.1093/mnras/stz1050},
  journal = {\mnras},
  author  = {Wilson, Thomas G. and Farihi, Jay and Gänsicke, Boris T. and Swan, Andrew},
  month   = jul,
  year    = {2019},
  pages   = {133--146},
}

@article{wright2010,
  title   = {The {Wide}-field {Infrared} {Survey} {Explorer} ({WISE}): {Mission} {Description} and {Initial} {On}-orbit {Performance}},
  volume  = {140},
  doi     = {10.1088/0004-6256/140/6/1868},
  journal = {\aj},
  author  = {Wright, Edward L. and Eisenhardt, Peter R. M. and Mainzer, Amy K. and Ressler, Michael E. and Cutri, Roc M. and Jarrett, Thomas and Kirkpatrick, J. Davy and Padgett, Deborah and McMillan, Robert S. and Skrutskie, Michael and Stanford, S. A. and Cohen, Martin and Walker, Russell G. and Mather, John C. and Leisawitz, David and Gautier, III, Thomas N. and McLean, Ian and Benford, Dominic and Lonsdale, Carol J. and Blain, Andrew and Mendez, Bryan and Irace, William R. and Duval, Valerie and Liu, Fengchuan and Royer, Don and Heinrichsen, Ingolf and Howard, Joan and Shannon, Mark and Kendall, Martha and Walsh, Amy L. and Larsen, Mark and Cardon, Joel G. and Schick, Scott and Schwalm, Mark and Abid, Mohamed and Fabinsky, Beth and Naes, Larry and Tsai, Chao-Wei},
  month   = dec,
  year    = {2010},
  pages   = {1868--1881},
}

@article{xing2025,
  author  = {{Xing}, Zepei and {Torres}, Santiago and {G{\"o}tberg}, Ylva and {Trani}, Alessandro A. and {Korol}, Valeriya and {Cuadra}, Jorge},
  title   = "{Combining REBOUND and MESA: dynamical evolution of planets orbiting interacting binaries}",
  journal = {\mnras},
  year    = 2025,
  month   = feb,
  volume  = {537},
  number  = {1},
  pages   = {285-292},
  doi     = {10.1093/mnras/stae2820},
}

@article{xu2014,
  title   = {The {Drop} during {Less} than 300 {Days} of a {Dusty} {White} {Dwarf}'s {Infrared} {Luminosity}},
  volume  = {792},
  doi     = {10.1088/2041-8205/792/2/L39},
  journal = {\apj},
  author  = {Xu, S. and Jura, M.},
  month   = sep,
  year    = {2014},
  pages   = {L39},
}

@article{xu2016,
  title   = {Evidence for {Gas} from a {Disintegrating} {Extrasolar} {Asteroid}},
  volume  = {816},
  doi     = {10.3847/2041-8205/816/2/L22},
  journal = {\apj},
  author  = {Xu, S. and Jura, M. and Dufour, P. and Zuckerman, B.},
  month   = jan,
  year    = {2016},
  pages   = {L22},
}

@article{xu2018,
  title     = {Infrared {Variability} of {Two} {Dusty} {White} {Dwarfs}},
  volume    = {866},
  doi       = {10.3847/1538-4357/aadcfe},
  journal   = {\apj},
  publisher = {IOP},
  author    = {Xu, Siyi and Su, Kate Y. L. and Rogers, L. K. and Bonsor, Amy and Olofsson, Johan and Veras, Dimitri and van Lieshout, Rik and Dufour, Patrick and Green, Elizabeth M. and Schlawin, Everett and Farihi, Jay and Wilson, Thomas G. and Wilson, David J. and Gänsicke, Boris T.},
  month     = oct,
  year      = {2018},
  pages     = {108},
}

@article{xu2018a,
  title   = {A dearth of small particles in the transiting material around the white dwarf {WD} 1145+017},
  volume  = {474},
  doi     = {10.1093/mnras/stx3023},
  journal = {\mnras},
  author  = {Xu, S. and Rappaport, S. and van Lieshout, R. and Vanderburg, A. and Gary, B. and Hallakoun, N. and Ivanov, V. D. and Wyatt, M. C. and DeVore, J. and Bayliss, D. and Bento, J. and Bieryla, A. and Cameron, A. and Cann, J. M. and Croll, B. and Collins, K. A. and Dalba, P. A. and Debes, J. and Doyle, D. and Dufour, P. and Ely, J. and Espinoza, N. and Joner, M. D. and Jura, M. and Kaye, T. and McClain, J. L. and Muirhead, P. and Palle, E. and Panka, P. A. and Provencal, J. and Randall, S. and Rodriguez, J. E. and Scarborough, J. and Sefako, R. and Shporer, A. and Strickland, W. and Zhou, G. and Zuckerman, B.},
  month   = mar,
  year    = {2018},
  pages   = {4795--4809},
}

@article{xu2019,
  title   = {Shallow {Ultraviolet} {Transits} of {WD} 1145+017},
  volume  = {157},
  doi     = {10.3847/1538-3881/ab1b36},
  number  = {6},
  journal = {\aj},
  author  = {Xu, Siyi and Hallakoun, Na’ama and Gary, Bruce and Dalba, Paul A. and Debes, John and Dufour, Patrick and Fortin-Archambault, Maude and Fukui, Akihiko and Jura, Michael A. and Klein, Beth and Kusakabe, Nobuhiko and Muirhead, Philip S. and Narita, Norio and Steele, Amy and Su, Kate Y. L. and Vanderburg, Andrew and Watanabe, Noriharu and Zhan, Zhuchang and Zuckerman, Ben},
  month   = jun,
  year    = {2019},
  pages   = {255},
}

@article{xu2020,
  title   = {Infrared {Excesses} around {Bright} {White} {Dwarfs} from {Gaia} and {unWISE}. {I}.},
  volume  = {902},
  doi     = {10.3847/1538-4357/abb3fc},
  journal = {\apj},
  author  = {Xu, Siyi and Lai, Samuel and Dennihy, Erik},
  month   = oct,
  year    = {2020},
  pages   = {127},
}

@article{xu2024,
  author  = {{Xu}, Siyi and {Yeh}, Sherry and {Rogers}, Laura. K. and {Steele}, Amy and {Dennihy}, Erik and {Doyle}, Alexandra E. and {Dufour}, P. and {Klein}, Beth L. and {Manser}, Christopher J. and {Melis}, Carl and {Wang}, Tinggui and {Weinberger}, Alycia J.},
  title   = "{Modeling Circumstellar Gas Emission around a White Dwarf Using CLOUDY}",
  journal = {\aj},
  year    = 2024,
  month   = may,
  volume  = {167},
  number  = {5},
  eid     = {248},
  pages   = {248},
  doi     = {10.3847/1538-3881/ad3737},
}

@article{york2000,
  title   = {The {Sloan} {Digital} {Sky} {Survey}: {Technical} {Summary}},
  volume  = {120},
  doi     = {10.1086/301513},
  journal = {\aj},
  author  = {York, Donald G. and Adelman, J. and Anderson, Jr., John E. and Anderson, Scott F. and Annis, James and Bahcall, Neta A. and Bakken, J. A. and Barkhouser, Robert and Bastian, Steven and Berman, Eileen and Boroski, William N. and Bracker, Steve and Briegel, Charlie and Briggs, John W. and Brinkmann, J. and Brunner, Robert and Burles, Scott and Carey, Larry and Carr, Michael A. and Castander, Francisco J. and Chen, Bing and Colestock, Patrick L. and Connolly, A. J. and Crocker, J. H. and Csabai, István and Czarapata, Paul C. and Davis, John Eric and Doi, Mamoru and Dombeck, Tom and Eisenstein, Daniel and Ellman, Nancy and Elms, Brian R. and Evans, Michael L. and Fan, Xiaohui and Federwitz, Glenn R. and Fiscelli, Larry and Friedman, Scott and Frieman, Joshua A. and Fukugita, Masataka and Gillespie, Bruce and Gunn, James E. and Gurbani, Vijay K. and de Haas, Ernst and Haldeman, Merle and Harris, Frederick H. and Hayes, J. and Heckman, Timothy M. and Hennessy, G. S. and Hindsley, Robert B. and Holm, Scott and Holmgren, Donald J. and Huang, Chi-hao and Hull, Charles and Husby, Don and Ichikawa, Shin-Ichi and Ichikawa, Takashi and Ivezić, Željko and Kent, Stephen and Kim, Rita S. J. and Kinney, E. and Klaene, Mark and Kleinman, A. N. and Kleinman, S. and Knapp, G. R. and Korienek, John and Kron, Richard G. and Kunszt, Peter Z. and Lamb, D. Q. and Lee, B. and Leger, R. French and Limmongkol, Siriluk and Lindenmeyer, Carl and Long, Daniel C. and Loomis, Craig and Loveday, Jon and Lucinio, Rich and Lupton, Robert H. and MacKinnon, Bryan and Mannery, Edward J. and Mantsch, P. M. and Margon, Bruce and McGehee, Peregrine and McKay, Timothy A. and Meiksin, Avery and Merelli, Aronne and Monet, David G. and Munn, Jeffrey A. and Narayanan, Vijay K. and Nash, Thomas and Neilsen, Eric and Neswold, Rich and Newberg, Heidi Jo and Nichol, R. C. and Nicinski, Tom and Nonino, Mario and Okada, Norio and Okamura, Sadanori and Ostriker, Jeremiah P. and Owen, Russell and Pauls, A. George and Peoples, John and Peterson, R. L. and Petravick, Donald and Pier, Jeffrey R. and Pope, Adrian and Pordes, Ruth and Prosapio, Angela and Rechenmacher, Ron and Quinn, Thomas R. and Richards, Gordon T. and Richmond, Michael W. and Rivetta, Claudio H. and Rockosi, Constance M. and Ruthmansdorfer, Kurt and Sandford, Dale and Schlegel, David J. and Schneider, Donald P. and Sekiguchi, Maki and Sergey, Gary and Shimasaku, Kazuhiro and Siegmund, Walter A. and Smee, Stephen and Smith, J. Allyn and Snedden, S. and Stone, R. and Stoughton, Chris and Strauss, Michael A. and Stubbs, Christopher and SubbaRao, Mark and Szalay, Alexander S. and Szapudi, Istvan and Szokoly, Gyula P. and Thakar, Anirudda R. and Tremonti, Christy and Tucker, Douglas L. and Uomoto, Alan and Vanden Berk, Dan and Vogeley, Michael S. and Waddell, Patrick and Wang, Shu-i. and Watanabe, Masaru and Weinberg, David H. and Yanny, Brian and Yasuda, Naoki and {SDSS Collaboration}},
  month   = sep,
  year    = {2000},
  pages   = {1579--1587},
}

@article{zackay2016,
  title   = {Proper {Image} {Subtraction}—{Optimal} {Transient} {Detection}, {Photometry}, and {Hypothesis} {Testing}},
  volume  = {830},
  doi     = {10.3847/0004-637X/830/1/27},
  journal = {\apj},
  author  = {Zackay, Barak and Ofek, Eran O. and Gal-Yam, Avishay},
  month   = oct,
  year    = {2016},
  pages   = {27},
}

@article{zuckerman1987,
  title   = {Excess infrared radiation from a white dwarf - an orbiting brown dwarf?},
  volume  = {330},
  doi     = {10.1038/330138a0},
  journal = {\nat},
  author  = {Zuckerman, B. and Becklin, E. E.},
  month   = nov,
  year    = {1987},
  pages   = {138--140},
}

@article{zuckerman2003,
  title   = {Metal {Lines} in {DA} {White} {Dwarfs}},
  volume  = {596},
  doi     = {10.1086/377492},
  journal = {\apj},
  author  = {Zuckerman, B. and Koester, D. and Reid, I. N. and Hünsch, M.},
  month   = oct,
  year    = {2003},
  pages   = {477--495},
}

@article{britt2003,
       author = {{Britt}, D.~T. and {Consolmagno}, G.~J.},
        title = "{Stony meteorite porosities and densities: A review of the data through 2001}",
      journal = {Meteoritics and Planetary Science},
         year = 2003,
        month = aug,
       volume = {38},
       number = {8},
        pages = {1161-1180},
          doi = {10.1111/j.1945-5100.2003.tb00305.x},
       adsurl = {https://ui.adsabs.harvard.edu/abs/2003M&PS...38.1161B}
}

@misc{akeson2019,
  title     = {The {Wide} {Field} {Infrared} {Survey} {Telescope}: 100 {Hubbles} for the 2020s},
  doi       = {10.48550/arXiv.1902.05569},
  publisher = {arXiv},
  author    = {Akeson, Rachel and Armus, Lee and Bachelet, Etienne and others},
  month     = feb,
  year      = {2019},
  note      = {arXiv:1902.05569},
}

@ARTICLE{liu2021a,
       author = {{Liu}, Jifeng and {Soria}, Roberto and {Wu}, Xue-Feng and {Wu}, Hong and {Shang}, Zhaohui},
        title = "{The SiTian Project}",
      journal = {Anais da Academia Brasileira de Ciencias},
         year = 2021,
        month = apr,
       volume = {93},
          eid = {20200628},
        pages = {20200628},
          doi = {10.1590/0001-3765202120200628},
archivePrefix = {arXiv},
       eprint = {2006.01844},
 primaryClass = {astro-ph.IM},
       adsurl = {https://ui.adsabs.harvard.edu/abs/2021AnABC..93..628L}
}

\begin{appendix}

\section{Data}

\subsection{Negative alert sample vetting}\label{sec:app_vetting}
Here we describe in more detail how the \num{4027} white dwarfs with three or more negative alerts (Fig.~\ref{fig:negalerts}) were vetted to arrive at the final sample. Within this sample we first identified $\approx$500 periodically eclipsing white dwarfs, which are presented in Van Roestel et al. 2026; most of these triggered multiple alerts in the first year of ZTF. We excluded these and visually inspected the light curves of the remaining candidates to identify long-timescale variability, irregular transits, and single dimming events.

The majority of the remaining candidates are not genuine dimming events. Many are cataclysmic variables \citep[e.g.][]{szkody2020,szkody2021}, which we rejected based on their known classification or infrared excess. Bogus signals show up as light curves with positive and negative alerts randomly scattered in time, alerts in only one band, or light curves with a linear trend; we inspected a number of these in more detail and found them to be most often the result of data calibration or subtraction artefacts, or of CCD artefacts such as bad pixel columns. Finally, high proper motion white dwarfs ($\gtrsim$\qty{150}{mas/yr}) move significantly between the science image and the reference image, since ZTF reference images have remained static since the start of the survey in 2018; depending on the magnitude and proper motion, these tend to trigger alerts in the later years of the survey.

Any object that showed potentially astrophysically real, non-periodic variability was investigated in more detail. We inspected the light curves of nearby stars to reject variability induced by the calibration of the data, and we checked whether the variability could be due to the star being located in two separate ZTF fields, which can in rare cases produce spurious variability. In addition, we used archival data (Sect.~\ref{sec:archivaldata}) including the \textit{Gaia} DR3 colour-magnitude diagram, and rejected objects located above the white dwarf track and/or showing excess red light, since these are likely cataclysmic variables undergoing accretion disc state changes.

\subsection{Archival light curves}\label{sec:app_lcs}


\textbf{LINEAR} The Lincoln Near-Earth Asteroid Research (LINEAR) survey was designed for astrometric discovery of moving objects such as Near-Earth Objects \citep[][]{stokes2000}. It used two \qty[]{1.0}{m} telescopes with a \qty[]{2}{\deg^2} field of view in New Mexico, USA, and operated from 2000 to 2013 \citep{palaversa2015}. We obtained 625 LINEAR epochs for ZTF~J2327+0019, spanning 2002 to late 2012, and extracted photometry using a custom PSF method (Drake~et~al., in prep.).

\textbf{CRTS} The Catalina Sky Survey began in 2004 and uses three telescopes to repeatedly survey the sky between declinations \qtyrange[]{-75}{+65}{}
degrees in search of NEOs and Potentially Hazardous Asteroids (PHAs)\footnote{\url{https://catalina.lpl.arizona.edu/}}. 
We obtained CRTS variable star photometry data spanning from 2006 to 2025 for six white dwarfs. See \citet{drake2009} for details. We use a custom PSF pipeline which is calibrated to PS1-r (Drake~et~al., in prep.). 

\textbf{PTF} The Palomar Transient Factory and the intermediate Palomar Transient Factory \citep{law2009,rau2009} operated from 2009 to 2017. It observed the northern sky with various cadences in the $g$ and $R$-bands. We obtained PTF data for five white dwarfs, mostly in $R$-band. For PTF data, we use data with \textsc{qaflags} and \textsc{photcalflag} both equal to zero\footnote{\url{https://www.ptf.caltech.edu/system/media_files/binaries/30/original/Objects_SourcesTable_cols_v3.html}}. 

\textbf{ATLAS} The ATLAS project \citep{tonry2018} uses multiple telescopes to observe the night sky several times per night. It began operations in 2015 and observes in cyan and orange bands. We used the ATLAS forced photometry service to obtain forced photometry on science images for all objects \citep{shingles2021}. When the ATLAS signal-to-noise ratio was too low even when the white dwarf was at its nominal brightness, we excluded those data from the analysis and light curve figures. We filtered the ATLAS photometry by removing points with \textsc{Chi/N}~>~10 and \textsc{m}~>~10.

\textbf{\textit{WISE/NEOWISE}}
The Wide-field Infrared Survey Explorer (\textit{WISE}) satellite \citep{wright2010}, re-initiated as the \textit{NEOWISE} mission \citep{mainzer2011}, has been carrying out an all-sky MIR survey in the $W1$ and $W2$ bands from 2014 to 2024. We used a customised code \citep{de2020} based on the ZOGY algorithm \citep{zackay2016} to perform image subtraction on the \textit{NEOWISE} images using the co-added images of the \textit{WISE} mission (obtained in 2010-2011) as reference images.

\subsection{High-speed photometry}\label{sec:app_highspeed}

\textbf{CHIMERA} We used the Caltech HIgh-speed Multi-color camERA \citep[CHIMERA,][]{harding2016} to observe ZTF~J0313+5206, ZTF~J0436+5728, ZTF~J1944\textminus2451, and ZTF~J2327+0019 (Figs.~\ref{fig:chimera_J0313}, \ref{fig:chimera_J0436}, \ref{fig:chimera_J1944}, and \ref{fig:hs_J2327}). CHIMERA is a dual-channel photometer that uses frame-transfer, electron-multiplying CCDs mounted on the Hale 200-inch (\qty{5.1}{m}) Telescope at Palomar Observatory (CA, USA). We used the conventional amplifier and 2x2 binning to reduce the readout noise and readout time. Each of the images was bias subtracted and divided by twilight flat fields\footnote{\url{https://github.com/caltech-chimera/PyChimera}}.

\textbf{ULTRACAM} We obtained high-speed light curves of ZTF~J1944\textminus2451 and ZTF~J2327+0019 with ULTRACAM in 2025 (Figs.~\ref{fig:ULTRACAM_J1944} and \ref{fig:hs_J2327}). ULTRACAM is a triple-arm camera that uses frame-transfer CCDs mounted on the \qty{3.6}{m} New Technology Telescope \citep[NTT;][]{dhillon2007} located on La Silla (Chile). We obtained data in the $ugr$ super SDSS filters and used 2x2 binning. Data reduction includes bias and flatfield corrections, and differential light curves were produced using the ULTRACAM pipeline.

\textbf{PRISM} We obtained two nights of data on ZTF~J1946+5543 with PRISM (Fig.~\ref{fig:PRISM_J1946}). PRISM is mounted on the \qty{1.8}{m} Perkins telescope located at Anderson Mesa Station (AZ, USA). We used the BG40 filter with exposure times of \qty{40}{s} and \qty{60}{s}. Differential light curves were created using the HiPERCAM photometry pipeline.

\subsection{Follow-up spectra and reduction}\label{sec:app_spectra}

\textbf{LRIS} For seven targets, we obtained a single ID-spectrum with the \qty{10}{m} Keck\,I Telescope (HI, USA) and the Low Resolution Imaging Spectrometer (LRIS; \citealt{oke1995,mccarthy1998}).
We used the R600 grism for the blue arm ($R\approx1100$) and the R600/7500 grating for the red arm ($R\approx1400$). The wavelength range is approximately \qtyrange{3200}{10000}{\angstrom}. A standard long-slit data reduction procedure was performed with the \textsc{Lpipe} pipeline\footnote{\url{http://www.astro.caltech.edu/~dperley/programs/lpipe.html}} \citep{perley2019}.

\textbf{DBSP} We obtained a spectrum of ZTF~J0313+5206 and ZTF~J1944\textminus2451 with the Double-Beam Spectrograph \citep[DBSP,][]{oke1982} mounted on the Palomar 200-inch telescope. The resolution is approximately $R\approx1500$. The data were processed using an average bias and a normalised flat-field frame obtained during the same night. Wavelength calibration was performed using calibration lamp spectra obtained at the start of the night. For the blue arm, FeAr and for the red arm, HeNeAr arc exposures were taken. Data were reduced using \textsc{DBSP\_DRP}\footnote{\url{https://github.com/finagle29/DBSP_DRP}}, a data reduction pipeline based on \textsc{PypeIt} \citep{prochaska2020}.

\textbf{SDSS} We obtained spectra from the Sloan Digital Sky Survey (SDSS) archive for ZTF~J0744+3803, ZTF~J1046+2021, and ZTF~J2327+0019. SDSS is an imaging and spectroscopic survey using the \qty{2.5}{m} wide-angle optical telescope at Apache Point Observatory in New Mexico, United States \citep{york2000}. The spectra typically have a total integration time of 45--60 minutes depending on the observing conditions. The wavelength coverage of the spectra is continuous from about \qty{3800}{\angstrom} to \qty{9200}{\angstrom} with a resolution of R$\approx$1800 \citep{uomoto1999}.

\textbf{SDSS-V} We obtained three spectra from SDSS-V DR19 of ZTF~J0313+5206, ZTF~J0744+3803, and ZTF~J2327+0019. The optical spectra of SDSS-V are obtained with the BOSS spectrographs ($R \approx 2000$). The exposure time is typically \qtyrange[]{1.5}{3.0}{h}.

\textbf{DESI} We obtained three spectra from DESI~DR1, of ZTF~J0015+3038, ZTF~J2327+0019, and ZTF~J0744+3803. DESI on the Mayall \qty{4}{m} telescope at Kitt Peak National Observatory (KPNO) is a multi-object spectroscopic instrument. The typical resolution is R$\approx$3000. The exposure times are between \qtyrange[]{600}{900}{s}.

\textbf{DeVeny} We obtained a spectrum of ZTF~J1946+5543 with the DeVeny spectrograph mounted on the \qty{4.3}{m} Lowell Discovery Telescope (AZ, USA). We used the 300/4000 grating, giving a resolution of $R\approx2000$ over the \qtyrange{3600}{8000}{\angstrom} wavelength range. The data were reduced using a standard \textsc{PypeIt} reduction \citep{prochaska2020}.

\textbf{LAMOST} Two targets, ZTF~J1046+2021 and ZTF~J2327+0019, were observed by the LAMOST survey \citep{cui2012}. The resolution of the LAMOST spectra is R$\approx$1800 and covers the 3800-9000\AA\ wavelength range. The exposure time is 75 to 90 minutes.

\FloatBarrier

\section{Extra Tables and Figures}\label{sec:appendix_B}
\begin{figure*}[ht]
     \centering
     \begin{subfigure}[t]{0.48\textwidth}
         \centering
         \includegraphics[]{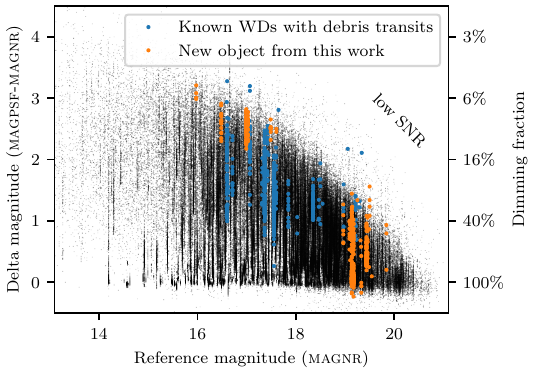}
         \caption{Difference between the measured alert magnitude (\textsc{magpsf}) and the measured magnitude on the reference image (\textsc{magnr}). The right y-axis shows the corresponding dimming fraction, with 100\% indicating the white dwarf completely vanished. This figure shows that for bright white dwarfs, alerts are generated for dimming events of just a few percent, while for faint white dwarfs alerts are only generated if the white dwarf fades significantly. The coloured dots show alerts by previously known and newly discovered white dwarfs with transiting debris.}
     \end{subfigure}\hfill
     \begin{subfigure}[t]{0.48\textwidth}
         \centering
         \includegraphics{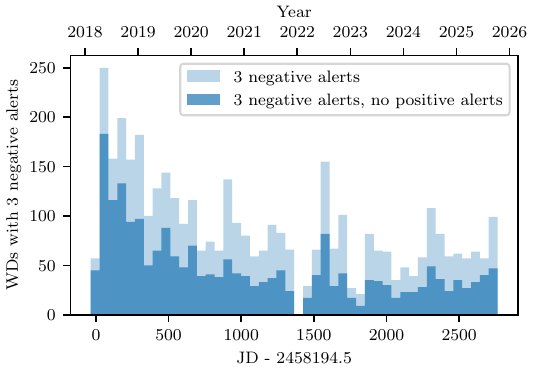}
         \caption{A histogram of when white dwarfs triggered a third negative alert. Bins are one month wide and show that typically $\approx$50 white dwarfs reach three negative alerts each month (\qty{4027} in total). The higher rate at the start is due to periodically variable white dwarfs, e.g. eclipsing white dwarfs. Later in the survey, white dwarfs with high proper motions start to trigger negative alerts. In addition, there are also subtraction artefacts that trigger bogus alerts.}
     \end{subfigure}
     \caption{An overview of negative ZTF alerts associated with white dwarf candidates from \citet{gentilefusillo2021a}.}
 \label{fig:negalerts}
 \end{figure*}

\begin{figure*}[htb]
  \setkeys{Gin}{width=\linewidth}
  \setlength\tabcolsep{2pt}
  \begin{tabularx}{\textwidth}{XXX}
    \includegraphics{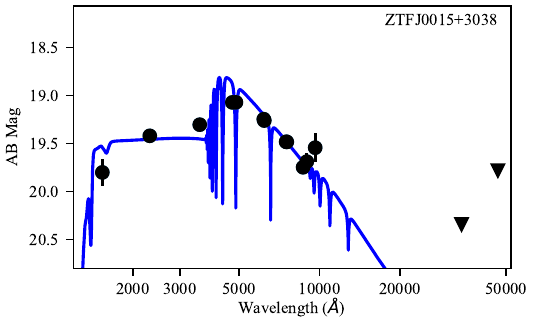} &
    \includegraphics{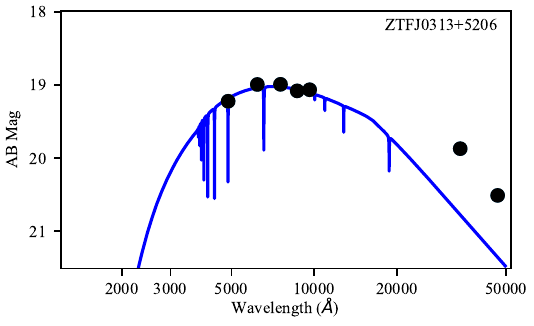} &
    \includegraphics{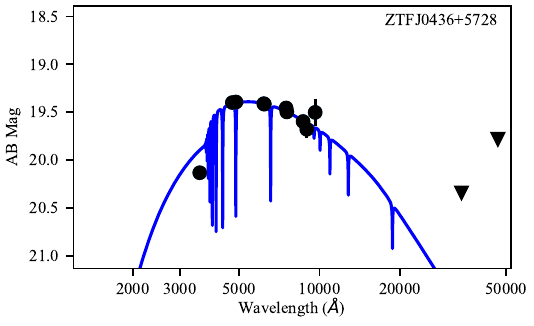} \\
    \includegraphics{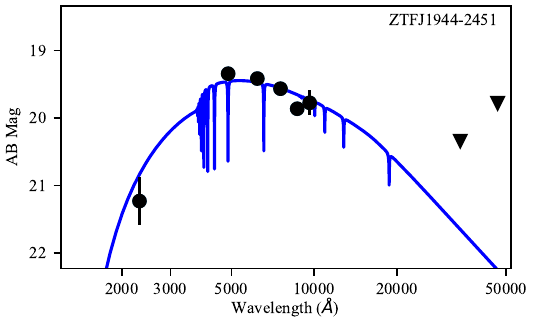} &
    \includegraphics{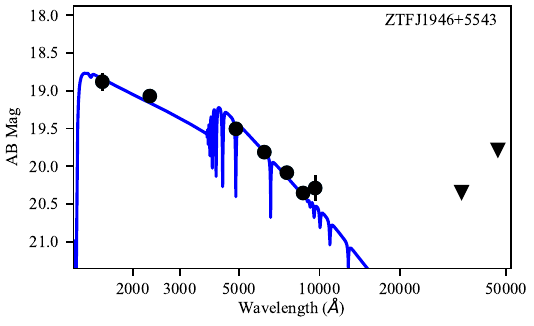} &
    \includegraphics{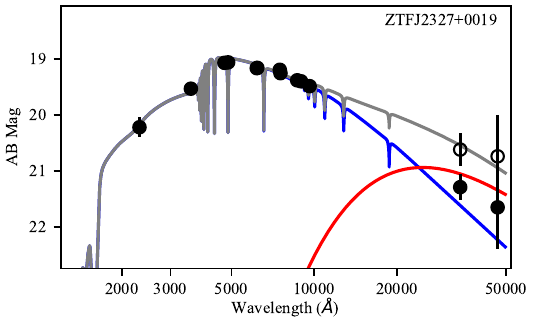} \\
    \includegraphics{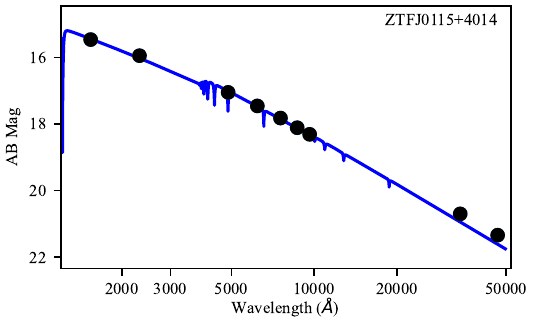} &
    \includegraphics{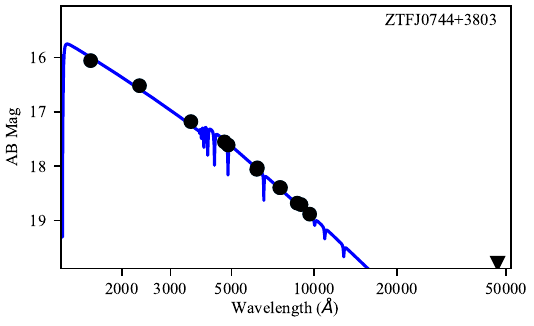} &
    \includegraphics{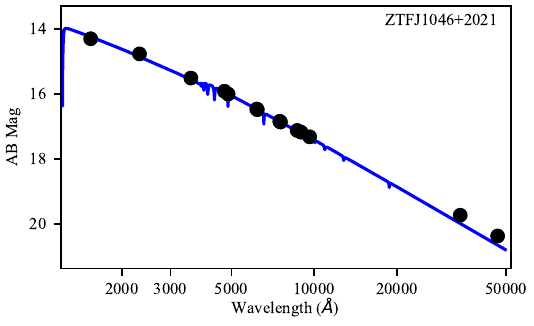}   \\
  \end{tabularx}
\caption{The spectral energy distribution of the sources presented in this paper with the best-fit SED model overplotted. The filled dots show archival measurements, and the triangles show upper limits. ZTF~J2327+0019 does not show any IR excess before the main obscuration event, while it does show a small amount of IR excess during the burst (open dots). The red line shows a simple dust disc model. Only a few of the other objects have \textit{WISE} IR detection which are all consistent with no-infrared excess. The apparent infrared excess for ZTF~J0313+5206 is likely due to a nearby star. Most objects only have upper limits in the IR that are not constraining.}
\label{fig:SEDs}
\end{figure*}

\begin{figure}
    \centering
    \includegraphics[]{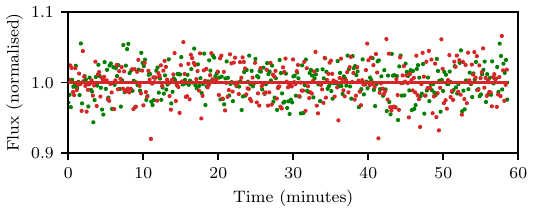}
    \caption{An hour long CHIMERA light curve of ZTF~J0313+5206 in $g$ and $r$ (green, red). No significant variability is detected in the light curve.}
    \label{fig:chimera_J0313}
\end{figure}

\begin{figure}
    \centering
    \includegraphics[]{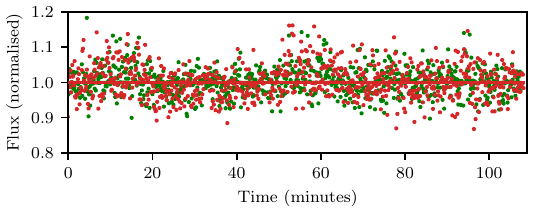}
    \caption{An 1.5 hour long CHIMERA light curve of ZTF~J0436+5728 in $g$ and $r$ (green, red). There seems to be a hint of variability, but this is likely not significant. }
    \label{fig:chimera_J0436}
\end{figure}

\begin{figure}
    \centering
    \includegraphics[]{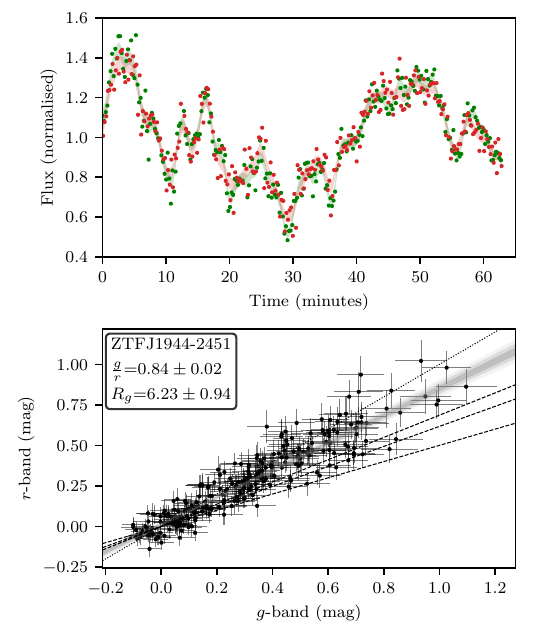}
    \caption{An hour long CHIMERA light curve of ZTF~J1944\textminus2451 in $g$ and $r$ (green, red). The coloured bands show independent Gaussian Process models with a Mat\'ern 3/2 kernel. The correlation timescale is \qtyrange[]{145}{165}{s}. The short variability timescale shows that debris crosses in front of the white dwarf quickly, which suggests a short orbital period on the order of hours. The bottom panel shows the correlation between $g$ and $r$ magnitudes (similar to Fig.~\ref{fig:colourtrends}), that suggests that the extinction in $g$ is slightly stronger than in $r$. The dotted lines show the expected trend for ISM-like dust with $R_V$=2.1,3.1, and 4.1}
    \label{fig:chimera_J1944}
\end{figure}

\begin{figure}
    \centering
    \includegraphics{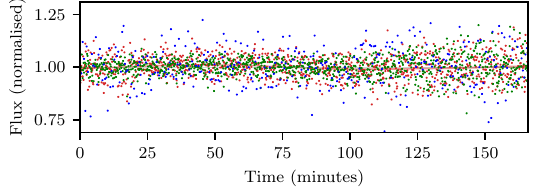}
    \caption{One night of high-speed ULTRACAM light curves of ZTF~J1944\textminus2451 in the $u$, $g$ and $r$ band (blue, green, red). Individual exposure times are \qty{10}{s} in the $g$ and $r$ bands, but the $u$ band exposure times are \qty{30}{s}. No obvious variability is detected in this data.}
    \label{fig:ULTRACAM_J1944}
\end{figure}

\begin{figure}
    \centering
    \includegraphics{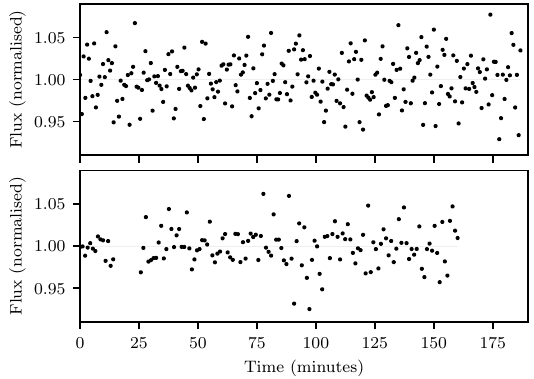}
    \caption{Two nights of high-speed PRISM light curves of ZTF~J1946+5543 with the BG40 filter. Individual exposure times are \qty{40}{s} in the first night, and \qty{60}{s} during the second night. The Gaussian process analysis shows there is no significant variability present in either light curve.}
    \label{fig:PRISM_J1946}
\end{figure}

\begin{figure}
    \centering
    \includegraphics{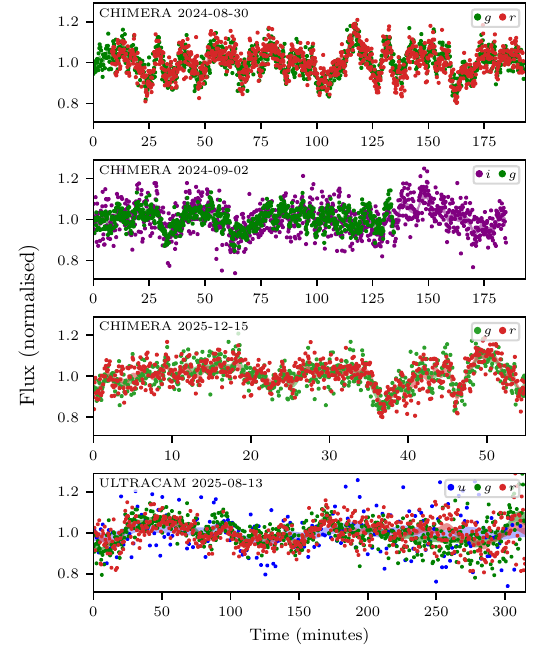}
    \caption{High-speed photometry of ZTF~J2327+0019. From top to bottom: CHIMERA on 2024-08-30 ($g$, $r$), CHIMERA on 2024-09-02 ($g$, $i$), CHIMERA on 2025-12-15 ($g$, $r$, using \qty{5}{s} exposures), and ULTRACAM on 2025-08-13 ($u$, $g$, $r$). Points show the data with the polynomial trend divided out; shaded bands show the Gaussian process model. The ULTRACAM points are binned to \qty{30}{s} ($g$, $r$) and \qty{90}{s} ($u$); the Gaussian process is fit to the unbinned data. The 2024 and 2025-12-15 CHIMERA data show irregular dips of 10--20\% on timescales as short as \qty{2}{min}, while the ULTRACAM data, taken when the white dwarf was back to its nominal brightness, show only minor fluctuations.}
    \label{fig:hs_J2327}
\end{figure}

\begin{figure*}[htb]
   \setkeys{Gin}{width=\linewidth}
   \setlength\tabcolsep{2pt}
   \begin{tabularx}{\textwidth}{XXX}
     \includegraphics[width=0.33\textwidth]{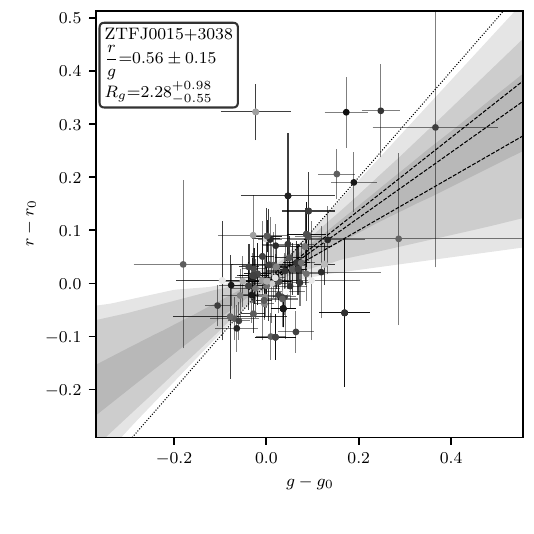} &
     \includegraphics[width=0.33\textwidth]{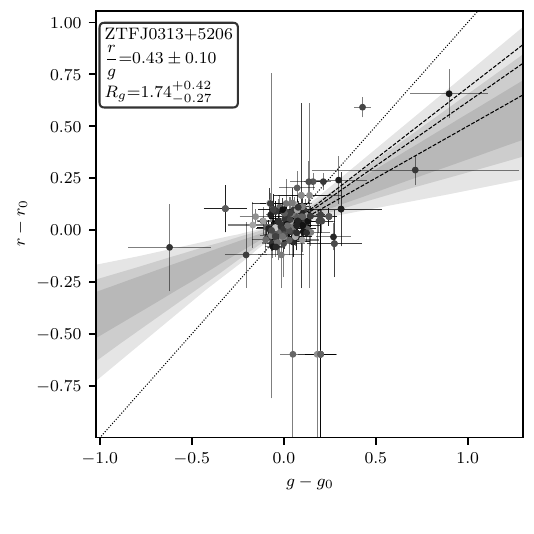} &
     \includegraphics[width=0.33\textwidth]{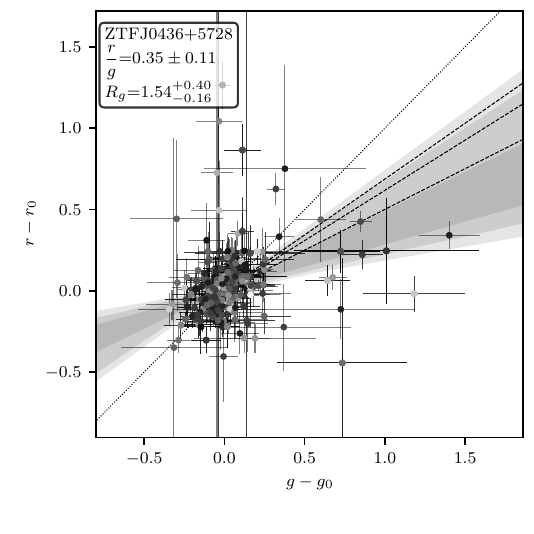} \\
     \includegraphics[width=0.33\textwidth]{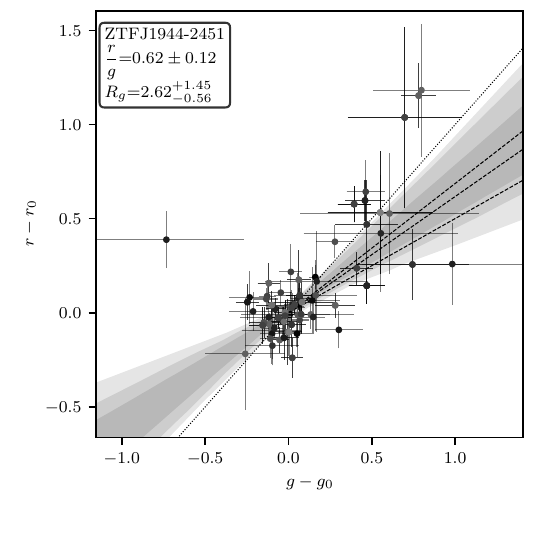} &
     \includegraphics[width=0.33\textwidth]{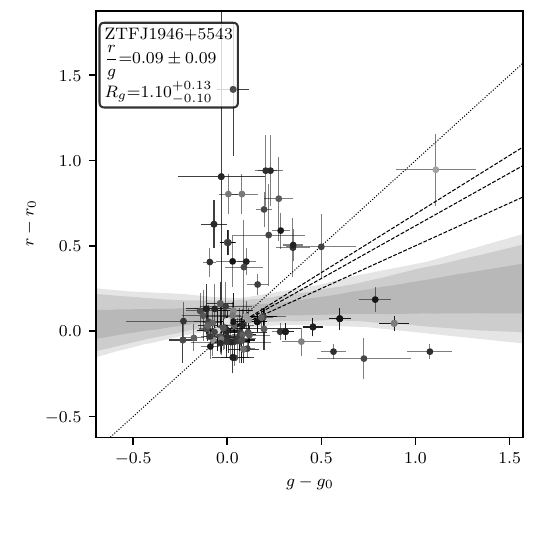}  &
     \includegraphics[width=0.33\textwidth]{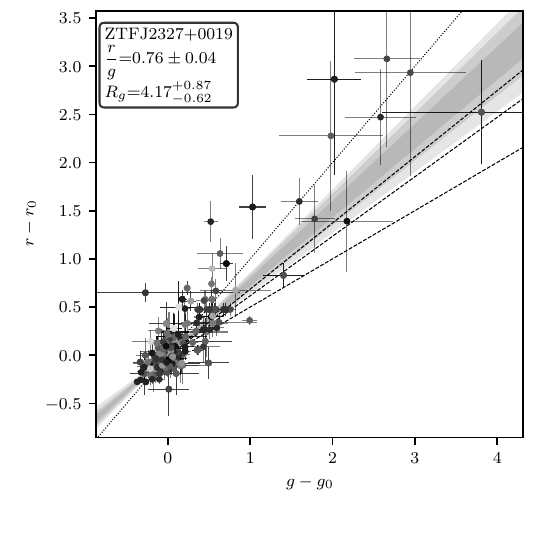} \\
     \includegraphics[width=0.33\textwidth]{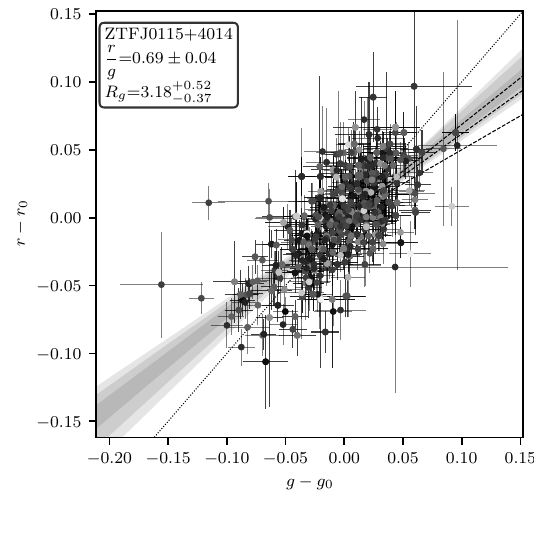} & 
     \includegraphics[width=0.33\textwidth]{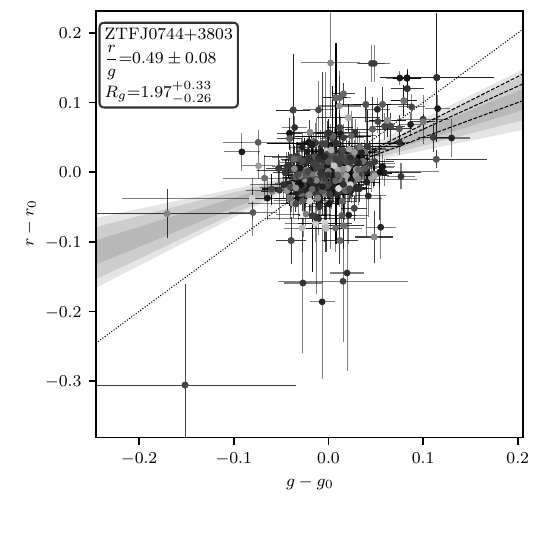} &
     \includegraphics[width=0.33\textwidth]{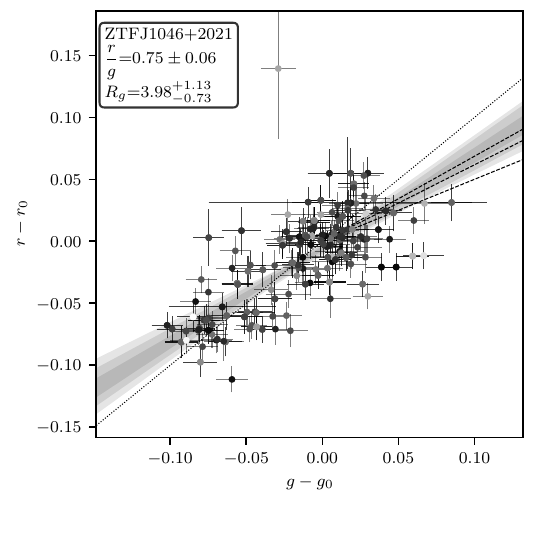}   \\
   \end{tabularx}
 \caption{The $g$ versus $r$ magnitudes of observations close in time. The greyscale indicates how close in time the observations were, with black points closer in time compared to white points. The dotted line shows the 1:1 line for perfectly grey extinction, while the dashed lines show extinction similar to interstellar dust with $R_V$ of 2.1, 3.1, and 4.1. The grey shaded region shows the 1, 2, and 3 standard deviation intervals of a linear fit to the data. Because white dwarfs with transiting debris can vary on timescales shorter than the typical time between $g$ and $r$ observations (typically 1--2 hrs), some of these figures are dominated by outliers.}
 \label{fig:colourtrends}
 \end{figure*}

Fig.~\ref{fig:negalerts} gives an overview of the negative ZTF alerts associated with white dwarf candidates, on which our target selection is based (Sect.~\ref{sec:targetselection}).
Table~\ref{tab:knownobject_overview} is the same as Table~\ref{tab:overview}, but lists the previously known white dwarfs that show debris transits. About half of the known systems were found in our search using ZTF alerts.
Fig.~\ref{fig:SEDs} shows the spectral energy distribution for each object. 
Figs.~\ref{fig:chimera_J0313}--\ref{fig:hs_J2327} show various high-speed photometry light curves.
Table~\ref{tab:spec_overview} shows an overview of all the follow-up data obtained, both spectroscopic and high-speed photometry.
In Table~\ref{tab:fit_pars}, we report the best-fit model values for the metallicity and light curves.
In Fig.~\ref{fig:colourtrends} we test for wavelength dependences to the variability observed by ZTF.
In Table~\ref{tab:J1944_parameters} we report the best-fit values for the spectrum of ZTF~J1944\textminus2451.

\begin{table*}[]
    \small 
    \centering
\caption{Properties of the 14 previously known white dwarfs with transiting debris, similar to Table~\ref{tab:overview}. The top part lists systems that we recovered in our search, and the bottom part lists known white dwarfs with transiting debris that we did not recover. Note that some do have detected ZTF alerts, indicated by the ZTF alert ID.}
    \renewcommand{\arraystretch}{1.2}
    \begin{tabular}{lllllllll}
    Name & ZTF alert ID & RA & Dec  & G-mag & $T_\mathrm{eff}$ & $\log g$ & Distance & Sp-type  \\
         &              & hh:mm:ss & dd:mm:ss  & Vega-mag & \unit{K} & \unit{cm.s^{-2}} & \unit{\parsec} &  \\
    \hline
    \hline
    \rule{0pt}{1ex} \\
    \multicolumn{9}{c}{Recovered known white dwarfs with debris transits} \\
    \hline
SDSS J0107+2107\tablefootmark{1} & ZTF18abvsplm & 01:07:49.30 & +21:07:44.48 & 19.11 & $\phantom{0}7590 \pm 800$ & $8.82\pm0.19$ & $\phantom{0}90.2^{+3.9}_{-3.5}$ & DAZ \\ 
ZTF~J0139+5245\tablefootmark{2} & ZTF18abkkzzw & 01:39:06.31 & +52:45:37.05 & 18.44 & $10170\pm 990$ & $7.97\pm0.25$ & $178^{+6}_{-6}$ & DA \\ %
ZTF~J0328\textminus1219\tablefootmark{1,4} & ZTF18abycutn & 03:28:33.66 & \textminus12:19:45.64 & 16.61 & $\phantom{0}8550 \pm 160$ & $8.46\pm0.06$ & $\phantom{0}43.3^{+0.2}_{-0.2}$ & DZ \\ 
ZTF~J0347\textminus1802\tablefootmark{1} & ZTF19abpnrti & 03:47:03.35 & \textminus18:02:53.51 & 17.39 & $13370 \pm 510$ & $8.88\pm0.04$ & $\phantom{0}76.4^{+0.8}_{-0.7}$ &DA \\ 
ZTF~J0923+4236\tablefootmark{1} & ZTF18aacdxxf & 09:23:11.39 & +42:36:33.70 & 17.14 & $13110 \pm 420$ & $8.25\pm0.04$ & $147^{+2}_{-2}$ & DAZ \\ 
WD~1145+017\tablefootmark{5} & ZTF19aaprzep & 11:48:33.54 & +01:28:59.40 & 17.43 & $15900\pm500$ & $8.10\pm0.11$ & $146^{+2}_{-2}$ & DBZ \\ 
SBSS~1232+563\tablefootmark{1,6} & ZTF23aaphnue & 12:34:32.58 & +56:06:42.84 & 18.04 & $11790\pm420$ & $8.30\pm0.06$ & $171^{+2}_{-2}$ & DBAZ \\ 
ZTF~J1302+1650\tablefootmark{3} & ZTF19aaczxaz & 13:02:17.67 & +16:50:08.43 & 19.18 & $18400 \pm 220$ & $8.19\pm0.04$ & $317^{+24}_{-25}$ & DBAZ \\ 
\rule{0pt}{1ex} \\
\multicolumn{9}{c}{Not recovered known white dwarfs with debris transits} \\
\hline
WD~J0923+7326\tablefootmark{3} & ZTF18acanduo & 09:23:51.44 & +73:26:23.68 & 18.24 & $13710\pm165$ & $8.25\pm0.04$ & $178^{+3}_{-4} $ & DAZ \\ 
WD~J1013\textminus0427\tablefootmark{3} & ZTF20aaelume & 10:13:06.34 & \textminus04:27:29.90 & 18.25 & $21900\pm50$ & $8.06\pm0.04$ & $305^{+22}_{-17}$ & DBAZe \\ 
WD~1054\textminus226\tablefootmark{7} & - & 10:56:38.63 & \textminus22:52:56.08 & 15.96 & $\phantom{0}7910\pm120$ & $8.05\pm0.08$ & $\phantom{0}36.11^{+0.05}_{-0.05}$ & DAZ \\ 
WD~J1237+5937\tablefootmark{3} & - & 12:37:54.23 & +59:37:27.29 & 20.12 & $\phantom{0}6480\pm75$ & $7.79\pm0.12$ & $192^{+21}_{-19}$ & DA \\ 
WD~J1650+1443\tablefootmark{3} & - & 16:50:50.01 & +14:43:22.46 & 18.80 & $10250\pm120$ & $8.22\pm0.14$ & $167^{+4}_{-4}$ & DAZ \\ 
WD~J1944+4557\tablefootmark{3} & ZTF18aazufsq & 19:44:31.88 & +45:57:53.22 & 19.36 & $20790\pm250$ & $8.27\pm0.05$ & $447^{+43}_{-39} $ & DAZ \\	 
\end{tabular}
\tablebib{The white dwarfs and their transits were first reported by:
(1)~\citet{guidry2021}; (2)~\citet{vanderbosch2020}; (3)~\citet{bhattacharjee2025};
(4)~\citet{vanderbosch2021}; (5)~\citet{vanderburg2015}; (6)~\citet{hermes2025};
(7)~\citet{farihi2022}.}
\label{tab:knownobject_overview}
\end{table*}

\begin{table}[]
     \centering
     \caption{An overview of all followup-data obtained, with spectra on top, and light curves at the bottom.}
     \footnotesize
     \begin{tabular}{llll}
         Object & Instrument & Date & Exp. time \\
         \hline
         ZTF~J0015+3038 & LRIS & 2023-05-23 & \phantom{0}600s \\
         ZTF~J0115+4014 & LRIS & 2021-02-12 & \phantom{0}300s \\
         ZTF~J0313+5206 & DBSP & 2023-09-14 & 1200s \\ 
         ZTF~J0436+5728 & LRIS & 2022-11-23 & 1200s \\
         ZTF~J0744+3803 & SDSS & 2010-10-11 & 4504s \\ 
         ZTF~J1046+2021 & SDSS & 2012-04-21 & 2702s \\
         ZTF~J1046+2021 & LAMOST & 2014-12-03 & 4500s \\
         ZTF~J1046+2021 & LAMOST & 2017-04-23 & 5400s \\
         ZTF~J1944\textminus2451 & DBSP & 2024-08-30 & 1800s \\
         ZTF~J1944\textminus2451 & LRIS & 2024-10-09 & 1200s \\
         ZTF~J1946+5543 & LRIS & 2026-05-12 & \phantom{0}900s \\
         ZTF~J1946+5543 & DeVeny & 2026-05-19 & 1800s \\
         ZTF~J2327+0019 & SDSS & 2001-10-17 & 3001s \\
         ZTF~J2327+0019 & LAMOST & 2014-11-21 & 5400s \\
         ZTF~J2327+0019 & LAMOST & 2018-10-11 & 1800s \\
         ZTF~J2327+0019 & DBSP & 2024-09-13 & 1800s \\
         ZTF~J2327+0019 & LRIS & 2024-10-09 & 1200s \\
         ZTF~J2327+0019 & LRIS & 2024-11-08 & 1200s \\
         ZTF~J2327+0019 & LRIS & 2025-06-30 & 1200s \\
         \hline
         ZTF~J0313+5206 & CHIMERA $g$+$r$ & 2024-12-30 & \phantom{0}350 x 10s \\
         ZTF~J0436+5728 & CHIMERA $g$+$r$ & 2022-11-30 & \phantom{0}800 x 8s\\
         ZTF~J1944\textminus2451 & CHIMERA $g$+$r$ & 2024-08-30 & \phantom{0}250 x 15s\\
         ZTF~J1944\textminus2451 & ULTRACAM $u$+$g$+$r$ & 2025-08-13 & 1022 x 10s\\

         ZTF~J1946+5543 & PRISM BG40 & 2025-10-21 & \phantom{0}250 x 40s\\
         ZTF~J1946+5543 & PRISM BG40 & 2025-10-22 & \phantom{0}139 x 60s\\
         ZTF~J2327+0019 & CHIMERA $g$+$r$ & 2024-08-30 & 1150 x 10s\\
         ZTF~J2327+0019 & CHIMERA $g$+$i$ & 2024-09-02 & 1100 x 10s\\
         ZTF~J2327+0019 & ULTRACAM $u$+$g$+$r$ & 2025-08-13 & 1938 x 10s\\
         ZTF~J2327+0019 & CHIMERA $g$+$r$ & 2025-12-15 & 650 x 5s\\
     \end{tabular}
     \label{tab:spec_overview}
 \end{table}

\begin{table*}[]
     \centering
     \caption{The modelled metallicity and light curve parameters. The first column shows the calcium abundance of the white dwarf, the next two columns show the ingress and egress timescale of the obscuration events, and latter two columns give the model depth and the deepest single detection. }
     \small
     \begin{tabular}{llllll}
         Object & $\log_{10} [\mathrm{Ca/H}]$ & $\tau_1$ (d) & $\tau_2$ (d) & Model depth & Max. depth  \\
         \hline
         ZTF~J0015+3038 & $\lesssim -5.5$ & $29\pm3$ & $\phantom{0}62\pm10$ & 25\% & \qty{40\pm8}{\%} \\ 
         ZTF~J0313+5206 & $\lesssim -7.0$ & $8.4\pm1.4$ & $\phantom{0}21\pm4$ & 39\% & \qty{60\pm4}{\%} \\ 
         ZTF~J0436+5728 & $\lesssim -5.5$ & 0 & $127\pm49$ & 74\% & \qty{87\pm4}{\%} \\
         ZTF~J1944\textminus2451 & see Table~\ref{tab:J1944_parameters} & $2.5\pm1.2$ & $437\pm75$ & 45\% & \qty{72\pm8}{\%} \\
         ZTF~J1946+5543 & $\lesssim -5.5$ & $18\pm4$ & $\phantom{0}28\pm4$ & 30\% & \qty{100\pm2}{\%} \\
         ZTF~J2327+0019 & $-6.34\pm0.15$ & $76\pm22$ & $317\pm114$ & 100\% & \qty{100}{\%} \\
     \end{tabular}
     \label{tab:fit_pars}
 \end{table*}

\begin{table}[]
    \centering
    \caption{The best fit DBZ model atmosphere parameters of white dwarf ZTF~J1944\textminus2451. }
    \begin{tabular}{ll}
    Model parameter & value \\
    \hline
    parallax & $3.66\pm0.14$ mas \\
    radial velocity & $59\pm4$ km s$^{-1}$ \\
    E(B-V) & $0.043\pm0.011$ mag \\
    $T_\mathrm{eff}$ & $9570\pm70$ K \\
    log(g) & $7.84\pm0.06$ cm s$^{-2}$\\
    log(Na/He) & $-6.96\pm0.18$ \\
    log(Mg/He) & $-5.76\pm0.04$ \\
    log(Ca/He) & $-6.98\pm0.05$ \\
    log(Fe/He) & $-5.98\pm0.04$ \\
    $\log{(\dot{M})}$  &  9.7--10.1 [\unit{g.s^{-1}}]\\
    $M_\mathrm{min}$ & $\approx 2 \times 10^{24}$ g
    \end{tabular}
    \label{tab:J1944_parameters}
\end{table}
\end{appendix}

\end{document}